\documentclass[aps,pra,twocolumn,groupedaddress,floatfix]{revtex4-1}

\usepackage{graphicx}
\usepackage{dcolumn}
\usepackage{bm}
\usepackage{amsmath}
\usepackage{xcolor}
\usepackage{braket}
\usepackage{float}

\usepackage{bm}
\usepackage[utf8]{inputenc}
\usepackage[T1]{fontenc}
\usepackage{hyperref}
\hypersetup{
breaklinks=true,
   colorlinks=true,
    linkcolor=blue,
    citecolor=blue,
    filecolor=magenta,
    urlcolor=cyan,
}

\usepackage{amsfonts}
\usepackage{blkarray}
\usepackage{amssymb}
\usepackage{multirow}
\usepackage{mathrsfs}

\newcommand{\cc}{{\cal C}}
\newcommand{\cg}{{\cal G}}
\newcommand{\cf}{{\cal F}}
\newcommand{\br}{{\bf r}}

\newcommand{\tz}{{\widetilde{\zeta}}}
\newcommand{\tcf}{{\widetilde{\cf}}}

\DeclareMathOperator{\sgn}{sgn}

\begin{document}

\title{Fractional quantum Hall physics and higher-order momentum correlations in few spinful 
fermionic contact-interacting ultracold atoms in rotating traps}

\author{Constantine Yannouleas}
\email{Constantine.Yannouleas@physics.gatech.edu}
\author{Uzi Landman}
\email{Uzi.Landman@physics.gatech.edu}

\affiliation{School of Physics, Georgia Institute of Technology,
             Atlanta, Georgia 30332-0430}

\date{14 June 2020}

\begin{abstract}
The fractional quantum Hall effect (FQHE) is theoretically investigated, with numerical and
algebraic approaches, in assemblies of a few spinful ultracold neutral fermionic atoms, interacting
via repulsive contact potentials and confined in a single rapidly rotating two-dimensional harmonic
trap. Going beyond the commonly used second-order correlations in the real configuration space, the
methodology in this paper will assist the analysis of experimental observations by providing
benchmark results for $N$-body spin-unresolved, as well as spin-resolved, momentum correlations
measurable in time-of-flight experiments with individual particle detection. Our analysis shows that
the few-body lowest-Landau-level (LLL) states with good  magic angular momenta exhibit inherent
ordered quantum structures in the $N$-body correlations, similar to those associated with rotating
Wigner molecules (WMs), familiar from the field of semiconductor quantum dots under high magnetic
fields. The application of a small perturbing stirring potential induces, at the ensuing avoided
crossings, formation of symmetry broken states exhibiting ordered polygonal-ring structures,
explicitly manifest in the single-particle density profile of the trapped particles. Away from the
crossings, an LLL state obtained from exact diagonalization of the microscopic Hamiltonian, found
to be well-described by a (1,1,1) Halperin two-component variational wavefunction, represents also a
spinful rotating WM. Analysis of the calculated LLL wavefunction enables a two-dimensional
generalization of the Girardeau one-dimensional 'fermionization' scheme, 
originally invoked for mapping of bosonic-type wave functions to those of spinless fermions.
\end{abstract}

\maketitle


\section{Introduction}

The discovery \cite{tsui82} of the fractional quantum Hall effect (FQHE) in extended (bulk)
electronic semiconductor samples of high purity, cooled down to low temperatures, and subjected to
high perpendicular magnetic fields gave rise to a new subfield in condensed-matter physics,
resulting in a large number of both experimental and theoretical investigations of correlated states
of interacting electronic systems exhibiting emergent topological phases of matter. Among the
theoretical approaches, we note in particular those based on the introduction of families of
variational wave functions in the lowest Landau level (LLL) (see, e.g., Refs.\
\cite{laug83.1,laug83.2,halp83,jain89,yann02}), following Laughlin's seminal publications
\cite{laug83.1,laug83.2}. 

The unprecedented experimental advances achieved recently in the area of trapped ultracold neutral
atoms generated intense interest in finite-size bosonic analogs
\cite{wilk00,coop03,popp04,barb06,baks07,coop08,hazz08,geme10} of the FQHE, being embodied in clouds
of a few ultracold atoms trapped in rotating harmonic traps, with the rotation acting as a synthetic
(rotational gauge) magnetic field.  

The expansion of the horizon of such LLL investigations to encompass the regime of a few ultracold
spinful fermionic atoms is a natural undertaking. Theoretical investigations of such an endeavor, 
presented in this paper, are further supported by a growing number of recent experimental advances 
\cite{joch11,joch12,berg19,prei19,bech20} in the deterministic control of assemblies of a few ($N$)
trapped $^6$Li atoms, and in particular by the anticipated implementation \cite{palm18} of a single
rapidly rotating two-dimensional (2D) harmonic trap able to project the few-body wave functions
within the LLL Hilbert space.

We use state-of-the-art computational tools based on exact diagonalization of the microscopic
Hamiltonian with the use of the full configuration interaction (CI) methodology, as was adapted to
two dimensions \cite{yann03,yues07,yann04,ront06,blun10}, in contrast to the familiar
three-dimensional (3D) CI chemistry formalism \cite{szabobook}; indeed, this approach has been
proved successful in previous studies of few bosons \cite{baks07} or electrons
\cite{yann03,yann11} in the LLL. Our study will assist the analysis of experimental observations 
by providing benchmark results for $N$-body (spin-unresolved, as well as spin-resolved) momentum 
correlations that can be measured directly with time-of-flight (TOF) protocols employing individual 
particle detection through fluorescent imaging in free space \cite{joch18,berg19,prei19,bech20}. 
Such research endeavors aim at revealing the microscopic structure of the correlated FQHE states 
(here for contact-interacting spinful fermions), adding, supplementing, and going beyond the 
information gained from studies of bulk properties, e.g., Hall resistance. In this respect, the 
approach in this paper, demonstrated earlier for a few bosons in Ref.\ \cite{baks07}, goes beyond 
the common theoretical analyses that are based on second-order correlations in the real 
configuration space \cite{laug83.2,yann03,barb06}. 

The main issues discussed and analyzed in this work include the following:\\
\indent 
(1) The formation of few-body LLL states with magic 
angular momenta \cite{girv83,maks90,yann07} exhibiting intrinsic ordered structures in the $N$-body 
correlations. Such ordered quantum structures, referred to commonly as rotating Wigner molecules 
(RWMs) \cite{maks96,yann03,yann07}, have been seen previously in the case of semiconductor quantum 
dots (electrons) under high magnetic fields. Here they are shown to appear even in the case of ultracold 
contact interacting fermionic atoms confined in a rotating trap.\\
\indent
(2) The application of a small perturbing stirring potential, $V_P$, to the rotating trap, as
described in experimental protocols \cite{popp04,hazz08,geme10}, where this perturbation enables
transition between good-total-$L$ states. In the presence of $V_P$, symmetry-broken states
(referred to as pinned Wigner molecules \cite{yann11,roma09}) emerge in the neighborhood of ensuing 
avoided crossings, exhibiting the ordered structures already at the lowest level of the first-order 
correlations, i.e., the single-particle densities.\\
\indent
(3) A CI-calculated LLL state, corresponding to 
Halperin's (1,1,1) variational wave function \cite{halp83}, is shown to provide an example of both
a rotating Wigner molecule and of a generalization to two dimensions of Girardeau's one-dimensional 
``fermionization'' scheme \cite{gira60,joch12}, originally invoked 
for designating the mapping of bosonic-type wave functions to those of spinless fermions.

The above theoretical predictions can be explicitly tested through analysis of experimentally
determined momentum correlations; that is, including up to $N$th-order correlation functions
obtained for $N$ fermionic atoms confined in the rotating trap via time-of-flight measurements. 

\subsection{Plan of paper}

The plan of the paper is given below. Following this Introductory section, we present in Sec. \ref{thpr}
theoretical preliminaries which aim at defining the problem, establishing notations, and giving a brief 
survey of the methodologies and techniques employed in this study. In more detail:  Sec.\ \ref{mbhm} 
presents the microscopic many-body Hamiltonian of ultracold fermionic atoms (here 4 $^6$Li) confined in 
a rapidly rotating (stirred up) trap, with or without a perturbation, $V_P$, that breaks the cylindrical 
symmetry of the trap. Sec.\ \ref{ci} describes the configuration interaction method used to obtain numerical
solutions of the Hamiltonian via exact diagonalization of the microscopic Hamiltonian, with illustrations of
the effect of the perturbation, resulting in avoided crossings, depending on the strength of $V_P$, between 
neighboring eigenstates of the unperturbed Hamiltonian; in most calculations demonstrated in this paper,
we consider a $V_P$ perturbing stirring potential of hexadecapolar symmetry  in the many-body Hamiltonian.

Sec.\ \ref{anal} discusses the tools of analysis 
used in this investigation, in particular, the spin-unresolved and spin-resolved correlation functions, that
is, single-particle, first-order, correlation function (i.e., CI-single-particle density), and higher-order 
(up to 4th-order) correlation functions in real coordinate space. Sec.\ \ref{analmom} describes these 
tools of analysis in the momentum space, that is, it discusses the
Fourier transforms of the real-space correlation functions, as measured in TOF measurements of particles 
propagating in free space after confinement removal. 

In Sec.\ \ref{spec}, we discuss the LLL spectra and the concept of magic angular momenta, including the
combined effects of the rotational and spin degrees of freedom; a group theoretical discussion of the 
geometrical-symmetry origins of the magic angular momenta sequences can be found in Appendix 
\ref{a1}. Sec.\ \ref{trcr} is devoted to analysis of the  properties of the ground-state in  
the spin sector ($S = 0$, $S_z = 0$) of the 4 $^6$Li trapped and rotating atoms while traversing  
an avoided crossing, originating from the symmetry-breaking perturbation $V_P$. This includes illustration 
of the formation of a pinned crystalline-ordered (square) symmetry-broken single-particle density, 
revealed in the single-particle density three-dimensional surface plots; see Sec.\ \ref{s0sz01st}. Away from 
the avoided crossing, the circular symmetry of the single-particle density is automatically reestablished,
and the crystalline order becomes intrinsic and hidden, but it can still be revealed in the $N$th-order 
(here 4th-order) correlation function (Sec.\ \ref{s0sz04th}). Second--order correlations are discussed in 
Sec.\ \ref{s0sz02nd}, and the spin structure of the ground state is analyzed in Sec.\ \ref{spnstr}. These 
results illustrate  the formation of a quantum ultracold rotating-Wigner molecule (UC-RWM) of square 
symmetry, and have been obtained for the case of a hexadecapolar stirring potential. In Sec.\ 
\ref{quad}, we  illustrate the formation of an UC-RWM for the case of $N=4$ for a quadrupolar trap
deformation. The resulting molecule is shown to be closely similar to the one obtained for the 
hexadecapolar perturbation, even though the  symmetry of the quadrupolar stirring potential does not 
coincide with the square symmetry.   

One of the main foci of this work is addressed in Sec. \ref{ferm}, namely examination of the  generalization 
of the Laughlin wave function by Halperin to include FQHE spinful (non-spin-polarized) configurations. To 
this end, we concentrate our discussion on the spin sector $(S=2, S_z=0)$ of the four $^6$Li ultracold atoms
in the rotating trap, and compare the predictions of our exact diagonalization CI calculations for the 
structure of the ground state in this sector with that of the Halperin (1,1,1) trial function. 
The presentation in Sec.\ \ref{ferm} includes three subsections: Sec.\ \ref{s2sz04th}: 
The 4th order correlation and the molecular configuration predicted for the ground state of above 
spin sector by the CI calculation. Sec.\ \ref{s2sz0comp}: Comparison between CI state and trial (1,1,1) 
Halperin wave function. Sec.\ \ref{s2sz02nd}: Examination of the limitations of analysis of the CI wavefunction 
in the ($S = 2$, $S_z = 0$) spin sector when using 2nd-order correlations [particularly for angular momenta
corresponding to the (1,1,1) Halperin state], showing the advantages offered by the $N$-body correlation 
function (here $N = 4$). 
 
 In Sec.\ \ref{fermanal}, we  discuss and illustrate  a fermionization analogy in two-dimensions, enabled by
derivation of appropriate analytic expressions for the calculated exact CI wave function. This is done for 
the angular momentum $L=6$ for $N=4$ fermions in Sec.\ \ref{fermn4}, and for $L=15$ for $N=6$ fermions
in Sec.\ \ref{fermn6}

In Sec.\ \ref{wigp}, we pause to discuss a comparison between the Wigner parameter $R_W$ --- specifying the 
interparticle interaction strength, and used to define the regime of formation of crystal-like-ordered 
geometric configurations (that is quantum Wigner-molecule formations) for confined particles 
interacting via sufficiently long-range interactions (such as Coulomb-interacting electrons in quantum 
dots, or the forces between trapped atoms interacting via dipolar interaction potentials) ---
and the parameter $R_\delta$ used as the strength of short-range contact interactions between trapped 
neutral ultracold atoms in fastly rotating traps (that is in the LLL regime). We summarize in Sec.\ \ref{summ}. 

\section{Theoretical Preliminaries}
\label{thpr}

\begin{figure}[t]
\includegraphics[width=7.5cm]{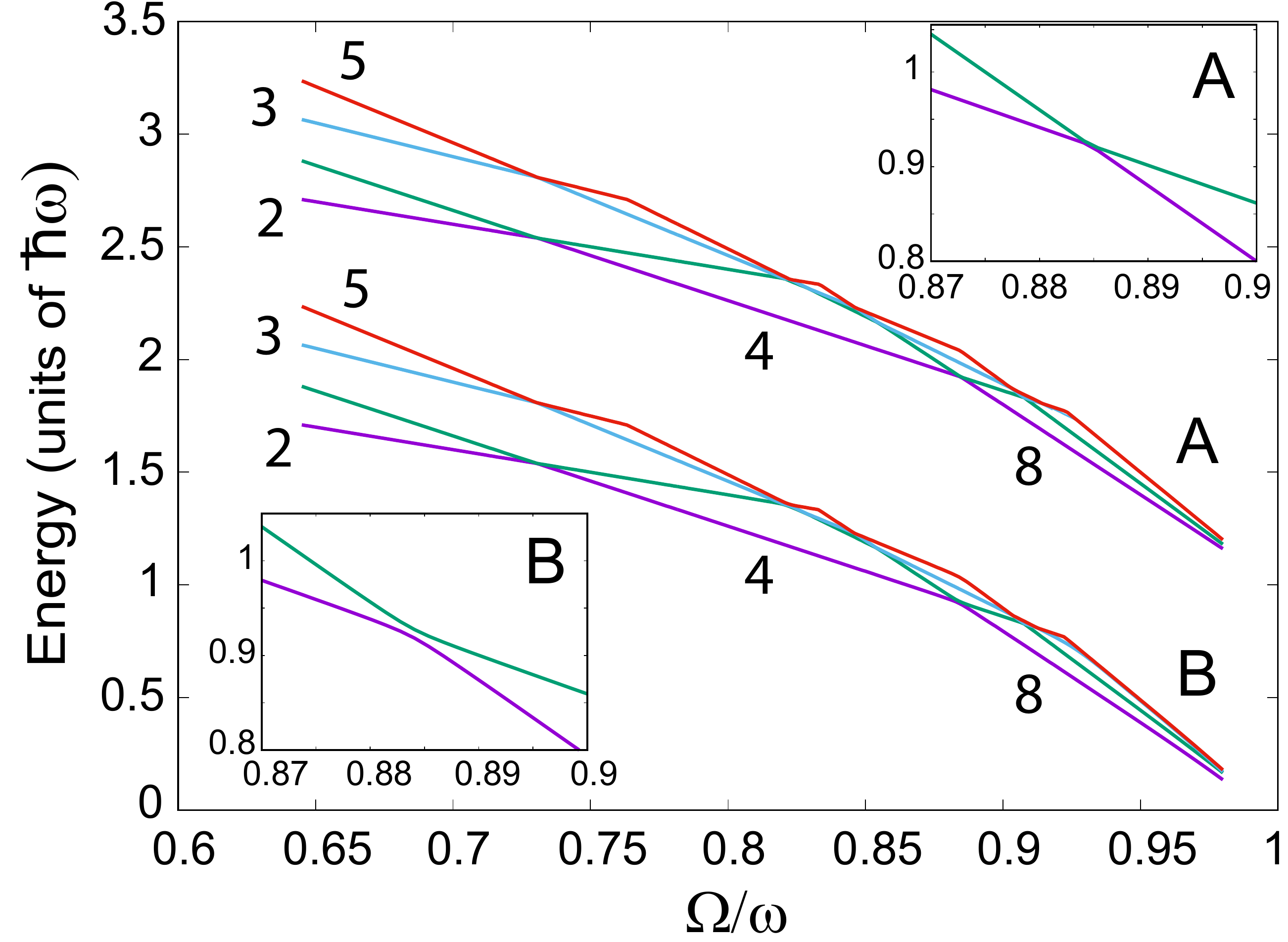}
\caption{Spectra of the four lowest-in-energy CI solutions of the many-body Hamiltonian $H_{\rm MB}$ 
[Eq.\ (\ref{hmb})] for $N=4$ fermions as a function of the ratio $\Omega/\omega$ (the constant 
energy $N\hbar\omega$ has been subtracted). The spin sector with ($S=0$, $S_z=0$) is displayed. 
$R_\delta=0.4$ and $m=4$ (hexadecapole trap deformation). Label A corresponds to $\cc=0.0001$ (weak 
$V_P$ perturbation). Label B corresponds to $\cc=0.004$ (strong $V_P$ perturbation). The energies 
associated with label A have been shifted upwards by 1$\hbar \omega$. The integers next to the 
curves denote the corresponding ideal good total angular momenta in the absence of any perturbation 
($V_P=0$). In the relative ground state (lowest-in-energy state within the spin sector), only the 
magic angular momenta 2, 4, and 8 appear (see Sec.\ \ref{spec}). The non-zero value of $\cc=0.004$ 
generates a visible (stronger) avoided crossing in the neighborhood of $\Omega/\omega \sim 0.884$, 
which is magnified in the inset labeled B. For the much smaller value of $\cc=0.0001$, this avoided 
crossing is minute, as seen from the inset labeled A (no energy shift). Overall, with naked eye, the
two spectra A and B are difficult to differentiate. However, the effect of a weak versus strong 
perturbation is pronounced on the properties (correlations) of the many-body wave functions when 
traversing the avoided crossing, as it is discussed in Sect.\ \ref{s0sz01st}. 
}
\label{avcr}
\end{figure}    

\subsection{Many-body Hamiltonian}
\label{mbhm}

The Fock-Darwin spectrum \cite{fock,darw} associated with the $(n,l)$ single-particle states in a rapidly 
rotating two-dimensional (2D) trap is given by \cite{note1} 
\begin{align}
\epsilon^{\rm FD}_{n,l}=\hbar [(2n+|l|+1) \omega - l \Omega], 
\label{spfd}
\end{align}
where $\omega$ is the trapping frequency of the harmonic confinement, $\Omega$ in the rotational frequency
of the trap, $n$ is the number of nodes, and $l$ denotes the single-particle angular momentum. 
The projection on the LLL imposes $n=0$ (Fock-Darwin single-particle states without radial nodes), and the 
associated many-body Hamiltonian without the perturbing contribution is \cite{note2}:
\begin{equation}
\frac{ H_{\rm LLL} }{\hbar \omega}=N + (1-\frac{\Omega}{\omega}) L
+2 \pi R_\delta \sum_{i < j}^{N} \delta({\bf r}_i-{\bf r}_j),
\label{H_lll}
\end{equation}
where $N$ is the number of particles, and $L$ denotes the total angular momentum, $L=\sum_{i=1}^N l_i$, 
along the axis perpendicular to the 2D trap plane. The energies in Eq.\ (\ref{H_lll}) are in units of 
$\hbar \omega$ and the lengths in units of the oscillator length $\Lambda=\sqrt{ \hbar/(M\omega) }$. Here,
the first and second terms express the LLL kinetic-energy contribution, $H_K$, whereas the third term 
represents the contact-interaction contribution, $H_{\rm int}$.

The dimensionless parameter 
\begin{align}
R_\delta=\frac{g}{2 \pi \Lambda^2 \hbar\omega}=\frac{gM}{2\pi\hbar^2}
\label{rdlt}
\end{align}
expresses the strength, $g$, of the coupling constant associated with an area $2\pi\Lambda^2$, 
relative to the zero-point energy, $\hbar\omega$, associated with the 2D harmonic trap; $M$ is the 
mass of the ultracold fermionic atoms. Naturally, the $\delta$-functions in Eq.\ (\ref{H_lll}) are 
two dimensional.

For the quasi-2D traps realized in experiments, the coupling constant $g$, as a function of the
3D scattering length $a_s$ and the oscillator length $l_z=\sqrt{\hbar/(M \omega_z)}$ in the tight
transverse direction, can be calculated numerically by considering the $s$ channel in a 
two-particle scattering problem \cite{shly00}. When $a_s$ is smaller than $l_z$, the analytic 
expression, $R_\delta=\sqrt{2/\pi}a_s/l_z$ \cite{note12}, can also be derived using the expression for the
coupling constant in Eq.\ (11) of Ref.\ \cite{shly00}; see also Eq.\ (1.122) in Ref.\ \cite{astr04}.  
Experimentally, the 3D scattering length can be varied over a wide range with the use of the 
Feshbach resonance; e.g., for $^6$Li atoms, see Refs.\ \cite{zuer12,juli14}. We note that
 the organization of the LLL trapped atoms in geometric structures of particular symmetries  (that is, 
 the  formation of LLL ultracold-atom Wigner molecules, discussed below) is independent of the 
 precise value  of $R_\delta$; for further discussion, see Sec.\ \ref{wigp}.

Another way for interpreting the parameter $R_\delta$ is that it equals the direct matrix element
[see Eqs.\ (\ref{psir}), (\ref{v1234}), and (\ref{dltme}) below]
\begin{align}
e_{\rm max}=\langle l_1=0,l_2=0|H_{\rm int}|l_3=0,l_4=0 \rangle=R_\delta.
\label{emax}
\end{align}
$e_{\rm max}$ is in units of $\hbar \omega$ and the subscript 'max' indicates that this energy 
represents the maximum repulsion that two fermions with opposite spins can attain in the LLL Hilbert
space. Since the energy gap between the lowest and the first-excited Landau levels is 
$2\hbar \omega$ (see the Appendix in Ref.\ \cite{yann07}), the condition for validating the 
projection of the few-body problem in the LLL is $R_\delta < 2$. 
In the following, for all 
calculations, we use a value of $R_\delta=0.4$ \cite{note13}.

Adding a small perturbation $V_P$, the total many-body Hamiltonian becomes
\begin{align}
H_{\rm MB}=H_{\rm LLL}+V_P.
\label{hmb} 
\end{align}

Traditionally, a $V_P$ perturbation or its effects have not been considered in the literature of 
the electronic FQHE (see, e.g., Refs.\ \cite{laug83.1,laug83.2,halp83,jain89,yann02}), with the 
exceptions of Refs.\ \cite{yann11} and \cite{roma09} in the context of disorder effects in the 
semiconductor sample and on the edge states in graphene quantum dots, respectively. A tunable 
$V_P$ perturbation representing a multipole deformation of the shape of the rotating harmonic trap
[see Eq.\ (\ref{vp}) below] has been proposed in Refs.\ \cite{popp04,hazz08} as the building block
of protocols for experimentally controlled assemblies of a few ultracold bosonic atoms enabling 
simulations of states characterized by well-known trial FQHE states, like the bosonic Laughlin 
ones. A proposal to use this type of perturbation in order to simulate well-known variational 
spinful FQHE states with ultracold $^6$Li atoms has been advanced in Ref.\ \cite{palm18}. To this 
effect, consideration of energy spectra and spatial correlations up to second order was 
sufficient. By considering higher-order correlations (both spatial and momentum ones) beyond the 
second order, and investigating the spontaneous symmetry breaking induced by the $V_P$ 
perturbation in the regions of the avoided crossings, we focus here on previously 
unexplored fundamental properties of the many-body LLL states of a finite-size assembly of spinful,
contact-interacting ultracold atoms.

For reasons of experimental convenience in transitioning from one LLL state to another, it has been shown 
\cite{popp04,hazz08,palm18} that the following perturbation (in second quantization), associated with a 
small multipole deformation of the rotating trap, is desirable:
\begin{align}
\frac{V_P}{\hbar \omega} = \cc \left( \sum_l \frac{ \sqrt{(l+m)!} }{ 2^{m/2} \sqrt{l!} }
a^\dagger_{l+m}a_l + H.c. \right),
\label{vp}
\end{align} 
where $m$ is the order of the multipole deformation, and $\cc$ is a dimensionless constant specifying the
strength of the deformation. This perturbation can be introduced as a stirring 
potential. It couples the many-body solutions of $H_{\rm LLL}$ that differ by $m$ units in their total
angular momenta $L$, and generates avoiding crossings, with an example given in Fig.\ \ref{avcr}.
Note that in showing the spectra of $H_{\rm MB}$, we limit ourselves to a particular spin sector; in 
Fig.\ \ref{avcr} for $N=4$, the spin sector is ($S=0$, $S_z = 0$), with the lowest-in-energy state
within the spin sector termed ``the relative ground state''.

\begin{figure}[t]
\includegraphics[width=7.5cm]{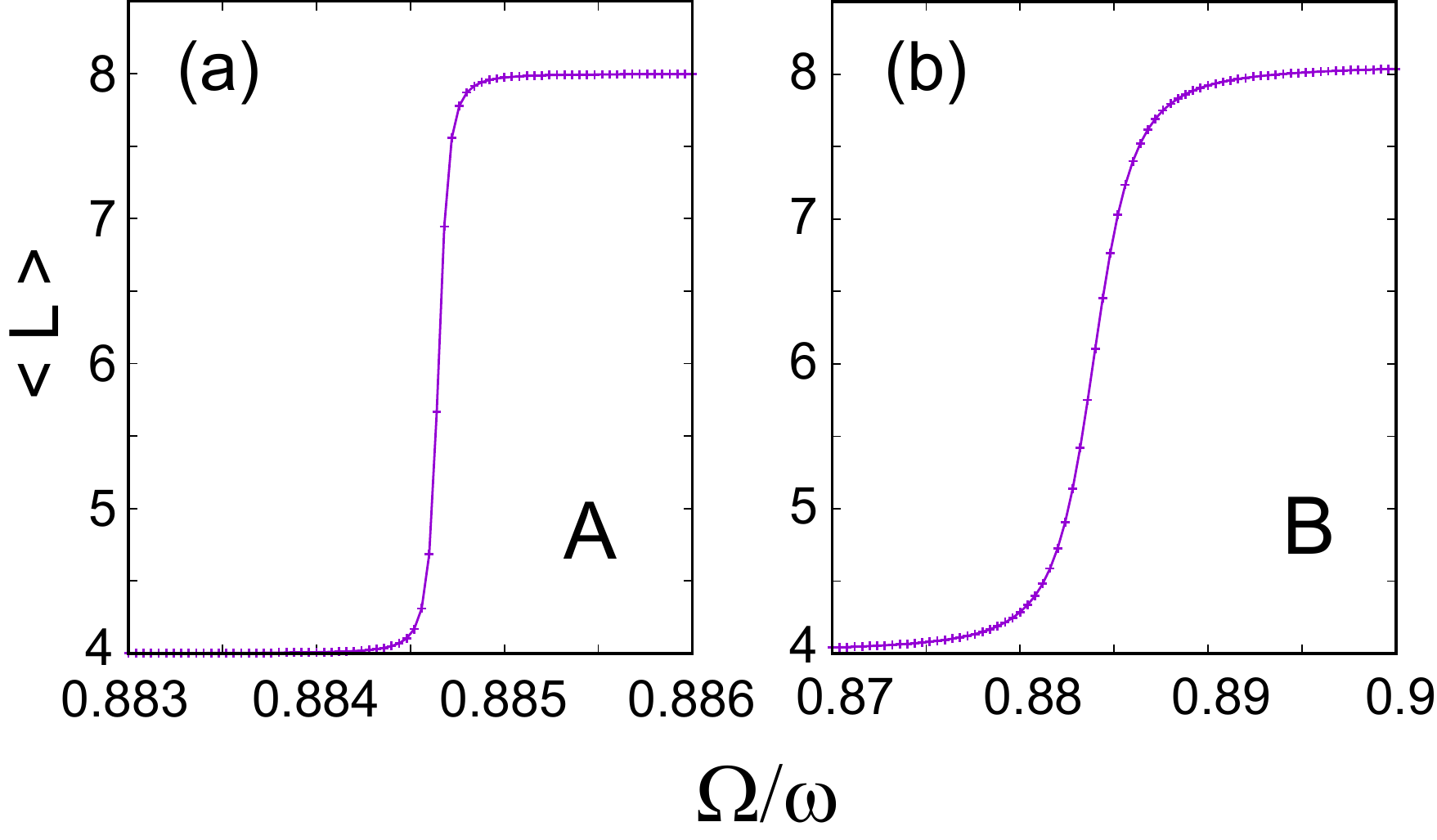}
\caption{Expectation values of the total angular momentum, $\langle L \rangle$, for $N=4$ fermions and
for the relative ground state in the spin sector ($S=0$, $S_z=0$).
(a) $\cc=0.0001$ corresponding to the avoided crossing highlighted in the inset labeled as $A$ in Fig.\ 
\ref{avcr}.
(b) $\cc=0.004$ corresponding to the avoided crossing highlighted in the inset labeled as $B$ in Fig.\ 
\ref{avcr}. Note the pronounced difference between (a) and (b) in the displayed ranges of trap rotational 
frequencies (horizontal axes). $m=4$ and $R_\delta=0.4$
}
\label{lxv}
\end{figure}

For a small value of the parameter $\cc$ (e.g., $\cc=0.0001$), $V_P$ couples mainly the two originally 
(when $V_P=0$) crossing states with good total $L$ and $L+m$, or $L-m$ and $L$. 
In this case, the expectation value, 
$\langle L \rangle$, of the total angular momentum along the avoided crossing exhibits a sharp stepwise
profile; see Fig.\ \ref{lxv}(a). A larger value of the parameter $\cc$ introduces additional couplings 
to $L\pm 2m$, $L\pm 3m$, etc..., states, which may become non-negligible, and simultaneously the
$\langle L \rangle$-profile along the avoided crossing broadens and exhibits a slower variation rate;
see Fig.\ \ref{lxv}(b) for the strong-coupling case of $\cc=0.004$.
 
\subsection{Configuration Interaction method}
\label{ci}
Denoting the spin degree of freedom by $\sigma$, in the CI method, one writes the many-body wave function 
$\Phi_{\rm CI} ({\bf r}_1\sigma_1, {\bf r}_2\sigma_2, \ldots , {\bf r}_N\sigma_N)$ as a linear
superposition of Slater determinants 
$\Psi({\bf r}_1\sigma_1, {\bf r}_2\sigma_2, \ldots , {\bf r}_N\sigma_N)$ that span the many-body
Hilbert space and are constructed out of the single-particle {\it spin-orbitals\/} 
\begin{equation}
\chi_j ({\bf r}) = \psi_j ({\bf r}) \alpha, \mbox{~~~if~~~} 1 \leq j \leq K,
\label{chi1}
\end{equation}
and 
\begin{equation}
\chi_j ({\bf r}) = \psi_{j-K} ({\bf r}) \beta, \mbox{~~~if~~~} K < j \leq 2K, 
\label{chi2}
\end{equation}
where $\alpha (\beta)$ denote up (down) spins. Namely, the wave function of the $q$th CI state is given by
\begin{equation}
\Phi_{\rm CI}^q ({\bf r}_1\sigma_1, \ldots , {\bf r}_N\sigma_N) = 
\sum_I c^q(I) \Psi_I({\bf r}_1\sigma_1, \ldots , {\bf r}_N\sigma_N),
\label{mbwf}
\end{equation}
where 
\begin{equation}
\Psi_I = \frac{1}{\sqrt{N!}}
\left\vert
\begin{array}{ccc}
\chi_{j_1}({\bf r}_1) & \dots & \chi_{j_N}({\bf r}_1) \\
\vdots & \ddots & \vdots \\
\chi_{j_1}({\bf r}_N) & \dots & \chi_{j_N}({\bf r}_N) \\
\end{array}
\right\vert,
\label{detexd}
\end{equation}
and the master index $I$ counts the number of arrangements $\{j_1,j_2,\ldots,j_N\}$
 under the restriction that $1 \leq j_1 < j_2 <\ldots < j_N \leq 2K$. Of course, 
$q=1,2,\ldots$ counts the excitation spectrum, with $q=1$ corresponding to the overall ground state
for each total spin projection $S_z$.

Because we restrict the Hilbert space in the LLL, the single-particle spatial orbitals are nodeless and 
they are given in polar coordinates by the expression
\begin{align}
\psi_l({\bf r})=\frac{1}{\sqrt{\pi l!}} r^l e^{i l \theta} e^{-r^2/2},
\label{psir}
\end{align} 
where $l \geq 0$ is the single-fermion angular momentum, and $r$ is in units of the oscillator length
$\Lambda=\sqrt{\hbar/(M\omega)}$. 

Next, the CI (exact) diagonalization of the many-body Schr\"{o}dinger equation 
\begin{equation}
H_{\rm MB} \Phi_{\rm CI}^q=E_{\rm CI}^q \Phi_{\rm CI}^q
\label{mbsch}
\end{equation}
transforms into a matrix diagonalizatiom problem, which yields the coefficients $c^q(I)$ and the CI 
eigenenergies $E_{\rm CI}^q$. Because the resulting matrix is sparse, we implement its numerical 
diagonalization employing the very efficient ARPACK solver \cite{arpack} of large-scale 
eigenvalue problems with implicitly restarted Arnoldi methods \cite{arno51}.

The matrix elements $\langle \Psi_I | H_{\rm MB} | \Psi_J \rangle$ between the basis determinants [see Eq.\ 
(\ref{detexd})] are calculated using the Slater rules \cite{szabobook}. Naturally, an important ingredient 
in this respect are the two-body matrix elements of the contact interaction,
\begin{align}
\begin{split}
& V_{1234}= \\
& \int_{-\infty}^{\infty} \int_{-\infty}^{\infty} d{\bf r}_i d{\bf r}_j
\psi^*_1({\bf r}_i) \psi^*_2({\bf r}_j)
\delta({\bf r}_i-{\bf r}_j)
\psi_3({\bf r}_i) \psi_4({\bf r}_j),
\end{split}
\label{v1234}
\end{align}
in the basis formed out of the single-particle spatial orbitals 
$\psi_i({\bf r})$, $i=1,2,\ldots,K$ [Eq.\ (\ref{psir})].  
When the lengths are expressed in units of $\Lambda$, these matrix elements are dimensionless and are 
given analytically by \cite{pape99}
\begin{align}
V_{1234}=\frac{1}{2\pi}\frac{\delta_{l_1+l_2,l_3+l_4}}{\sqrt{l_1! l_2! l_3! l_4!}}
\frac{(l_1+l_2)!}{2^{l_1+l_2}}.
\label{dltme}
\end{align}

The Slater determinants $\Psi_I$ [see Eq.\ (\ref{detexd})] conserve the third projection $S_z$,  
but not the square $\hat{\bf S}^2$ of the total spin. However, because $\hat{\bf S}^2$ commutes with the 
many-body Hamiltonian, the nondegenerate 
CI solutions are automatically eigenstates of $\hat{\bf S}^2$ with eigenvalues
$S(S+1)$. After the diagonalization, these eigenvalues are determined by
applying $\hat{\bf S}^2$ onto $\Phi_{\rm CI}^q$ and using the relation
\begin{equation}
\hat{{\bf S}}^2 \Psi_I = 
\left [(N_\uparrow - N_\downarrow)^2/4 + N/2 + \sum_{i<j} \varpi_{ij} \right ] \Psi_I,
\label{s2spin}
\end{equation}
where the operator $\varpi_{ij}$ interchanges the spins of fermions $i$ and 
$j$ provided that their spins are different; $N_\uparrow$ and $N_\downarrow$ denote 
the number of spin-up and spin-down fermions, respectively.

\subsection{Tools of analysis: Real configuration space}
\label{anal}

The tools of analysis used in this paper are the single-particle densities (1st-order correlations),
the spin-{\it un\/}resolved and spin-resolved 2nd-order correlations, as well as the higher-order $N$-body 
correlations (4th-order for $N=4$ fermions).

The spin-unresolved CI single-particle densities (1st-order correlation functions) are given by
\begin{align}
\rho_{\rm CI}({\bf r})=^1\cg(\br)=
\langle \Phi_{\rm CI}| \sum_{i=1}^N \delta({\bf r}_i-{\bf r}) | \Phi_{\rm CI} \rangle.
\label{spd}
\end{align}
Here and in the following, it is understood that evaluation of expectation values over the many-body
wave function $\Phi_{\rm CI}({\bf r}_1\sigma_1, \ldots , {\bf r}_N\sigma_N)$ involves integration
over all the particles' coordinates (including the spin ones).

We note that, in the case of a single Slater determinant, the above definition yields the simple 
formula of summation over the modulus square of the single-particle spatial orbitals.

The spin-unresolved 2nd-order correlations (pair correlations) are specified as
\begin{align}
^2\cg( {\bf r}, {\bf r}_0 )=
\langle \Phi_{\rm CI} | \sum_{i \neq j} \delta({\bf r}-{\bf r}_i) \delta({\bf r}_0-{\bf r}_j)
| \Phi_{\rm CI} \rangle,
\label{2ndcun}
\end{align}
whereas the definition of the spin-resolved 2nd-order correlations (pair correlations) includes the
spin degree of freedom as follows,
\begin{align}
^2\cg_{\sigma\sigma_0}( {\bf r}, {\bf r}_0 )=
\langle \Phi_{\rm CI} | \sum_{i \neq j} \delta({\bf r}-{\bf r}_i) \delta({\bf r}_0-{\bf r}_j)
\delta_{\sigma\sigma_i} \delta_{\sigma_0\sigma_j} | \Phi_{\rm CI} \rangle.
\label{2ndc}
\end{align}

The spin-resolved $^2\cg_{\sigma,\sigma_0}$ is also referred to as conditional probability distribution (CPD)
\cite{yann07,yann11}
because it gives the spatial probability distribution for finding a second fermion with spin projection
$\sigma$ under the condition that a first fermion with spin projection $\sigma_0$ is fixed at 
${\bf r}_0$; $\sigma$ and $\sigma_0$ can be either up $(\uparrow)$ or down $(\downarrow)$.
The first and second-order correlations defined above are calculated using the Slater rules \cite{note10} 
for the matrix elements between determinants of one-body and two-body operators, respectively. 

More importantly, here, we use in addition higher-order correlations, and in particular the $N$-body 
correlations (4th-order for $N=4$ fermions which are the focus of this paper). To motivate our discussion, 
we start first with the case of 4 fully polarized fermions ($S=2$ and $S_z=2$), whose spatial part is 
equivalent to the case of spinless fermions. For this case the CI wave function can be written as
\begin{align}
\begin{split}
& \Phi_{\rm CI}( {\bf r}_1 \alpha(1),{\bf r}_2 \alpha(2),{\bf r}_3 \alpha(3),{\bf r}_4 \alpha(4) )= \\
& F ( {\bf r}_1,{\bf r}_2,{\bf r}_3,{\bf r}_4 ) \alpha(1)\alpha(2)\alpha(3)\alpha(4),
\end{split}
\label{phip}
\end{align}
and the 4th-order correlation function is given simply by the modulus square of the spatial part, i.e., 
\begin{align}
\begin{split}
^4\cg_{\rm CI}( {\bf r}_1, {\bf r}_2, {\bf r}_3, {\bf r}_4 ) =
|F ( {\bf r}_1, {\bf r}_2, {\bf r}_3, {\bf r}_4) |^2.
\end{split}
\label{ngp}
\end{align}

The cases of non-spin-polarized configurations are more complicated,
involving both spin-resolved and spin-unresolved correlations. In general, in the case of $N$ spinful 
fermions (with $N=N_\uparrow+N_\downarrow$), the CI wave function $\Phi_{\rm CI}$ contains 
$K=N!/(N_\uparrow N_\downarrow)$ primitive spin functions of the form
\begin{equation}
\zeta_i(N_\uparrow,N_\downarrow)=\alpha \alpha...\beta \beta.
\end{equation}   
To be specific, for the case of $N=4$ fermions with $N_\uparrow=N_\downarrow=2$ ($S_z=0$), 
there are $K=6$ such spin primitives, namely
\begin{align}
\begin{split}
&\zeta_1=\alpha(1) \alpha(2) \beta(3) \beta(4),\\
&\zeta_2=\alpha(1) \alpha(3) \beta(2) \beta(4),\\
&\zeta_3=\alpha(1) \alpha(4) \beta(2) \beta(3),\\
&\zeta_4=\alpha(2) \alpha(3) \beta(1) \beta(4),\\
&\zeta_5=\alpha(2) \alpha(4) \beta(1) \beta(3),\\
&\zeta_6=\alpha(3) \alpha(4) \beta(1) \beta(2),
\end{split}
\label{sppr}
\end{align}
where the arguments from 1 to 4 in the $\alpha$'s and $\beta$'s correspond to particle indices.  

Considering the 4 spin orbitals $u^I_1=\phi^I_{l_1}\alpha$, $u^I_2=\phi^I_{l_2}\alpha$, 
$u^I_3=\phi^I_{l_3}\beta$, and $u^I_4=\phi^I_{l_4}\beta$ of the $I$-th determinant in the CI expansion [Eq.\
(\ref{mbwf})], which (for a given determinant) are the same for all six $\zeta$'s listed in Eq.\ (\ref{sppr}),
the many-body CI wave function for $N_\uparrow=N_\downarrow=2$ can be rewritten as
\begin{align}
\Phi_{\rm CI}= \sum_{i=1}^6 \cf_i({\bf r}_1,{\bf r}_2,{\bf r}_3,{\bf r}_4) \zeta_i,
\label{phiz}
\end{align}
where 
\begin{align}
\begin{split}
\cf_i=\sum_I c(I) 
{\rm Det}^\uparrow[\phi^I_{l_1}({\bf s}_1^i),\phi^I_{l_2}({\bf s}_2^i)]
{\rm Det}^\downarrow[\phi^I_{l_3}({\bf s}_{3}^i),\phi^I_{l_4}({\bf s}_4^i)],
\end{split}
\end{align}
where $c(I)$ are the coefficients of the CI expansion and 
$({\bf s}_1^i,{\bf s}_2^i)$ and $({\bf s}_3^i,{\bf s}_4^i)$ 
coincide with the spatial coordinates associated with the particle indices for the up and down spins in the 
$\zeta_i$ spin primitives defined in Eq.\ (\ref{sppr}). For example, for $i=5$, one has
\begin{align}
\begin{split}
& {\bf s}_1^5 \rightarrow {\bf r}_2\\
& {\bf s}_2^5 \rightarrow {\bf r}_4\\
& {\bf s}_3^5 \rightarrow {\bf r}_1\\
& {\bf s}_4^5 \rightarrow {\bf r}_3
\end{split}
\label{s1234}
\end{align}

The spin-unresolved 4th-order correlation is then given by 
\begin{align}
^4\cg^{\rm un}_{\rm CI}=\sum_{i=1}^6 \cf_i^*\cf_i.
\label{spundef}
\end{align}

The spin resolved 4th-order correlations are defined as a partial summation over the spin-primitive
index $i$ [Eq.\ (\ref{sppr})]. For example, the probability of finding the fourth fermion with spin down 
[$\beta(4)$ in any $\zeta_i$ spin primitive] at a position ${\bf r}$, given the positions 
of the first three fermions with unresolved spins, is:
\begin{align}
^4\cg^{\rm res,1}_{\rm CI}=\cf_1^*\cf_1+\cf_2^*\cf_2+\cf_4^*\cf_4.
\label{res1}
\end{align}

Other spin-resolved 4th-order correlations are possible: for example, finding the fourth fermion with spin
down at position ${\bf r}$, given the positions of the first 3 fermions with the 2nd fermion having a spin 
up and the 1st and 3rd ones with unresolved spins is given by
\begin{align}
^4\cg^{\rm res,2}_{\rm CI}=\cf_1^*\cf_1+\cf_4^*\cf_4.
\label{res2}
\end{align}

\begin{figure}[t]
\includegraphics[width=7.5cm]{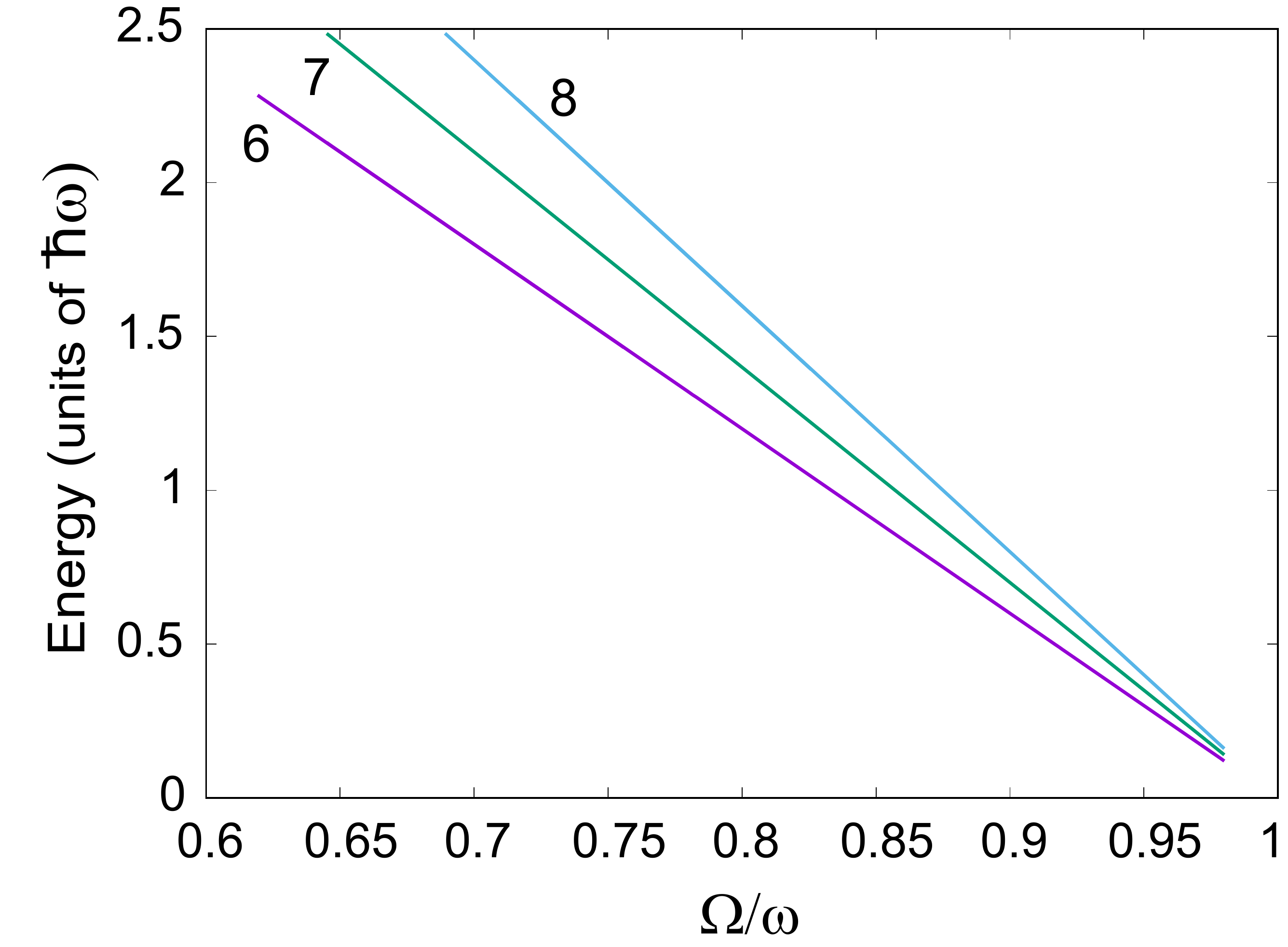}
\caption{Spectra of the three lowest-in-energy CI solutions of the many-body Hamiltonian $H_{\rm MB}$
[Eq.\ (\ref{hmb})] for $N=4$ fermions as a function of the ratio $\Omega/\omega$ (the constant energy
$N\hbar\omega$ has been subtracted). The spin sector with ($S=2$, $S_z=0$) is displayed.
The order of the multipole perturbation [see Eq.\ (\ref{vp})] $m=4$,
and $R_\delta=0.4$. Because there are no crossings, only the value $\cc=0.0001$ (weak $V_P$ 
perturbation) is plotted. The relative ground state has $L=6$. 
}
\label{sps2}
\end{figure}

\begin{figure*}[t]
\includegraphics[width=17cm]{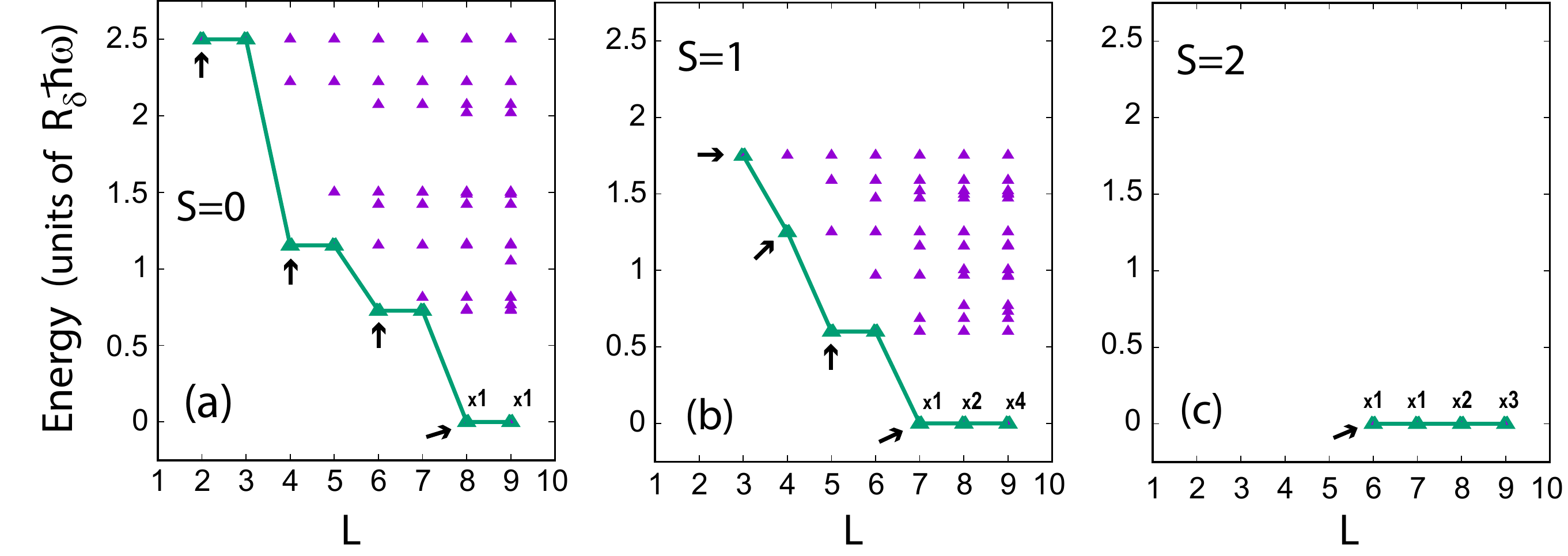}
\caption{
CI-calculated, partial LLL energy spectra for $N=4$ fermions deriving from the diagonalization of the 
contact-interaction term only; see third term of the $H_{\rm LLL}$ Hamiltonian in Eq.\ (\ref{H_lll}). 
(a) The spectrum of the ($S=0, S_z=0$) sector. (b) The spectrum of the ($S=1, S_z=0$) sector. 
(c) The spectrum of the ($S=2, S_z=0$) sector. The CI spectra were calculated for $S_z=0$; however, note
that these LLL spectra are independent of the precise value of $S_z$. The arrows indicate magic
angular momenta. The symbols x$n$, with $n$ being an integer, denote the degeneracy of the 
vanishing-energy states in each spin sector. The horizontal axis represents the total angular 
momentum $L$. Energies in units of $R_\delta \hbar \omega$.}
\label{fintspec}
\end{figure*}

\subsection{Tools of analysis: Momentum space}
\label{analmom}

To channel our discussion about momentum-space correlation functions, we recall again that, 
usually, a CI calculation (or other exact diagonalization schemes used for 
solution of the microscopic many-body Hamiltonian) yields a many-body wave function (e.g., $\Phi_{\rm CI}$) 
in position coordinates $({\bf r}_1\sigma_1, {\bf r}_2\sigma_2, \ldots , {\bf r}_N\sigma_N)$; see 
Eq.\ (\ref{mbwf}), which for the case of $N=4$ fermions can take the form in Eq.\ (\ref{phiz}).

The CI wave functions, $\Phi_{\rm CI}$, are particularly conducive for carrying out their mapping into 
the momentum-space ones, $\Phi_{\rm CI}^{\cal M}$; naturally the momentum space is spanned
by the coordinates $({\bf k}_1\sigma_1, {\bf k}_2\sigma_2, \ldots , {\bf k}_N\sigma_N)$, with 
${\bf k}_j={\bf p}_j/\hbar$. Indeed it is sufficient to replace each LLL single-particle real-space orbital 
$\psi_j(\br)$ in the basis determinants $\Psi_I$ [Eq. (\ref{detexd})] by its 2D Fourier transform, which 
is given by [compare to the real-space function in Eq.\ (\ref{psir})] 
\begin{align}
\psi_l({\bf k})=\frac{i^l}{\sqrt{\pi l!}} k^l e^{i l \varphi} e^{-k^2/2},
\label{psik}
\end{align} 
where $k$ is in units of the inverse of the oscillator length, $1/\Lambda$; see definition after 
Eq.\ (\ref{psir}).

Having obtained the many-body wave function in real space, all and each formula (in Sec.\ \ref{anal})
specifying the tools of analysis in real space (1st, 2nd, and $N$th-order correlations) can be immediately 
translated in momentum space by simply replacing $\br_j \rightarrow {\bf k}_j$ and $\Phi_{\rm CI} 
\rightarrow \Phi_{\rm CI}^{\cal M}$. In addition, Eq.\ (\ref{psik}) shows that,
apart from a phase $i^l$, the LLL orbitals in momentum space retain the same form as the corresponding 
ones in configuration space, with the following substititions: 
$(r, \theta) \longleftrightarrow (k, \varphi)$ 
and $\Lambda \longleftrightarrow 1/\Lambda$. Consequently: (i) All the expressions and final results, 
including the figure plots, for the 1st, 2nd, and 4th correlations calculated in real space represent also
corresponding results in momentum space, the only change being the units of the axes ($1/\Lambda$ versus
$\Lambda$). (ii) The TOF measurements in the far field \cite{coop03,altm04} act as a microscope that 
magnifies directly the {\it in situ} many-body wave function. 

In deterministic time-of-flight measurements, the $N$ trapped ultracold 
atoms expand subsequent to a sudden turn-off of 
the trapping potential, and a snapshot of the free-space traveling $N$ atomic particles is taken in the far
field after a time $t_{\rm TOF}$. This step is repeated several thousand times and the compilation of the 
ensuing snapshots reproduces the modulus square of the Fourier transform of the {\it in situ\/} many-body 
wave function \cite{prei19}. 
$t_{\rm TOF}$ is taken long enough so that the size of the compiled ensemble is much larger 
than its original (confined) size. The TOF far-field real-space coordinates of the particles at time 
$t_{\rm TOF}$ are given by $\br_j = \hbar {\bf k}_j t_{\rm TOF}/M$, with $j = 1, 2 \dots, N$, where 
$\hbar {\bf k}_j$ is the single-particle momentum at the source (the 
confining trap). From the above, we note that during the 
expansion the interatomic interactions can be neglected, whereas prior to the expansion the interactions 
play a key role in determining the properties of the trapped correlated LLL state \cite{coop03,altm04}. In 
this way, analyses of TOF measurements allow determination of the properties of the many-body state of the 
confined system via analyses of the all-order (1 to $N$) momentum-space correlation functions. 
These momentum correlation functions are indeed the focus of our study. As aforementioned, for the LLL case
investigated here, the single-particle Fourier transform in Eq.\ (\ref{psik}) retains the same functional 
form on ${\bf k}$ as does $\psi_l(\br)$ in Eq.\ (\ref{psir}) on $\br$. As a result, apart of units, the
{\it in situ} real-space and momentum correlations coincide, and the TOF measurements act as a microscope 
of the {\it in situ\/} many-body wave function.

\section{LLL spectra and magic angular momenta}
\label{spec}

A primary tool for the classification and for gaining a deeper understanding of the geometric 
aspects of the intrinsic correlations in the LLL many-body wave functions is the concept of magic 
angular momenta, introduced and extensively utilized in the treatments of semiconductor quantum 
dots \cite{girv83,maks90,maks96,ruan95,seki96,yann03,yann03.2,yann07,dai07,shi07}. 

It is pertinent to note here analogies between the ultracold-atoms case and that of electrons
confined in the above-mentioned semiconductor quantum dots.
Indeed, in such parabolically confined (i.e., with a harmonic external
potential), finite 2D strongly-interacting correlated electron-gas structures, the emergence of 
intrinsic quantum crystalline-like (or molecular-like) features (so-called Wigner molecules, WMs) 
is traditionally revealed through analysis of 2nd-order correlations in the CI 
\cite{yann03,yann07,maks96}, or center-of-mass separable \cite{yann00}, many-body wave functions, 
associated with spontaneous symmetry breaking at the mean-field unrestricted Hartree-Fock level 
\cite{yann99,roma04,cava07}. At zero magnetic field, formation of such ordered structures has been
shown to be driven by competition between the electron-crystallization that minimizes the
long-range repulsive coulomb inter-electron interaction and the opposing effect due to the
increase in the zero-point kinetic energy that accompanies crystallization (that is, the reduced
electron positional uncertainty that occurs due to the localization of the electron at the induced
crystalline site) \cite{yann99,yann07}. On the other hand, an applied magnetic field is 
acting as an independent factor inducing WM formation \cite{maks96,yann02,yann04,yann07};
see also Sec.\ \ref{wigp} below. The 
predicted occurrence of such WM electron structures in 2D electron dots under magnetic-field-free 
conditions, and in the presence of applied magnetic fields (where, as aforementioned, the 
magnetic-field-induced rotating molecular structures have been termed as rotating Wigner molecules 
\cite{yann02,yann07,note3}), have been confirmed experimentally 
\cite{elle06,maks06,kall08,mint18,ilan13,hoen14}. Here we establish a broader viewpoint by showing
that such RWMs emerge also in the case of ultracold fermionic neutral atoms (e.g., $^6$Li atoms) 
interacting via short-range contact interactions and confined in a rotating harmonic trap (that is
emulating a magnetic field via implementation of a synthetic gauge).  

An early \cite{girv83,maks90} recognized signature of magic angular momenta was their forming sets
of energetically advantageous states (referred to also as cusp states) in the LLL spectra of a few
2D fully spin-polarized electrons. According to subsequent 
\cite{girv83,maks90,maks96,ruan95,seki96,yann03,yann03.2,yann07,dai07,shi07} findings from CI 
calculations in the field of semiconductor quantum dots (a few electrons confined in a harmonic 
potential), the 2D electrons under a perpendicular high magnetic field localize relative to each 
other and form ordered ring-like configurations $(n_1,n_2,...,n_r)$ (with $\sum_i^r n_i=N$). Such 
ordered ring configurations are not visible in the CI single-particle densities, which are 
azimuthally (rotationally) uniform, but are revealed by using higher-order correlations 
\cite{yann07}. Furthermore, the CI total angular momenta must be compatible (i.e., satisfy) the 
$C_n$ point group, etc., symmetries associated with the ring configurations 
\cite{maks96,ruan95,seki96,yann03,yann03.2,yann07,dai07,shi07}.
Similar magic angular momenta appear also in the LLL spectra and CI solutions for ultracold bosonic 
atoms \cite{baks07}. The present paper demonstrates that magic angular momenta are pertinent as well
to ultracold fermions in rapidly rotating harmonic traps.

Going beyond the fully spin-polarized case, previous investigations have found 
\cite{maks96,yann03.2,dai07,shi07} that the magic angular momenta depend in a nontrivial way on the 
value $S$ of the total spin. In particular, for the case of $N=4$ fermions (which is the focus of 
this paper), the associated ring-like configuration is a square 
(denoted as (0,4) \cite{note11}),
 
and the series of magic angular momenta are as follows \cite{dai07,shi07}:
\begin{align}
\begin{split}
& S=0 \; \rightarrow L=4n \;\; {\rm or} \;\; L=4n+2, \\
& S=1 \; \rightarrow L=4n-1 \;\; {\rm or} \;\; L=4n ;\; {\rm or} \;\; 4n+1, \\
& S=2 \; \rightarrow L=4n+2, 
\end{split}
\label{mam}
\end{align}
where $n=0,1,2,3,\ldots$. 

The relative ground states of $H_{\rm LLL}$ in each spin sector [see definition below Eq. (6)] are associated with magic angular momenta. 
Indeed, the 
values of $L=2$, 4, and 8 in Fig.\ \ref{avcr}, corresponding to the ground states in the spin sector 
($S=0$, $S_z=0$), belong to the $S=0$ series listed in Eq.\ (\ref{mam}); note that not all terms in the series given in Eq. (30) 
correspond to the sequence of ground states of the total Hamiltonian 
$H_{\rm MB}$. Furthermore, in the spin sector
($S=2$, $S_z=0$), the relative ground state has the magic angular momentum $L=6$ in agreement with the
series in Eq.\ (\ref{mam}); see Fig.\ \ref{sps2}.

To further elaborate on the relation between magic angular momenta and LLL spectra, we display in Fig.\ 
\ref{fintspec} the restricted LLL spectra in each spin sector corresponding to the diagonalization of the
contact-interaction term only, that is, to the last term in Eq.\ (\ref{H_lll}). These spectra are plotted as
a function of the total angular momentum $L$; for each value of $L$, a tower of excited states is shown 
(upward standing triangles above the yrast-band line that connects the lowest-energy triangles). 
These excited LLL states display a highest-energy bound at $2.5 R_\delta \hbar \omega$. 
The number of states in each tower increases with 
increasing $L$, and every newly appearing energy at a given $L$ repeats itself at larger $L$'s. 
In this figure, the lowest-energies for each $L$ (forming the socalled yrast band) are highlighted by 
passing a line through them. For the ($S=0, S_z=0$) and ($S=1, S_z=0$) spin sectors, the yrast bands 
involve successively lower energies and eventually they collapse to a horizontal line at vanishing energy. 
For the ($S=2, S_z=0$), only the horizontal segment at zero enery is present. 
The zero-energy horizontal segment in Fig.\ \ref{fintspec} is a property connected to the zero range of the 
contact interaction; it is absent in the case of the long-range Coulomb interaction 
\cite{yann03,shi07,yann07}.

In Fig.\ \ref{fintspec}, the magic angular momenta according to Eq.\ (\ref{mam}) are marked by an
arrow. They are preceded by a sharp drop in energy relative to the previous angular momentum (as
long as the previous $L$ is non-magic or nonvanishing); as a result the associated LLL states are
often referred to as cusp states \cite{yann07,jainbook}.

For a discussion on the  group theoretical relationship between the geometric structure of a Wigner
molecule made of spinful fermions, and the corresponding magic angular momenta sequences for
different spin states of the WM, we refer the reader to the illustrative example given in 
Appendix \ref{a1}.

\begin{figure*}[t]
\includegraphics[width=17cm]{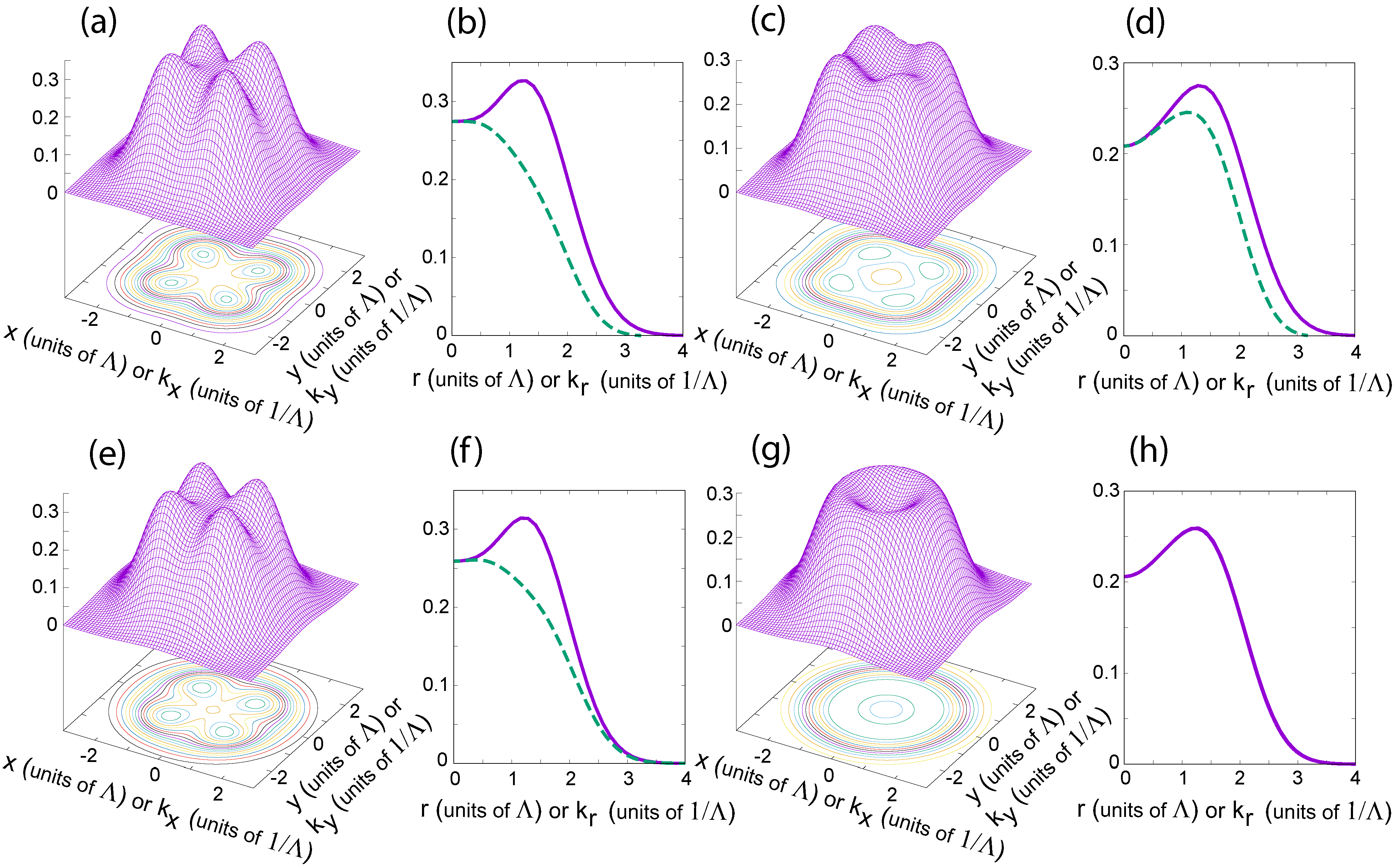}
\caption{CI single-particle densities (both in real space and momentum space) 
of the relative ground state of $N=4$ fermions in the spin sector with ($S=0$, $S_z=0$).
3D surfaces are plotted in (a), (c), (e), and (g). Corresponding cuts through the 
origin along the diagonals (solid line, violet) and perpendicular to the sides (dashed line, green) of the
square configuration are displayed in (b), (d), (f), and (h), respectively. In (h) both curves overlap. 
(a,b) Calculation for $\cc=0.004$ (strong perturbation) at the point 
$\Omega/\omega=0.8855$. The expectation value of the total angular momentum is $\langle L \rangle=7.189$, 
indicating that the plotted case is a state with broken rotational symmetry. 
(c,d) Calculation for $\cc=0.004$ (strong perturbation) at the point $\Omega/\omega=0.90$.
The expectation value of the total angular momentum is $\langle L \rangle=8.0364$, 
closer to integer 8, but the broken rotational symmetry is still present. 
(e,f) Calculation for $\cc=0.0001$ (weak perturbation) at the point $\Omega/\omega=0.8847$. 
The expectation value of the total angular momentum is $\langle L \rangle=7.330$, 
and the single-particle density exhibits strong breaking of the rotational symmetry. 
(g,h) Calculation for $\cc=0.0001$ (weak perturbation) at the point $\Omega/\omega=0.90$. 
The expectation value of the total angular momentum is $\langle L \rangle=8.00002$, 
very close to integer 8, and the rotational symmetry is practically reestablished.
$R_\delta=0.4$ and the order of the multipole trap deformation $m=4$.
Because of the properties of the Fourier transform of the LLL orbitals [see Eq.\
(\ref{psik})], both real-space and momentum densities are given by the same 3D numerical surface.
For the spatial densities, the lengths along the $x$, $y$, and $r$ axes are given in units of $\Lambda$,
and the vertical axes are in units of $1/\Lambda^2$. For the momentum densities, the momenta along the
$k_x$, $k_y$, and $k_r$ axes are given in units of $1/\Lambda$, and the vertical axes are in units of 
$\Lambda^2$.
}
\label{fspd}
\end{figure*}    

\section{The spin sector ($S=0$, $S_z=0$): Traversing the avoided crossing}
\label{trcr}

In this Section, the properties of the relative-ground-state wave functions in the spin sector 
($S=0$, $S_z=0$) and for $N=4$ fermions will be investigated in detail along the avoided crossings 
highlighted in the insets (labeled as A and B) of Fig.\ \ref{avcr}. The tools used in this analysis 
are the single-particle densities (1st-order correlations) and the $N$-body correlations (4th-order for 
$N=4$ fermions) defined in Sec.\ \ref{anal}.

\subsection{Single-particle densities}
\label{s0sz01st}

In Fig.\ \ref{fspd}, we plot the CI single-particle density for two different strengths of the 
perturbation $V_P$ [$\cc=0.004$, top row (a-d) and $\cc=0.0001$, bottom row (e-h)] and for two 
different values of the rotational frequency across the avoided crossing highlighted in the insets 
of Fig.\ \ref{avcr}, i.e., near the midpoint at $\Omega/\omega=0.8855$ [Figs.\ \ref{fspd}(a,b)], or 
$\Omega/\omega=0.8847$ [Figs.\ \ref{fspd}(e,f)], and after the crossing at $\Omega/\omega=0.90$ 
[Figs.\ \ref{fspd}(c,d,g,h)]. As was the case in Fig.\ \ref{avcr}, $R_\delta=0.4$ and $m=4$
(hexadecapole multipolarity); for the case of a quadrupolar trap deformation ($m=2$), see Sec.\
\ref{quad} below. We note  again that, in the LLL,  the formation of geometric structures of the
particles in the trap and their  symmetries, does not depend on the precise value of $R_\delta$ (see
also discussion in Sec. \ref{wigp} below).

As seen from Figs.\ \ref{fspd}(a-d), the value $\cc=0.004$ is rather large and results in a symmetry
broken solution even at the point $\Omega/\omega=0.90$. Indeed, at this point, the associated 
expectation value of the total angular momentum, $\langle L \rangle= 8.0364$, is still rather 
different from the integer value of 8. On the contrary, the small value $\cc=0.0001$ yields 
$\langle L \rangle= 8.00002$ at $\Omega/\omega=0.90$, and the corresponding many-body wave function 
preserves the rotational symmetry for all practical purposes; see the single-particle density in 
Fig.\ \ref{fspd}(g). 

This behavior conforms with the fact that the many-body wave functions, $\Phi_{\rm gs}$'s, 
associated with Figs.\ \ref{fspd}(a,b), \ref{fspd}(c,d), and \ref{fspd}(e,f) contain significant 
contributions of basis determinants with total angular momenta other than $L=8$, i.e., $L=4$ and 
$L=12$; see TABLES STI  - STIV in the Supplemental Material (SM) \cite{supp}. 
In contrast, the many-body 
wave function, $\Phi_{\rm gs}$, associated with Figs.\ \ref{fspd}(g,h) consists mainly of basis 
determinants each with total angular momentum $L=8$ (with one exception of a basis determinant of 
$L=4$ having a very small weight); see TABLE STIV in the SM \cite{supp}.

It is remarkable that in all cases of symmetry breaking portrayed by the 3D surfaces in 
Figs.\ \ref{fspd}(a,c,e), the same underlying Wigner-molecule, square-ring configuration emerges. 
This (0,4) ring configuration is further highlighted by plotting the corresponding cuts through the
origin along the diagonals (solid line, violet) and perpendicular to the sides (dashed line, green) 
of the square configuration in Figs.\ \ref{fspd}(b,d,f,h), respectively. In Fig.\ \ref{fspd}(h), the
reestablishment of rotational symmetry is reflected in the fact that both the solid and dashed cuts 
do overlap. Furthermore, the demonstrated here effect of $V_P$ at the avoided crossing upon the CI 
single-particle density is so profound and disproportionate to the smallness of the perturbation 
[compare, e.g., Figs.\ \ref{fspd}(e) and \ref{fspd}(g)] that it is appropriate to characterize the 
present results as a numerical example of the 'flea on the elephant' concept 
\cite{land20,jona81,simo85}, developed in mathematical treatments of the phenomenon of spontaneous 
symmetry breaking \cite{ande52}.

\begin{figure}[t]
\includegraphics[width=7.0cm]{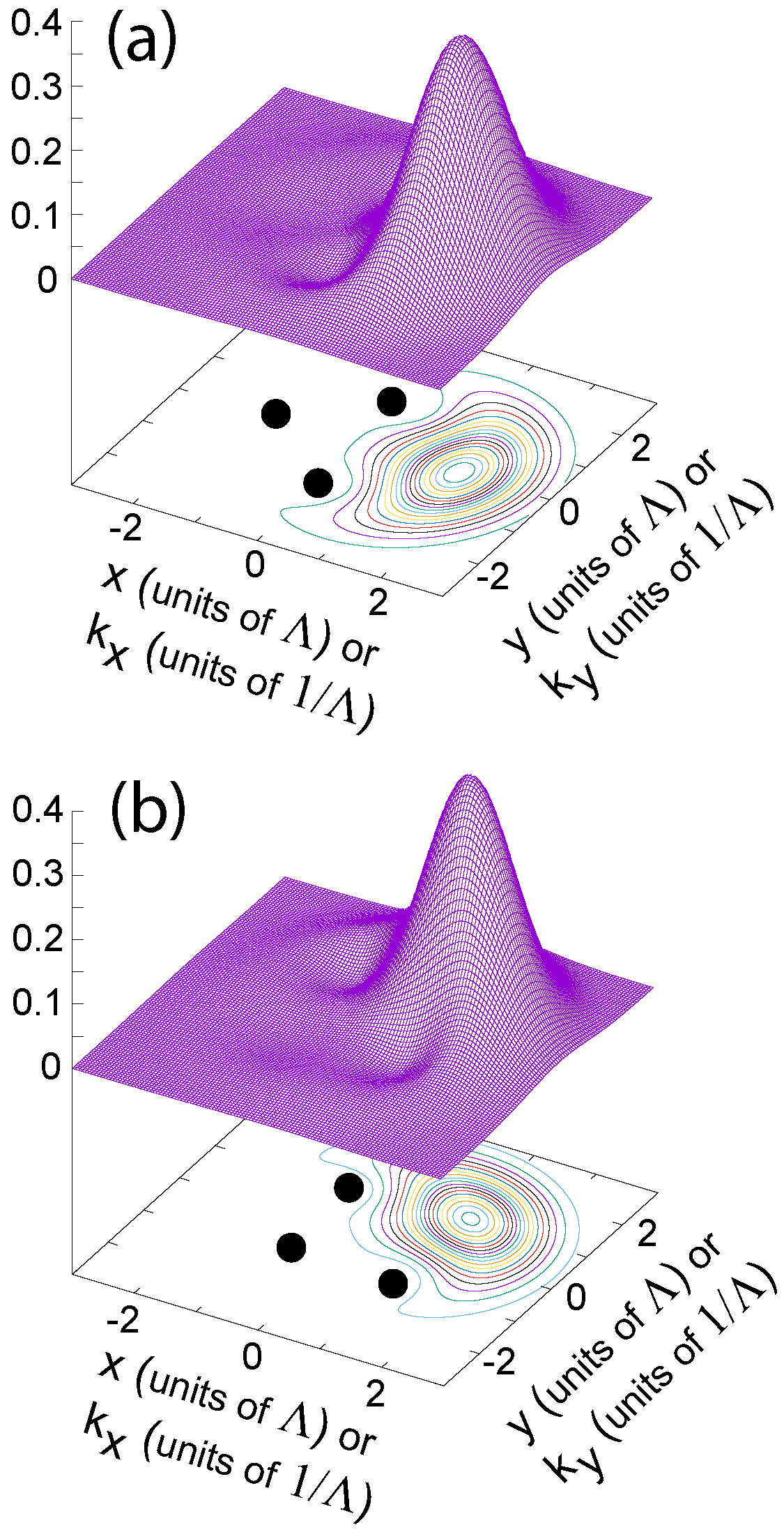}\\
\caption{Unresolved 4th-order correlations (both in real space and momentum space) 
of the relative CI ground state in spin sector ($S=0$, $S_z=0$) for $N=4$ fermions at the trap 
angular frequency $\Omega/\omega=0.90$. The strength of the $V_P$ perturbation
is weak with $\cc=0.0001$. $R_\delta=0.4$, and the order of the multipole trap deformation $m=4$.
These correlations correspond to the 2D rotationally symmetric single-particle density in Fig.\ 
\ref{fspd}(g). The three fixed points (denoted by the solid dots) are placed at a radius 
$r_0=1.22$ $\Lambda$. The reference angle $\Theta=0$ in (a) and $\Theta=\pi/4$ in (b).  
Because of the properties of the Fourier transform of the LLL orbitals [see Eq.\ (\ref{psik})], both 
real-space and momentum 4th-order correlations are given by the same 3D numerical surface.
For the spatial correlations, the lengths along the $x$ and $y$ axes are given in units of $\Lambda$,
and the vertical axes are in units of $1/(\pi^4\Lambda^8)$. For the momentum correlations, the momenta 
along the $k_x$ and $k_y$ axes are given in units of $1/\Lambda$, and the vertical axes are in units of
$\Lambda^8/\pi^4$. In the case of the momentum correlations, $r_0$ here is replaced by $k_r^0=1.22$ $1/\Lambda$.
Corresponding mappings between the fixed points of the real-space and momentum-space
correlations apply also for Figs.\ 7, 9(b), 10, and 11 below. 
}
\label{fng}
\end{figure}

\subsection{4th-order correlations associated with the symmetry-preserving $L=8$ relative 
ground state}
\label{s0sz04th}

A deeper understanding of the interplay, portrayed in Fig.\ \ref{fspd}, between symmetry-broken 
solutions and those that preserve the 2D rotational symmetry is gained by considering the 4th-order 
correlations defined in Sec.\ \ref{anal}. Fig.\ \ref{fng} displays the spin-unresolved 4th-order 
correlations [see Eq.\ (\ref{spundef})] associated with Fig.\ \ref{fspd}(g), i.e., for $\cc=0.0001$ 
at the point $\Omega/\omega=0.90$. The most natural way to analyze this quantity, that depends on 4 
variables, is to fix three variables and plot $^4\cg_{\rm gs}$ as a function of the fourth variable.
Motivated by the molecular ring configuration [usually denoted as (0,4)] of the broken-symmetry 
single-particle densities, we place the three fixed variables at the points 
$r_0 \exp(j\pi/2+\Theta)$ (with $j=1,2,3$), where $r_0=1.22$ $\Lambda$ is the radius of the maxima 
of the 4 humps in Fig.\ \ref{fspd}(e), the angle $\pi/2$ reflects the square arrangement of these 
four humps, and $\Theta$ is an arbitrary reference angle. Two values of $\Theta=0$ [Fig.\ 
\ref{fng}(a)] and $\Theta=\pi/4$ [Fig.\ \ref{fng}(b)] were used. In both cases, Fig.\ \ref{fng} 
shows that the conditional probability of finding the fourth fermion at a given point 
is localized around the apex point that completes the square of the (0,4) ring configuration. 

Naturally, the fact that the intrinsic (0,4) molecular configuration contained in the 
$^4\cg_{\rm gs}$ correlation is independent of the reference angle $\Theta$ is consistent with the 
uniform (2D rotationally symmetric) single-particle density in Fig.\ \ref{fspd}(g); it is also the 
property that suggests the characterization of the associated many-body state as a 
``rotating Wigner molecule'' \cite{note3}, in contrast to the term ``pinned Wigner molecule'' 
suggested by the broken-symmetry single-particle densities in Figs.\ \ref{fspd}(a), \ref{fspd}(c), 
and \ref{fspd}(e).\\
~~\\

\begin{figure*}[t]
\includegraphics[width=16cm]{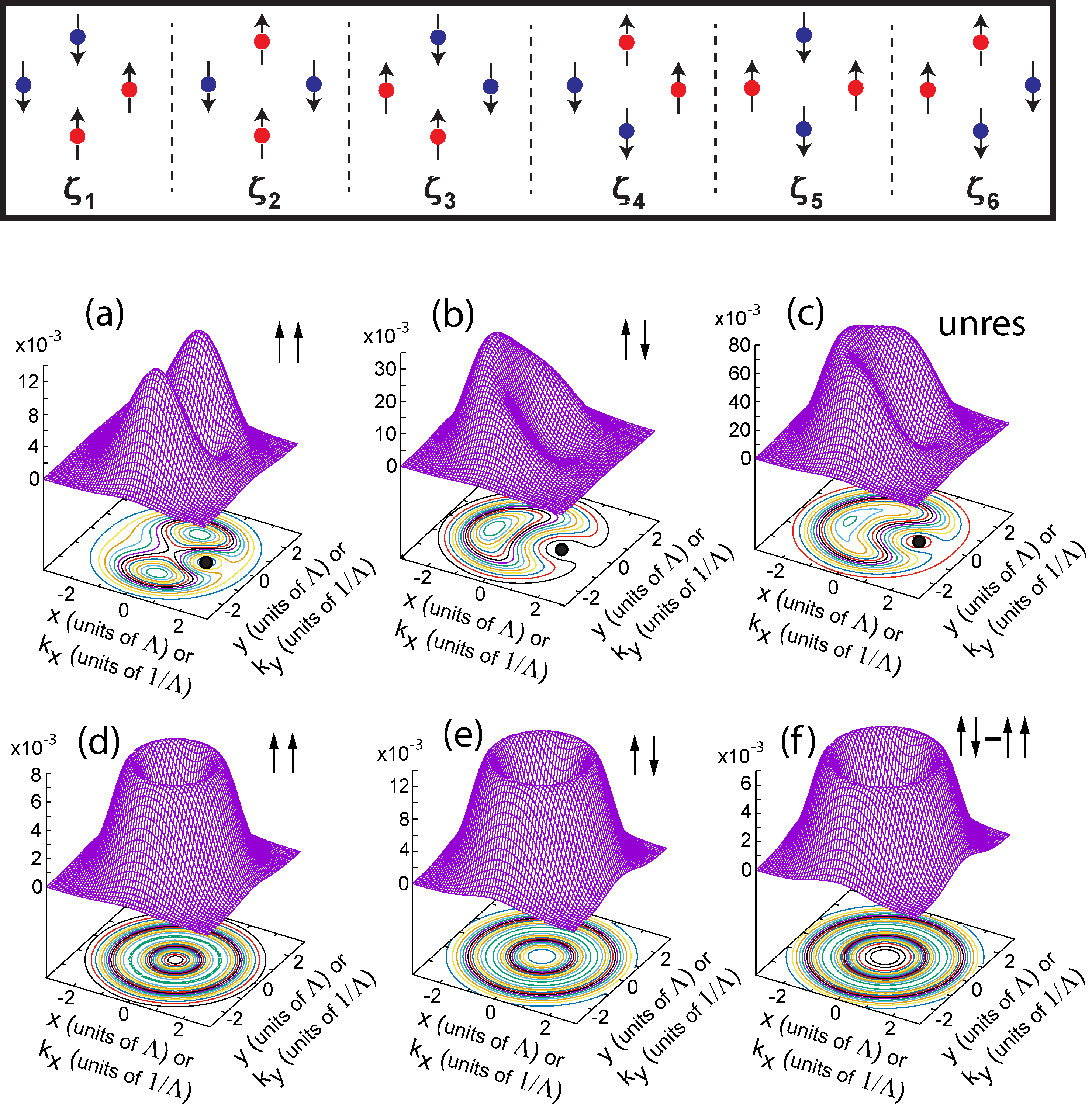}
\caption{$\uparrow\uparrow$ (a,d), $\uparrow\downarrow$ (b,e), 
and spin-unresolved (c) 2nd-order correlations (both in real space and momentum space) of the 
relative CI ground state in spin sector ($S=0$, $S_z=0$) with $L=8$ for $N=4$ fermions at the trap 
angular frequency $\Omega/\omega=0.90$. The strength of the $V_P$ perturbation is weak, using 
$\cc=0.0001$ with $m=4$ and $R_\delta=0.4$.  These correlations correspond to the 2D rotationally 
symmetric single-particle density in Fig.\ \ref{fspd}(g). (d) The difference 
$^2\cg_{\uparrow\downarrow} - ^2\cg_{\uparrow\uparrow}$. The fixed points (see text) are placed at 
$(x_0=1.22$ $\Lambda,y_0=0)$ for the three top panels, and at the origin $(x_0=0,y_0=0)$ for the 
three bottom panels. Because of the properties of the Fourier transform of the LLL orbitals [see 
Eq.\ (\ref{psik})], both real-space and momentum correlations are given by the same 3D numerical 
surface. For the space correlations, the lengths along the $x$ and $y$ axes are given in units of 
$\Lambda$, and the vertical axes are in units of $1/\Lambda^4$. For the momentum correlations, the 
momenta along the $k_x$ and $k_y$ axes are given in units of $1/\Lambda$, and the vertical axes are 
in units of $\Lambda^4$. In (a)-(c), the fixed point is denoted by a solid dot.
The inset provides a graphical representation of the six $\zeta_i$'s spin primitives 
[see Eq.\ (\ref{sppr})], when associated with the Wigner-molecule square geometry.  
}
\label{f2ndg}
\end{figure*}

\subsection{2nd-order correlations associated with the symmetry-preserving $L=8$ relative 
ground state}
\label{s0sz02nd}

As described in Sec.\ \ref{anal}, 2nd-order correlations are a complementary tool in obtaining 
information regarding the intrinsic structure of the many-body wave function in the absence of 
symmetry breaking. The 2nd-order correlations [see Eqs.\ (\ref{2ndcun}) and (\ref{2ndc})] for the 
symmetry-preserving relative ground state in the ($S=0$, $S_z=0$) spin sector at 
$\Omega/\omega=0.90$ with $m=4$ and $\cc=0.0001$ [corresponding to the single-particle density in 
Fig.\ \ref{fspd}(g)] are displayed in Fig.\ \ref{f2ndg}. Specifically, taking the fixed point at 
$\br_0=(1.22 \Lambda, 0)$, Figs.\ \ref{f2ndg}(a) and \ref{f2ndg}(b) portray the up-up, 
$^2\cg_{\uparrow\uparrow}$, and up-down, $^2\cg_{\uparrow\downarrow}$, spin-resolved 
2nd-order correlations, respectively, whereas Fig.\ \ref{f2ndg}(c) portrays the spin-unresolved one. 
Figs.\ \ref{f2ndg}(d) and \ref{f2ndg}(e) portray the up-up and up-down 2nd-order correlations, 
respectively, but with the fixed point taken to be at the origin. Lastly, Fig.\ \ref{f2ndg}(f) 
displays the difference between the two spin-resolved correlations, 
$^2\cg_{\uparrow\downarrow} - ^2\cg_{\uparrow\uparrow}$, when the fixed point is taken at the origin.

In addition to reproducing the relative single-particle localization of the four fermions in a square
configuration (discussed in Sec.\ \ref{s0sz04th} using 4th-order correlations), the 2nd-order correlations 
in Figs.\ \ref{f2ndg}(a) and \ref{f2ndg}(b) can assist in the determination of the underlying spin structure 
of the corresponding many-body wave function. To this end, the six $\zeta_i$'s spin primitives [see Eq.\ 
(\ref{sppr})], associated with the Wigner-molecule square geometry are depicted graphically in the inset of 
Fig.\ \ref{f2ndg}.

According to Fig.\ \ref{f2ndg}(a), when the fixed spin-up fermion is placed at one corner of the square, the
most probable locations of the other spin-up fermion are the two adjacent corners of the square, but not the
opposite corner along the diagonal. This behavior is consistent with the graphical depictions of $\zeta_1$ and 
$\zeta_3$ (or $\zeta_4$ and $\zeta_6$). In addition, it is straightforward to see that the $\uparrow\downarrow$ 
2nd-order correlation in Fig.\ \ref{f2ndg}(b) is consistent with the same graphics for $\zeta_1$ and $\zeta_3$
(or $\zeta_4$ and $\zeta_6$). Indeed, the most probable locations for the down-spin fermions 
are all three remaining corners, including the one across the diagonal. The higher probability at the corner
across the diagonal is accounted for by the fact that this corner appears in both graphics as probable 
location of the down fermions. Taking into consideration that $^2\cg_{\downarrow\downarrow} = 
^2\cg_{\uparrow\uparrow}$ and $^2\cg_{\downarrow\uparrow} = ^2\cg_{\uparrow\downarrow}$, one can conclude
that the dominant contributions in the many-body wave function contain the spin configuration
\begin{align}
 \zeta_1 - \zeta_3-\zeta_4+\zeta_6,
\label{dmsp}
\end{align} 
the minus signs resulting from the requirement that the spin function in Eq.\ (\ref{dmsp}) must be an
eigenstate of the total spin with $S=0$.

The spin-unresolved $^2\cg_{\rm unres}$ in Fig.\ \ref{f2ndg}(c) is the sum of all four spin-resolved 
correlations $^2\cg_{\uparrow\uparrow}$, $^2\cg_{\downarrow\downarrow}$, $^2\cg_{\uparrow\downarrow}$, and
$^2\cg_{\downarrow\uparrow}$, a fact that is reflected in the difference in the scales for the vertical 
axes going from Fig.\ \ref{f2ndg}(a) to Fig.\ \ref{f2ndg}(c); note that there are two spin-down, but only 
one spin-up, other fermions for any given spin-up fermion. Furthermore, although weakened, due to the 
overlap of the different components that are added up, the (0,4) square ring-like geometry is recognizable 
in Fig.\ \ref{f2ndg}(c), as well. 

Comparing the bottom row of panels in Fig.\ \ref{f2ndg} [i.e., panels (d), (e), and (f)] with the top row 
of panels in the same figure, it is apparent that placing the fixed point $\br_0$ at the origin misses 
crucial information concerning the many-body wave function, that is, it misses both the presence of the 
spin function displayed in Eq.\ (\ref{dmsp}), as well as the emergence of a square-ring Wigner-molecule 
configuration.     

\subsection{Spin structure of the symmetry-preserving $L=8$ relative ground state}
\label{spnstr}

Motivated by the analysis of the 2nd-order correlations in Sec.\ \ref{s0sz02nd}, showing that the spin 
function displayed in Eq.\ (\ref{dmsp}) must be an important component of the many-body CI wave function 
$\Phi_{\rm CI}^{L=8,S=0,S_z=0}$, it is instructive to interrogate whether the complete spin function of this
state can be determined from the miscroscopic CI wave function. To this end, we use the $c(J)$ 
coefficients (rounded to the fourth decimal point) listed in TABLE STIV in the SM \cite{supp}; 
naturally, we neglect the two orders-of-magnitude smaller $c(2)$ coefficient. With the above, the
CI wave function can be approximated by the sum of 15 Slater determinants (specified in TABLE STIV
by the single-particle angular momenta $l_i$, with $i=1,\ldots,4$), whose CI coefficients obey the following 
relations
\begin{align}
\begin{split}
& c(1)=c(16)=2c(4)=-2c(6)=-2c(9)=2c(11)=c_1, \\ 
& c(3)=c(15)=c_2,\\
& 2c(8)=2c(14)=-c_2,\\
& c(5)=c(13)=c_3, \\
& c(7)=c(12)=c_4, \\
& c(10)=c_5.
\label{rela}
\end{split}
\end{align}

From TABLE STIV in the SM \cite{supp}, one can extract numerical values for the 
5 constants $c_i$ (with $i=1,\ldots,5$)
in Eq.\ (\ref{rela}). However, as we will discuss below, the spin structure is independent of specific
numerical values. Note that the coefficients grouped together in each line of Eq.\ (\ref{rela}) are
associated with given (non-ordered) sets of single-particle angular momenta $l_i$, i.e., with the six sets
(0,1,3,4), (0,2,2,4), (1,2,2,3), (0,3,2,3), (1,2,1,4), and (1,3,1,3), respectively.

Using the relations (\ref{rela}) and the 15 Slater determinants in TABLE STIV in the SM \cite{supp}, 
and employing the MATHEMATICA algebraic language \cite{math}
we can write the CI wave function in the form of Eq.\ (\ref{phiz}). 
The analytic expressions of the space parts, $\cf_i(z_1,z_2,z_3,z_4)$ (with $i=1,\ldots,6$ and 
$z=x+iy=r e^{i\theta}$) are lengthy to be explicitly written in the text. However, the interested reader 
will find them as MATHEMATICA scripts in the Supplemental Material \cite{supp}.   

Before proceeding with the analysis, we recall here the form of the six spin eigenfunctions $\tz_i$ (with
$i=1,\ldots,6$) having both good total spin $S$ and a good spin projection $S_z$. These spin eigenfunctions
can be obtained by solving a 4-site Heisenberg Hamiltonian with the four spins arranged in a closed 
rectangular configuration, as was done in Appendix B of Ref.\ \cite{yann16}. Taking all four Heisenberg 
exchange constants to be equal, the spin eigenfunctions (B13)-(B18) in Ref.\ \cite{yann16} simplify to the 
following (relevant to the present paper) expressions:
\begin{align}
\tz_1= \frac{1}{\sqrt{12}}(\zeta_1+\zeta_3+\zeta_4+\zeta_6-2\zeta_2-2\zeta_5),\;\;\;S=0,
\label{tz1}
\end{align}
\begin{align}
\tz_2= \frac{1}{2}(\zeta_1-\zeta_3-\zeta_4+\zeta_6),\;\;\;S=0,
\label{tz2}
\end{align}
\begin{align}
\tz_3= \frac{1}{\sqrt{2}}(\zeta_6-\zeta_1),\;\;\;S=1,
\label{tz3}
\end{align}
\begin{align}
\tz_4= \frac{1}{\sqrt{2}}(\zeta_5-\zeta_2),\;\;\;S=1,
\label{tz4}
\end{align}
\begin{align}
\tz_5=\frac{1}{\sqrt{2}}(\zeta_4-\zeta_3),\;\;\;S=1,
\label{tz5}
\end{align}
\begin{align}
\tz_6= \frac{1}{\sqrt{6}}(\zeta_1+\zeta_2+\zeta_3+\zeta_4+\zeta_5+\zeta_6),\;\;\;S=2,
\label{tz6}
\end{align}
where the $\zeta_i$'s (with $i=1,\ldots,6$) were defined in Eq.\ (\ref{sppr}).

Solving the system of Eqs.\ (\ref{tz1}-\ref{tz6}) to obtain the spin primitives, $\zeta_i$ (with $i=1,\ldots,6$),
as a function of the spin eigenfunctions, $\tz_j$ (with $j=1,\ldots,6$), one can rearrange Eq.\ (\ref{phiz}) for
the many-body wave function as follows:
\begin{align}
\Phi_{\rm CI}=\sum_{i=1}^6 \cf_i \zeta_i = \sum_{i=1}^6 \tcf_i \tz_i,
\label{phiz2} 
\end{align}
where 
\begin{align}
\tcf_1= \frac{1}{\sqrt{12}}(\cf_1+\cf_3+\cf_4+\cf_6-2\cf_2-2\cf_5),\;\;\;S=0,
\label{tcf1}
\end{align}
\begin{align}
\tcf_2= \frac{1}{2}(\cf_1-\cf_3-\cf_4+\cf_6),\;\;\;S=0,
\label{tcf2}
\end{align}
\begin{align}
\tcf_3= \frac{1}{\sqrt{2}}(\cf_6-\cf_1),\;\;\;S=1,
\label{tcf3}
\end{align}
\begin{align}
\tcf_4= \frac{1}{\sqrt{2}}(\cf_5-\cf_2),\;\;\;S=1,
\label{tcf4}
\end{align}
\begin{align}
\tcf_5=\frac{1}{\sqrt{2}}(\cf_4-\cf_3),\;\;\;S=1,
\label{tcf5}
\end{align}
\begin{align}
\tcf_6= \frac{1}{\sqrt{6}}(\cf_1+\cf_2+\cf_3+\cf_4+\cf_5+\cf_6),\;\;\;S=2.
\label{tcf6}
\end{align}
We note that the arrangement of the $\tcf_i$'s in Eqs.\ (\ref{tcf1})-(\ref{tcf6}) coincide with that of the
$\zeta_i$'s in Eqs.\ (\ref{tz1})-(\ref{tz6}).

Using the analytic expressions \cite{supp} for the $\cf$'s, one can verify that 
$\tcf_3=\tcf_4=\tcf_5=\tcf_6=0$,
which is a confirmation of the fact that the CI wave function under consideration has total spin $S=0$. 
Consequently, one obtains the following general form for the $L=8$ ground state in the spin sector 
($S=0$, $S_z=0$):
\begin{align}
\Phi_{\rm CI}^{L=8,S=0,S_z=0}=\tcf_1 \tz_1 + \tcf_2 \tz_2.
\label{tcf1p2}
\end{align}  

From the analysis of 2nd-order correlations in Sec.\ \ref{s0sz02nd}, it follows that the contribution of the
first term in Eq.\ (\ref{tcf1p2}) must be less important than that of the second term. Indeed, this can further
be confirmed by choosing the four spatial coordinates to adhere to a square arrangement, i.e., by taking 
$z_1=z_0$, $z_2=z_0 e^{i\pi/2}$, $z_2=z_0 e^{i\pi}$, and $z_2=z_0 e^{i3\pi/2}$, with the point $z_0$ being
arbitrary. In this case, the spin structure of $\Phi_{\rm CI}^{L=8,S=0,S_z=0}$ agrees with Eq.\ (\ref{dmsp}),
i.e., one finds
\begin{align}
\begin{split}
&\tcf_1=0,\\
&\tcf_2 = \frac{3c_1 -3 \sqrt{6}c_2 +2\sqrt{2}c_3 +2\sqrt{3}c_4 -4c_5}{3 \pi^2 \sqrt{4!} } 
z_0^L e^{-2 z_0^* z_0},
\end{split}
\label{tcf1p2res}
\end{align}
where of course $L=8$ here.

We further note that both $\tcf_1$ and $\tcf_2$ may contain the associated Vandermonde determinant as 
a factor, like an assumption \cite{fert96} used earlier in the description of quantal versions of 
skyrmions. Naturally, the Fock antisymmetrization here is guaranteed by the fact that 
$\Phi_{\rm CI}^{L=8,S=0,S_z=0}$ is the sum of Slater determinants.

\begin{figure}[t]
\includegraphics[width=7cm]{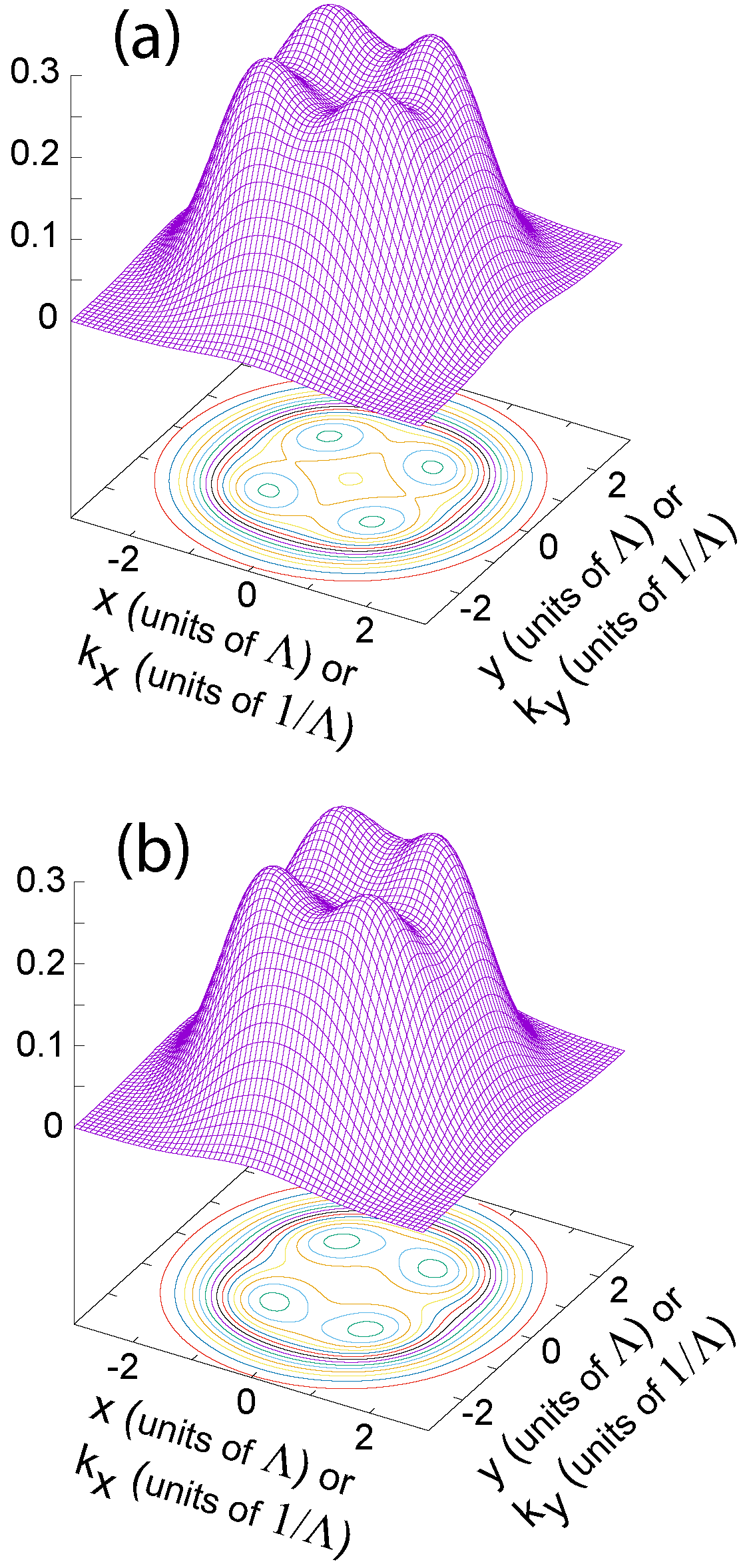}
\caption{
CI single-particle densities (both in real space and momentum space)
of the relative ground state of $N=4$ fermions in the spin sector with ($S=0$, $S_z=0$), but
considering a quadrupolar perturbation ($m=2$), in contrast to the hexadecapole perturbation in
Fig.\ \ref{fspd}. $R_\delta=0.40$. 3D surfaces are plotted. (a) Calculation for $\cc=0.0002$ 
(weak perturbation) at the point $\Omega/\omega=0.884651$. The expectation value of the total 
angular momentum is $\langle L \rangle=7.8501$, indicating that the plotted case is a state with broken rotational symmetry. (b) Calculation for $\cc=0.004$ (strong perturbation) at the point 
$\Omega/\omega=0.885$. The expectation value of the total angular momentum is 
$\langle L \rangle=7.8252$. For the spatial densities, the vertical axes are in units of $1/\Lambda^2$. For the momentum densities, the vertical axes are in units of $\Lambda^2$.
}
\label{fspdm2}
\end{figure}

\subsection{Pinning the Wigner molecule with a quadrupolar perturbation $(m=2)$}
\label{quad}

The 'flea on the elephant' behavior \cite{land20,jona81,simo85} played by the small perturbation in 
the emergence of the pinned and symmetry-broken WM was described in Sec.\ \ref{s0sz01st}. This 
behavior can be further illustrated by considering a $V_P$ with a multipolarity incommensurate 
to the intrinsic hexadecapole ($m=4$) multipolarity of the square Wigner molecule, associated with
$N=4$ fermions. To this effect, a quadrupolar multipolarity (i.e., $m=2$) is most relevant, because 
it may facilitate experimental endeavors.

In this context, Fig.\ \ref{fspdm2} displays the CI single-particle densities for $N=4$ LLL 
fermions at rotational frequencies located inside the region of the avoided crossing from angular 
momentum $L=4$ to $L=8$, but with the exact diagonalization of the many-body Hamiltonian 
(including $V_P$) performed with $m=2$  in Eq.\ (\ref{vp}).
It is seen that for a weak perturbation ($\cc = 0.0002$) the configuration of 
the Wigner molecule remains unaltered, exhibiting its intrinsic square geometry; see Fig.\ 
\ref{fspdm2}(a). For stronger perturbations (e.g., $\cc = 0.004$), the Wigner molecule starts 
feeling the details of the external perturbation and, naturally, it exhibits a slight rectangular 
deformation from the perfect square configuration; see Fig.\ \ref{fspdm2}(b).\\
~~~~~~~\\

\section{The spin sector ($S=2$, $S_z=0$): An analog of the (1,1,1) Halperin state}
\label{ferm}

Focusing now on the spin sector ($S=2$, $S_z=0$), we note that all 
the eigenvalues $E_{\rm int}$ associated with the contact-interaction term of the $H_{\rm LLL}$ 
Hamiltonian [third term in Eq.\ (\ref{H_lll})] are vanishing
[see Fig.\ \ref{fintspec}(c)], so that the curves in Fig.\ \ref{sps2} are non-intersecting straight lines, 
converging to zero for $\Omega/\omega=1$. The relative ground state has a total angular momentum
$L=6$, which is of interest because it coincides with the angular momentum of the trial wave function 
[denoted as (1,1,1)] proposed by Halperin \cite{halp83} for spinful fermions as a generalization of the 
celebrated Laughlin wave function \cite{laug83.2} (applicable only for the case of fully spin-polarized 
fermions). Indeed the general $(p,p,q)$ Halperin wave function, where $p$ and $q$ are positive
integers, is given by 
\cite{halp83,girv96,tong16}
\begin{align}
\begin{split}
\Upsilon&_{(p,p,q)}(z,w)= \\
&\prod_{i<j}^{N_\uparrow} (z_i-z_j)^p \prod_{k<l}^{N_\downarrow} (w_k-w_l)^p
\prod_{i,k}^{N_\uparrow,N_\downarrow} (z_i-w_k)^q.
\end{split}
\label{halp}  
\end{align}
In Eq.\ (\ref{halp}), $z_i=r_i e^{i\theta_i}$ and $w_k=r_k e^{i\theta_k}$ are the space coordinates
(here in units of $\Lambda$) in the complex plane
for the spin-up and spin-down fermions, respectively. Note further that the trivial Gaussian factors,
$\exp[-\sum_{i=1}^{N_\uparrow} z_i^* z_i/2]\exp[-\sum_{k=1}^{N_\downarrow} w_k^* w_k/2]$, 
have been omitted in Eq.\ (\ref{halp}). The total angular momentum associated with the
wave function $\Upsilon_{(p,p,q)}(z,w)$ is \cite{note4}
\begin{align}
L_{(p,p,q)}=p \frac{N_\uparrow ({N_\uparrow}-1)}{2}+p \frac{N_\downarrow ({N_\downarrow}-1)}{2}+
q {N_\uparrow} {N_\downarrow},
\label{lalp}
\end{align}
which indeed for ${N_\uparrow}={N_\downarrow}=2$ and $p=q=1$ gives $L_{(1,1,1)}=6$.

Of significance is the fact that the  original proposal for the $\Upsilon(z,w)$ wave functions did not 
include the spin variables. 
Below we will investigate the connection of the (1,1,1) Halperin wave function to the CI many-body wave 
function which is the relative ground state in the ($S=2$, $S_z=0$) spin sector \cite{note5};
recall that the relative ground state is the lowest-in-energy state within each spin sector.
Furthermore using this connection we will demonstrate a two-dimensional case of {\it mapping\/} from spinful 
to spinless fermions that is analogous to the fermionization mapping in one dimension 
\cite{gira60}.

\begin{figure}[t]
\includegraphics[width=7.0cm]{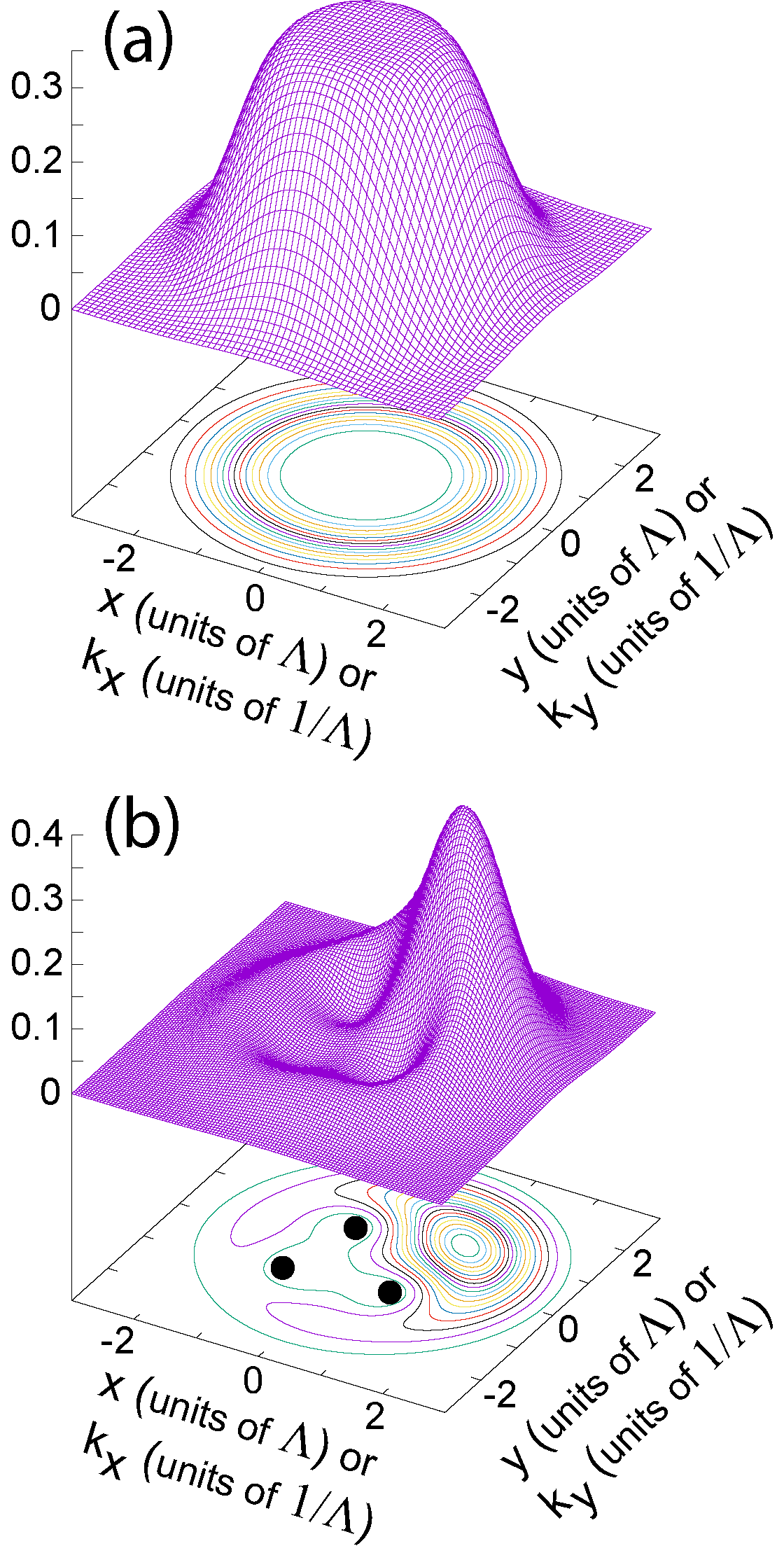}
\caption{Structure of the relative CI ground state (in both real space and momentum space) in the spin sector 
($S=2$, $S_z=0$) for $N=4$ fermions at the trap angular frequency $\Omega/\omega=0.90$. 
(a) Single-particle density. (b) Spin unresolved 4th-order correlation.
The strength of the $V_P$ perturbation is weak with $\cc=0.0001$. 
$R_\delta=0.4$, and the order of the multipole trap deformation $m=4$. 
Because of the properties of the Fourier transform of the LLL orbitals [see Eq.\ (\ref{psik})], 
both real-space and momentum densities and correlations are given by the same 3D numerical surface.
For the spatial quantities, the lengths along the $x$ and $y$ axes are given in units of $\Lambda$, and the
vertical axes are in units of $1/\Lambda^2$ for the density and $1/(\pi^4\Lambda^8)$ for the 4th-order
correlation. For the momentum quantities, the momenta along the $k_x$ and $k_y$ axes are given in units 
of $1/\Lambda$, and the vertical axes are in units of $\Lambda^2$ for the density and $\Lambda^8/\pi^4$ 
for the the 4th-order correlation.
The 2D single-particle density in (a) is rotationally symmetric. In (b), the three fixed 
points (denoted by solid dots) are placed at a radius $r_0=0.90$ $\Lambda$, whereas the azimuthal angle
between them is $\pi/2$, and the reference angle $\Theta=\pi/4$.  
}
\label{fs2}
\end{figure}

\begin{table}[b]
\caption{\label{ts2sz0} 
Dominant coefficients, $c(I)$, in the CI expansion of the relative LLL ground state (with $L=6$) 
in the ($S=2$, $S_z=0$) spin sector. The CI expansion ($I=1,2,\ldots,I_{\rm total}$) consists of 
$I_{\rm total}=1296$ basis determinants. The index $J$ is introduced to relabel the dominant 
coefficients. The dominance criterion was $|c(I)| > 10^{-3}$.}
\begin{ruledtabular}
\begin{tabular}{r|c|r|c|c}
$I$&$J$&$c(J)$    &  $(l_1\uparrow,l_2\uparrow,l_3\downarrow,l_4\downarrow)$ & 
$\sum_{i=1}^4 l_i $ \\ \hline
16  & 1 & -0.4082472 & (0,1,2,3)  & 6 \\
46  & 2 &  0.4082472 & (0,2,1,3)  & 6 \\
81  & 3 & -0.4082472 & (0,3,1,2)  & 6 \\
291 & 4 & -0.4082472 & (1,2,0,3)  & 6 \\
326 & 5 &  0.4082472 & (1,3,0,2)  & 6 \\
541 & 6 & -0.4082472 & (2,3,0,1)  & 6 \\
\end{tabular}
\end{ruledtabular}
\end{table}
     
\subsection{The 4th order correlation and the molecular configuration}
\label{s2sz04th}

First in Fig.\ \ref{fs2} we display the single-particle density [Fig.\ \ref{fs2}(a)] and the 
corresponding spin-unresolved 4th-order correlation [Fig.\ \ref{fs2}(b)] for the  CI state with
$S=2$, $S_z=0$ and $L=6$. It is seen that the single-particle density is rotationally symmetric, but an
intrinsic square geometrical configuration appears in the unresolved 4th-order correlation. This is similar
to the behavior found for the CI relative ground state in the spin sector ($S=0$, $S_z=0$) at the point
$\Omega/\omega=0.90$. Common to these two states is the fact that the corresponding angular momenta, i.e.,
$L=6$ and $L=8$, respectively, are magic ones compatible with the $C_4$ point group symmetry; see Eq.\
(\ref{mam}).

\subsection{Comparison between CI state and trial (1,1,1) Halperin wave function}
\label{s2sz0comp}

Furthermore, in TABLE \ref{ts2sz0}, we list the dominant CI coefficients, $c(I)$, and the spin 
orbitals, $(l_1\uparrow, l_2\uparrow, l_3\downarrow, l_4\downarrow)$, entering into the associated 
basis of Slater determinants (see Sec. \ref{ci}). The criterion used for  selection of the most 
dominant determinants in the CI solution was $|c(I)| > 10^{-3}$. The CI calculation used
1296 basis determinants with all possible total angular momenta from 2 to 30. From TABLE 
\ref{ts2sz0}, it is apparent that only six determinants with $L=6$ and equal weighting coefficients,
$|c(J)|$, contribute to the CI LLL state with $S=2$, $S_z=0$; indeed 
$\sum_{i=1}^6 |c(J)|^2=0.99999475$, i.e., the corresponding normalization constant differs from 
unity only in the sixth decimal point.

Taking into consideration that the numerical value of the $|c(J)|$'s in TABLE \ref{ts2sz0} equals 
$1/\sqrt{6}$, up to the sixth decimal point, and that the LLL single-particle orbitals [with lengths
in units of $\Lambda$, the harmonic confinement oscillator length, see text below Eq.\ (\ref{psir})]
are written as $z^l\exp[-z^*z/2]/\sqrt{\pi l!}$, one can verify the following algebraic identity
\begin{align}
\begin{split}
& \Phi_{\rm CI}^{S=2,\; S_z=0} \\
& = \sum_{J=1}^6 \frac{\sgn(J)}{\sqrt{4!} \sqrt{6} \sqrt{2!3!}\pi^2} \\
& ~~~\times {\rm Det}[z_1^{l_1(J)}\alpha(1), z_2^{l_2(J)}\alpha(2), 
z_3^{l_3(J)}\beta(3), z_4^{l_4(J)}\beta(4)] \\
& = - \frac{1}{ 2 \sqrt{3 \times 4!} \pi^2 }
\left( \prod_{i<j}^4 (z_i-z_j) \right) \frac{ \sum_{J=1}^6 \zeta_J } { \sqrt{6} }, 
\end{split}
\label{iden}
\end{align}
where $\sgn(J)$ is the $+$ or $-$ sign of the $c(J)$ coefficients according to TABLE \ref{ts2sz0}.
The $\zeta_J$, $J=1,2,\ldots,6$ are defined in Eq.\ (\ref{sppr}), and we omitted the trivial 
Gaussian factors. 

Renaming the spatial coordinates of the spin-up fermions as $z_3 \rightarrow w_1$ and $z_4 \rightarrow w_2$,
one sees immediately that the space part of the $\Phi_{\rm CI}^{S=2,\; S_z=0}$ wave function [see 
Eq.\ (\ref{iden})] coincides with the (1,1,1) Halperin function, i.e., with the expression for $\Upsilon(z,w)$
in Eq.\ (\ref{halp}) when $p=q=1$. We recall here the possibility that in certain instances the LLL CI wave 
function may be expressed exactly in analytical form, as it has been noted in earlier publications 
\cite{pape01,wilk00.2,yann10} for the case of LLL ground states of a few spinless bosons in the range 
$0 \leq L \leq N$. As a notable counterexample, we mention here the 
disagreement between the Moore-Read trial wave function \cite{mr91}, which consists mainly of a (0,5) ring
configuration, and the CI wave function, which contains mainly a (1,4) ring configuration, in the case of
the LLL ground state for $N=5$ spinless bosons and $L=8$ \cite{note6}. 

\begin{figure}[t]
\includegraphics[width=7.0cm]{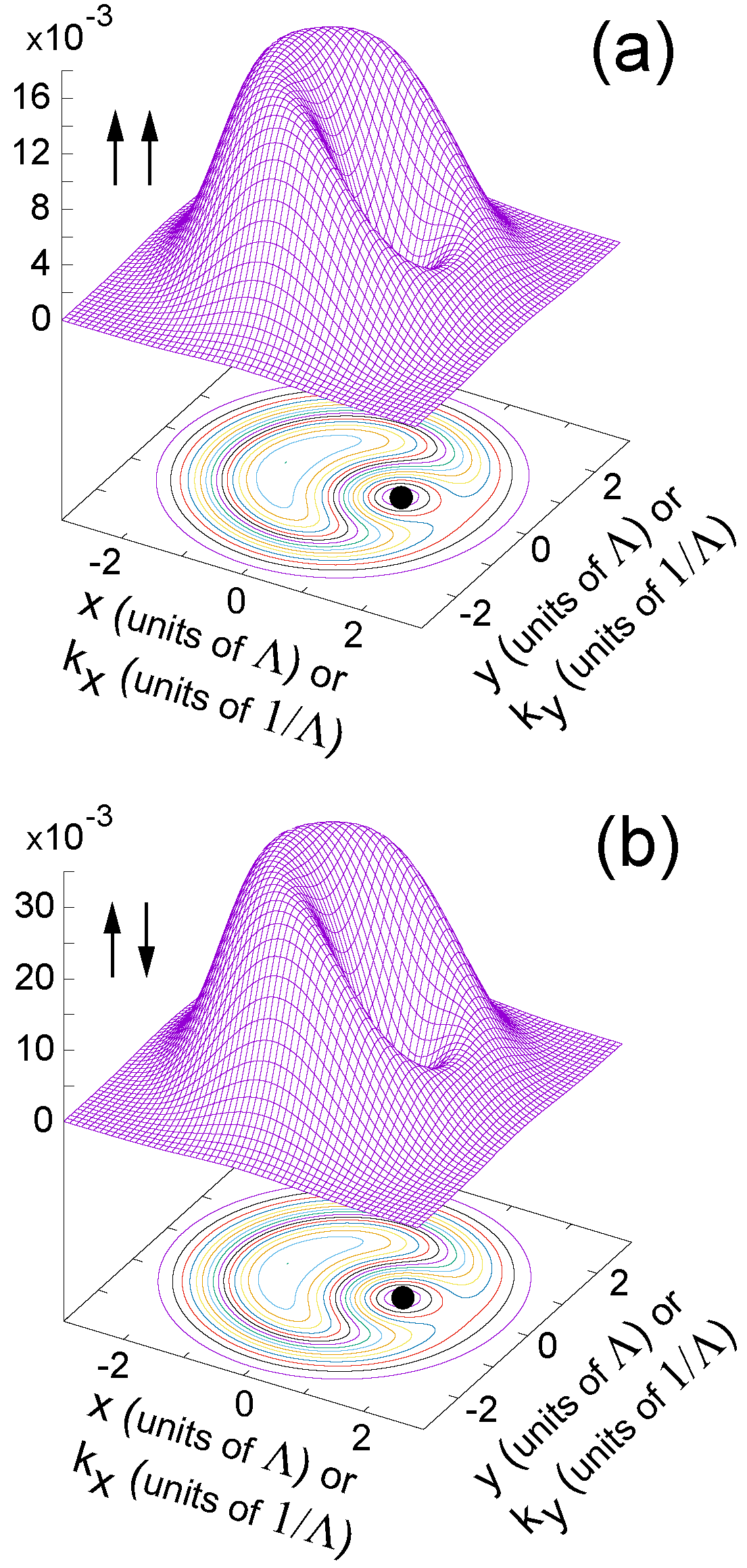}
\caption{CI spin-resolved 2nd-order correlations of the relative ground state in the spin sector
($S=2$, $S_z=0$) for $N=4$ fermions at the trap angular frequency $\Omega/\omega=0.90$.
(a) $\uparrow \uparrow$ correlation. (b) $\uparrow \downarrow$ correlation.
The strength of the $V_P$ perturbation is weak with $\cc=0.0001$.
$R_\delta=0.4$, and the order of the multipole trap deformation $m=4$.
The fixed point (denoted by a solid dot) was placed at a radius $r_0=0.90$ $\Lambda$. 
Because of the properties of the Fourier transform of the LLL orbitals [see Eq.\ (\ref{psik})],
both real-space and momentum correlations are given by the same 3D numerical surface.
For the spatial correlations, the lengths along the $x$ and $y$ axes are given in units of $\Lambda$,
and the vertical axes are in units of $1/\Lambda^4$.
For the momentum correlations, the momenta along the $k_x$ and $k_y$ axes are given in units
of $1/\Lambda$, and the vertical axes are in units of $\Lambda^4$. 
Note the different scales between (a) and (b).
}
\label{fs2nd}
\end{figure}

\subsection{What about the 2nd-order correlations?}
\label{s2sz02nd}

Unlike the approach in this paper, and a handful of earlier publications \cite{baks07,yann10},
the 2nd-order correlations have been traditionally considered sufficient (see, e.g., Refs.\
\cite{laug83.2,maks96,barb06,shi07,popp04,palm18}) for analyzing the intrinsic structure of the
highly-correlated LLL states. The case of the ($S=2$, $S_z=0$) CI LLL state for $N=4$ fermions and
$L=6$ shows that the above supposition does not hold in general. Indeed, in Fig.\ \ref{fs2nd},
we display the up-up ($\uparrow \uparrow$) and up-down ($\uparrow \downarrow$)
spin-resolved 2nd-order correlations for this CI state [which corresponds to the (1,1,1) Halperin wave
function]. Note that there is a 1-to-2 ratio between the $\uparrow \uparrow$ and the ($\uparrow \downarrow$)
2nd-order correlations, because, for each spin-up fermion, there are one spin-up and two spin-down 
additional fermions.

As seen from Fig.\ \ref{fs2nd}, only the existence of the zero probability for finding two fermions at the
same position is visible. Any signature of the square-ring intrinsic molecular structure 
has been washed away in Fig.\ \ref{fs2nd} due to the averaging performed through the double 
integrations over the coordinates of the third and fourth particles; see the definition of the 2nd-order 
correlations in Eq.\ (\ref{2ndc}). Revealing the intrinsic Wigner-molecule structure using 2nd-order 
correlations requires higher total angular momenta, as shown in Ref.\ \cite{baks07} for the analogous 
cases of LLL bosons. However, it appears that the experimental window \cite{palm18} for a few rapidly 
rotating ultracold fermions is restricted to the range of small $L$'s, up to values in the neighborhood of
$L_{(1,1,1)}$, corresponding to the (1,1,1) Halperin states. Consequently, we conclude that consideration 
of the $N$-body correlations offers, as shown in this paper, essential additional information regarding the 
CI wave functions.

\section{A fermionization analog in two dimensions}
\label{fermanal}

\subsection{The $L=6$ state for $N=4$ fermions}
\label{fermn4}

The derivation of the exact CI analytic expression in Eq.\ (\ref{iden}) enabled us to make another 
important comparison. It is well known that the fully spin-polarized fermionic LLL CI state with
$L=N(N-1)/2=L_{(1,1,1)}^{N_\uparrow=N_\downarrow}$ consists of only one Slater determinant constructed 
with the single-particle orbitals
$z^0\alpha$, $z^1\alpha$,...,$z^{N-1}\alpha$ (again the Gaussian factors are omitted). For the case of 
$N=4$ fermions, this state is written as (considering that the space part is a Vandermonde determinant):
\begin{align}
\begin{split}
& \Phi_{\rm CI}^{S=2,\; S_z=2} \\
& = \frac{1}{ \sqrt{4!}\sqrt{2!3!}\pi^2 } {\rm Det}[\alpha(1),z_2\alpha(2),z_3^2\alpha(3),z_4^3\alpha(4)] \\
& = \frac{1}{ 2 \sqrt{3 \times 4!} \pi^2 } 
\left( \prod_{i<j}^4 (z_i-z_j) \right) \alpha(1) \alpha(2) \alpha(3) \alpha(4).
\end{split}
\label{s2sz2}
\end{align}       

One sees that, apart from a sign, the space parts of the spinful $\Phi_{\rm CI}^{S=2,\; S_z=0}$ and 
the fully spin-polarized $\Phi_{\rm CI}^{S=2,\; S_z=2}$ wave functions are the same. This mapping 
between a nonpolarized many-body wave function representing {\it repulsively interacting\/} fermions 
and that of fully spin-polarized (equivalent to spinless) {\it non-interacting\/} fermions is reminiscent of 
the well-known mapping \cite{gira60} in one dimension between the wave function of $N$ hard 
bosons, i.e., bosons with strong interparticle contact interaction, and that of $N$  
non-interacting and spinless fermions; it can be viewed as a generalization of the ``fermionization''
concept \cite{gira60} to two dimensions 
(indeed fermions with different spin projection can coexist in the same position like two bosons).   

We note that, because of the exchange hole, the contact interaction becomes inoperative in the case 
of fully polarized (or spinless) fermions, and as a result this fermionization mapping demonstrates 
that the intrinsic crystalline correlations portrayed in Fig.\ \ref{fs2}(b) can be generated, as a 
limiting case to the quantum Wigner molecule, by the Pauli exclusion principle alone \cite{note7}. 

\begin{table}[t]
\caption{\label{ts3sz0} 
The 20 dominant coefficients, $c(I)$, in the CI expansion of the relative LLL ground state for $N=6$
and  $L=15$ in the ($S=3$, $S_z=0$) spin sector. The CI expansion ($I=1,2,\ldots,I_{\rm total}$) 
consists of $I_{\rm total}=7056$ basis determinants. The index $J$ is introduced to relabel the dominant 
coefficients. The dominance criterion was $|c(I)| > 10^{-3}$.}
\begin{ruledtabular}
\begin{tabular}{r|c|r|c|c}
$I$&$J$&$c(J)$    &  $(l_1\uparrow,l_2\uparrow,l_3\uparrow,l_4\downarrow, l_5\downarrow, l_6\downarrow)$ & 
$\sum_{i=1}^6 l_i $ \\ \hline
65    & 1 & -0.22360479 & (0,1,2,3,4,5)  & 15 \\
139  & 2 &  0.22360479 & (0,1,3,2,4,5)  & 15 \\
219  & 3 & -0.22360479 & (0,1,4,2,3,5)  & 15 \\
302  & 4 &  0.22360479 & (0,1,5,2.3.4)  & 15 \\

628    & 5 & -0.22360479 & (0,2,3,1,4,5)  & 15 \\
708    & 6 &  0.22360479 & (0,2,4,1,3,5)  & 15 \\
791    & 7 & -0.22360479 & (0,2,5,1,3,4)  & 15 \\
1123  & 8 & -0.22360479 & (0,3,4,1,2,5)  & 15 \\

1206  & 9   &  0.22360479 & (0,3,5,1,2,4)  & 15 \\
1541  & 10 & -0.22360479 & (0,4,5,1,2,3)  & 15 \\
2371  & 11 &  0.22360479 & (1,2,3,0,4,5)  & 15 \\
2451  & 12 & -0.22360479 & (1,2,4,0,3,5)  & 15 \\

2534  & 13 &  0.22360479 & (1,2,5,0,3,4)  & 15 \\
2866  & 14 &  0.22360479 & (1,3,4,0,2,5)  & 15 \\
2949  & 15 & -0.22360479 & (1,3,5,0,2,4)  & 15 \\
3284  & 16 &  0.22360479 & (1,4,5,0,2,3)  & 15 \\

4120  & 17 & -0.22360479 & (2,3,4,0,1,5)  & 15 \\
4203  & 18 &  0.22360479 & (2,3,5,0,1,4)  & 15 \\
4538  & 19 & -0.22360479 & (2,4,5,0,1,3)  & 15 \\
5377  & 20 &  0.22360479 & (3,4,5,0,1,2)  & 15 \\
\end{tabular}
\end{ruledtabular}
\end{table}

\subsection{The $L=15$ state for $N=6$ fermions}
\label{fermn6}

The fermionization analog discussed in Sec.\ \ref{fermn4}, i.e., the precise mapping between the 
space parts of the nonpolarized ($S=N/2$, $S_z=0$) state with $L=N(N-1)/2$ and the corresponding 
fully polarized ($S=N/2$, $S_z=N/2$) one, is not limited to the $N=4$ case. Here, we elaborate on
another example concerning the more complex LLL state of $N=6$ fermions with an angular momentum
$L=15$.

As a first step, we establish that the space part of the CI relative ground state for $N=6$ fermions
with $L=15$ and ($S=3$, $S_z=3$) is very well approximated by the Halperin (1,1,1) trial function.
To this end, in TABLE \ref{ts3sz0} we list the 20 dominant Slater determinants, out of a total of
7056 ones with various total $L$'s that comprise the basis employed in the actual CI calculation; 
indeed $\sum_{i=1}^{20} |c(J)|^2=0.9999820$, i.e., the corresponding normalization constant differs 
from unity only in the fifth decimal point.

From TABLE \ref{ts3sz0}, two observations are crucial, that is: (1) The coefficients $c(J)$ of 
these 20 Slater determinants are all equal and their absolute value approximates  
$1/\sqrt{20}=0.22360680$ to the 5th decimal point, and (2) Only the six single-particle angular 
momenta 0, 1, 2, 3, 4, and 5 appear in these dominant Slater determinants, resulting for
all of them in a total angular momentum $L=15$.

Taking into consideration that the numerical value of the $|c(J)|$'s in TABLE \ref{ts3sz0} equals
$1/\sqrt{20}$, up to the fifth decimal point, and that the LLL single-particle orbitals [with 
lengths in units of $\Lambda$, the harmonic confinement oscillator length, see text below Eq.\ 
(\ref{psir})] are written as $z^l\exp[-z^*z/2]/\sqrt{\pi l!}$, one can verify the following 
algebraic identity (using the MATHEMATICA scripts given in the Supplemental Material \cite{supp}):
\begin{widetext}
\begin{align}
\begin{split}
\Phi_{N=6,\;{\rm CI}}^{S=3,\; S_z=0} 
& = \sum_{J=1}^{20} \frac{\sgn(J)}{\sqrt{6!} \sqrt{20}}
{\rm Det}[z_1^{l_1(J)}\alpha(1), z_2^{l_2(J)}\alpha(2), z_3^{l_3(J)}\alpha(3), 
z_4^{l_4(J)}\beta(4), z_5^{l_5(J)}\beta(5), z_6^{l_6(J)}\beta(6),] \\
& = - \frac{1}{ \sqrt{2!3!4!5! } \sqrt{6!}  \pi^3 }
\left( \prod_{i<j}^6 (z_i-z_j) \right) \frac{ \sum_{J=1}^{20} Z_J } { \sqrt{20} },
\end{split}
\label{iden2}
\end{align}
\end{widetext}
where we omitted the trivial Gaussian factors. $\sgn(J)$ is the $+$ or $-$ sign of the $c(J)$ 
coefficients according to TABLE \ref{ts3sz0}. The $Z_J$'s, $J=1,2,\ldots,20$ are the spin primitives
associated with 6 spins having an $S_z=0$ total spin projection. They are defined explicitly in
Appendix  \ref{zetan6}.

The $L=15$ state corresponding to $N=6$ fully polarized and non-interacting fermions is written as 
(considering that the space part is a Vandermonde determinant):
\begin{align}
\begin{split}
& \sqrt{2!3!4!5!}\sqrt{6!} \pi^3  \Phi_{ N=6,\; {\rm CI}}^{S=3,\; S_z=3} \\
& = {\rm Det}[\alpha(1),z_2\alpha(2),z_3^2\alpha(3),z_4^3\alpha(4),z_5^4\alpha(5),z_6^5\alpha(6)] \\
& = 
\left( \prod_{i<j}^6 (z_i-z_j) \right) \alpha(1) \alpha(2) \alpha(3) \alpha(4) \alpha(5) \alpha(6).
\end{split}
\label{s3sz3}
\end{align}  

As was the case for $N=4$ LLL fermions, one sees from Eqs.\ (\ref{iden2}) and (\ref{s3sz3}) that, 
apart from a sign, the space parts of the spinful  $\Phi_{ N=6,\; {\rm CI} }^{S=3,\; S_z=0}$ and of 
the fully spin-polarized $\Phi_{ N=6,\;{\rm CI} }^{S=3,\; S_z=3}$ wave functions are the same, offering 
another example of the concept of "fermionization" in two dimensions discussed in Sec.\ \ref{fermn4}. 

\begin{figure}[t]
\includegraphics[width=7.5cm]{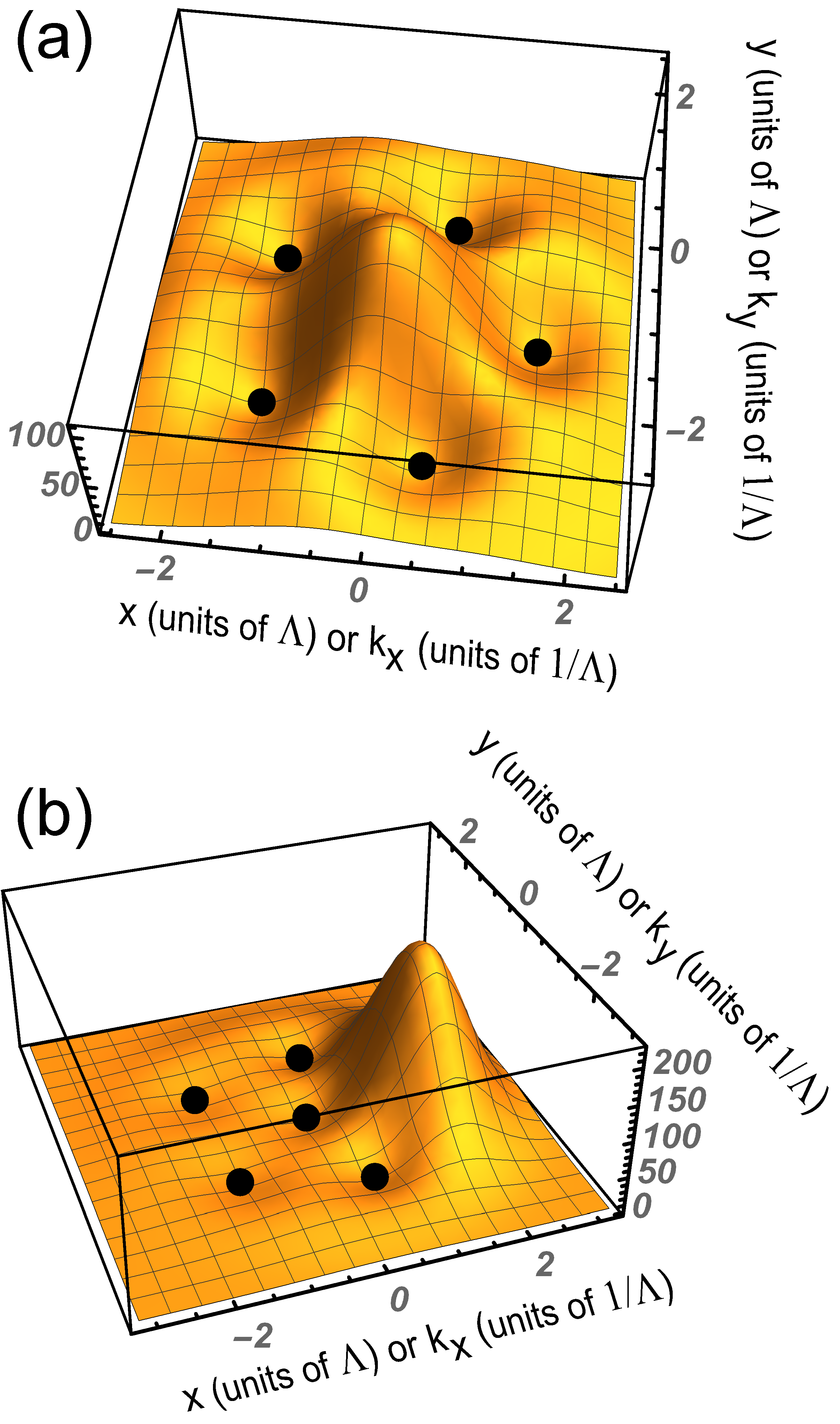}
\caption{
The $^6\cg(x,y)$ (with $x+iy=z$) 6th-order spin-unresolved correlation [see Eq. (\ref{gg6n6})] of $N=6$ 
LLL fermions for the relative ground state with $L=15$  in the spin sector $S=3$ (with $S_z=0$ or $S_z=3$).
It exhibits a (1,5) intrinsic geometrical configuration. (a) The fixed points (highlighted by
solid dots) are placed at $z_2^0=r_0$,
$z_3^0=r_0 e^{2\pi i /5}$, $z_4^0=r_0 e^{4\pi i/5}$, $z_5^0=r_0 e^{6\pi i/5}$, and $z_6^0=r_0 e^{8\pi i/5}$. 
(b) The fixed points (highlighted by solid dots) 
are placed at $z_2^0=0$, $z_3^0=r_0 e^{2\pi i/5}$, $z_4^0=r_0 e^{4\pi i/5}$, 
$z_5^0=r_0 e^{6\pi i/5}$, and  $z_6^0=r_0 e^{8\pi i/5}$. $r_0=1.6$ $\Lambda$. The units of the vertical 
axes are arbitrary, but the same for both frames. The same surface portrays also the associated
momentum correlations $^6{\cal G}(k_x,k_y)$,
with corresponding mappings for the fixed points, i.e., $r_0 \rightarrow k_r^0 = 1.6$ $1/\Lambda$.}
\label{fgg6n6}
\end{figure}

To complete the inquiry concerning the $L=15$ relative ground state for $N=6$ LLL fermions
in the spin sector $S=3$ (with $S_z=0$ or $S_z=3$), we investigate its intrinsic symmetries. 
To this end, we display in Fig.\ \ref{fgg6n6} the associated spin-unresolved sixth-order correlation 
function. Instead of using the numerical result of the CI calculation, we can take advantage of the 
identity in Eq.\ (\ref{iden2}) and plot the algebraic expression 
\begin{align}
\begin{split}
& ^6\cg(z,z_2^0,z_3^0,z_4^0,z_5^0,z_6^0)\\
& \propto (z-z_2^0)(z-z_3^0)(z-z_4^0)(z-z_5^0)(z-z_6^0) \\ 
& \times (z^*-z_2^{0*})(z^*-z_3^{0*})(z^*-z_4^{0*})(z^*-z_5^{0^*})(z^*-z_6^{0^*}).
\end{split}
\label{gg6n6}
\end{align}  
The superscript 0 indicates a fixed point. Two cases are plotted in Fig.\ \ref{fgg6n6}: in Fig.\ 
\ref{fgg6n6}(a) the five fixed points are located on a circle of radius $r_0=1.6$ $\Lambda$ and form 
a regular pentagon, whereas in Fig.\ \ref{fgg6n6}(b), one of the five fixed points was moved from the 
corner of the pentagon to the origin. In Fig.\ \ref{fgg6n6}(a), it is apparent that the sixth fermion lies 
at the origin, while in Fig.\ \ref{fgg6n6}(b), the sixth fermion completes the apices of the regular 
pentagon. The above results for the 6th-order correlation function clearly portray the formation of a 
concentric  (1,5) arrangement of the emergent UC-RWM in the LLL in the case of six trapped 
fermionc atoms, described by the wave function in Eq.\ (\ref{iden2}).  

We note that the (1,5) ring geometrical structure found here to be associated with $N=6$ LLL fermions 
(both contact-interacting and non-interacting) is familiar \cite{yann07}, as one of two competing 
configurations from the case of $N=6$ Coulomb-interacting confined electrons in the field of 2D 
semiconductor quantum dots, the other configuration being a (0,6) arrangement. \\
~~~~~\\ 

\section{Discussion: The role of the Wigner parameter}
\label{wigp}

The dimensionless parameter $R_\delta$ [defined in Eq.\ (\ref{rdlt})] enters naturally in the many-body 
LLL Hamiltonian in Eq.\ (\ref{H_lll}). This applies also for the many-body Hammiltonians for ultracold
atoms in non-rotating traps, i.e., when $\Omega=0$; see, e.g., Ref.\ \cite{roma04}. We note that, in the 
absence of a magnetic field and for a finite number of $N$ trapped electrons in 2D semiconductor quantum 
dots, a corresponding parameter \cite{yann00,yann99,yann07} (usually referred to as the Wigner parameter) 
is defined as 
\begin{align}
R_W=Q/(\hbar \omega),
\label{rw}
\end{align}
where $Q=e^2/(\kappa l_0)$ is the Coulomb repulsive energy between two electrons at a distance
equal to the oscillator strength $l_0=\sqrt{\hbar/(m^*_e\omega)}$, $\kappa$ is the dielectric constant of the
semiconducting medium, $m^*_e$ is the effective mass of the electron, and $\omega$ is the frequency of the
parabolic (harmonic) 2D potential confinement. 

In the case of a high applied magnetic field $B$ (LLL Hilbert space), 
$l_0$ in Eq.\ (\ref{rw}) is replaced by the magnetic 
length and $\omega$ is replaced by the cyclotron frequency, that is, $l_0 \rightarrow l_B$ and 
$\omega \rightarrow \omega_c=eB/(m^*c)$, with $l_B=\sqrt{\hbar/(m^*\omega_c)}$. As is the case with the
contact interaction -- i.e., the fact (discussed in Sec.\ \ref{spec}) that the LLL spectrum associated solely 
with the interaction term, $H_{\rm int}$ [third term in Eq.\ (\ref{H_lll})], scales with $R_\delta$ -- 
the LLL spectrum associated solely with the long-range Coulomb repulsion scales also with $R_W$. 
As a result, the values $R_\delta$ and $R_W$ do not influence the intrinsic structure of the LLL many-body 
wave functions. [Note that the eigenstates of $H_{\rm int}$ are also eigenstates of the LLL kinetic-energy 
Hamiltonian, $H_K$; see second term in Eq.\ (\ref{H_lll}).] 
The independence of WM formation from the 
precise value of $R_\delta$ is further demonstrated in Fig.\ \ref{af1} in Appendix \ref{a3}, where a
different value $R_\delta=0.2$ was used.

The only effect of the magnitude of $R_\delta$ is
to determine the precise value of $\Omega/\omega$ where the crossings in Fig.\ \ref{avcr} occur. 
In contrast, for vanishing and small magnetic fields, or for a non-rotating trap, the emergence of the Wigner
molecular structures does depend on the value of $R_W$ and $R_\delta$, respectively, requiring values of 
these parameters larger than unity 
\cite{yann00,yann99,yann07,yues07,ront06,roma04,cava07,blun10,note9}. 

The apparent above inconsistency concerning the qualitative role of the Wigner parameter
motivates the following deeper insight. Indeed both the $R_W$ and $R_\delta$ parameters at $B=0$ and 
$\Omega=0$, respectively, express the ratio
\begin{align}
{\cal R}= \frac{\Delta E_{\rm int}}{\Delta E_{\rm sp}}, 
\label{calr} 
\end{align}
where $\Delta E_{\rm int}$ is a representative amount of repulsive energy and $\Delta E_{\rm sp}$ is 
an average energy spacing in the single-particle spectrum. For $B=0$, or $\Omega=0$, the $\hbar\omega$ used 
in Eqs.\ (\ref{rdlt}) and (\ref{rw}) reflects indeed the average energy gap between the single-particle 
states of the familiar 2D harmonic oscillator. In the case of the Landau-level spectrum 
(Fock-Darwin oscillator \cite{fock,darw,yann07}), $2 \hbar\omega$, or $\hbar\omega_c$, 
represents the energy spacing between Landau levels. However, the relevant many-body Hilbert space
is restricted in the LLL where the energy gap 
between the single-particle states vanishes due to the well-known infinite degeneracy of the Landau levels;
this is also referred to as single-particle kinetic-energy quenching. Thus with
respect to the pertinent dimensionless parameter that controls Wigner-molecule formation in the LLL, the 
denominator in Eq.\ (\ref{calr}) must be taken to be presisely zero, which results in all instances in an 
infinite value for ${\cal R}$. Interestingly, the single-outcome value of $R\rightarrow +\infty$ implies 
that the LLL many-body case is preset for favoring the emergence of Wigner molecules, independently of the 
strength or the type of the two-body interaction. In fact, in addition to the Coulombic and 
contact-interaction cases, this qualitative prediction has been confirmed by numerical calculations in the 
case of few fully spin-polarized LLL fermions interacting via a dipole-dipole potential
\cite{lewe07}. 

\section{Summary}
\label{summ}

The development and employment of both computational, numerical (two-dimensional configuration 
interaction \cite{yann03,yann07,yann04}) and algebraic (MATHEMATICA \cite{math}), state-of-the-art
methodological approaches were shown here to bring forth advanced tools (e.g., all-order momentum 
correlations) that boost and refine our ability to in-depth interrogate the complex many-body 
physics underlying the fractional quantum-Hall effect in assemblies of a few ultracold neutral 
fermionic atoms, interacting via repulsive contact potentials and confined in a single rapidly 
rotating two-dimensional harmonic trap. We considered spinful fermionic atom assemblies, where in 
addition to the two-dimensional orbital degree of freedom, each orbital within a degenerate Landau
level state has also spin degrees of freedom. Detailed results were given for the illustrative 
example of four spinful ultracold fermions in a rapidly rotating trap (a case anticipated to be 
among the first to be experimentally explored in the near future).

As pointed out earlier \cite{geme10,palm18,popp04,hazz08}, rotating assemblies of a few ultracold 
atoms have become particularly promising for exploring the LLL physics due to experimental 
difficulties in reaching sufficiently dilute regimes (low filling fractions) with a large number 
of atoms \cite{schw04,zwie05} in rotating traps; in the former experiment \cite{schw04} high 
rotational rates of a BEC cloud of $^{87}$Rb atoms resulted in formation of ordered Abrikosov 
vortices, and similarly for the case of a large-number BEC cloud of $^7$Li  atoms \cite{zwie05}. 
In this context, the raised level of understanding brought forth by consideration of the $N$-body 
correlations, compared to studies limited to examination of merely the 2nd-order ones, appears to 
be pivotal for making further progress in this field. This is the case in particular because the 
experimental window of fermionic LLL states is restricted to the lowest range of total angular 
momenta, up to values in the neighborhood of $L_{(1,1,1)}$ associated with the (1,1,1) Halperin 
wave function. Indeed for spin-balanced assemblies (with $N$ particles), the total angular 
momentum value is $L_{(1,1,1)} = N(N-1)/2$, which is smaller than  the value of $N(N-1)$ for the 
bosonic, and $3N(N-1)/2$ for the fermionic, Laughlin states. 

Our analysis showed that the few-body LLL states with magic angular momenta exhibit intrinsic 
ordered quantum structures in the $N$-body correlations, similar to those associated with rotating
Wigner molecules \cite{yann03,yann07}, familiar from the field of semiconductor quantum dots under
high magnetic fields. 

The application of a small perturbing stirring potential $V_p$ [specifically a multipole 
deformation of the trap; see Eq.\ (\ref{vp})] induces, in the neighborhood of the ensuing avoided 
crossings in the global LLL energy spectra [see Fig.\ \ref{avcr} associated with the ($S=0$, 
$S_z=0$) spin sector], states with broken rotational symmetry (i.e., without good total angular 
momenta, referred to accordingly as pinned Wigner molecules). These structures, exhibit 
molecular-type (or crystalline-type) configurations which are manifested already at the lowest 
level of first-order correlations (i.e., in the single-particle CI spin-unresolved densities;
see Fig.\ \ref{fspd}). This behavior portrays characteristics reminiscent of the 'flea on the 
elephant' concept \cite{land20,jona81,simo85}, familiar from the mathematical treatment of 
spontaneous symmetry breaking phenomena \cite{ande52}. 

Furthermore, our analysis identified a CI LLL state in the $(S=2$, $S_z=0$) spin sector, which was
shown to be well-described by a Halperin (1,1,1) two-component orbital variational wave function. 
Analysis of this CI LLL wavefunction enabled a two-dimensional generalization of the Girardeau 
one-dimensional 'fermionization' scheme \cite{gira60}, originally invoked for the mapping of 
bosonic-type wave functions to those of spinless fermions.

We stress that our systematic comparative analysis and investigations led us to conclude that in 
order to uncover the intrinsic geometrical structural characteristics of the symmetry-preserving 
ultracold rotating Wigner molecules that form in the rotating traps and exhibit magic angular 
momenta, it is imperative to carry out analysis that goes beyond second-order correlations in the 
real configuration space. To assist the design and analysis of experimental observations in 2D 
traps, we illustrate these findings through benchmark theoretical predictions for all-order 
spin-unresolved, as well as spin-resolved, all-order momentum correlations. These can be indeed 
directly measured \cite{berg19,prei19,clem19} with time-of-flight protocols employing individual 
particle detection in the far-field region.

Our conclusions regarding all-order momentum correaltions apply to the correlated FQHE states 
formed in ultracold neutral atom assemblies trapped in rotating traps on which we focused in the 
current study, as well as to future investigations, including interrogations of quantum magnetism 
in finite 2D systems (extending previous studies on 1D-trapped ultracold atoms 
\cite{yann16,deur15}), hole-pairing in 2D-plaquettes \cite{bran17}, and Mott-insulator to 
superfluid quantum phase transitions in finite 2D ultracold atom systems \cite{yann20}.

\section{Acknowledgments}
This work has been supported by a grant from the Air Force Office of Scientific Research (AFOSR, 
USA) under Award No. FA9550-15-1-0519. Calculations were carried out at the GATECH Center for
Computational Materials Science.

\appendix

\section{Magic angular momenta for the $N=3$ WM with spinful fermions}
\label{a1}

To enhance the brief historical overview in Sec.\ \ref{spec}, and to illustrate the role of the 
underlying geometric picture, we sketch here the derivation for the spin-dependent magic angular
momenta in the simpler case of three localized fermions arranged in an intrinsic configuration of
an equilateral triangle \cite{note8}. For $N=3$ fermions, both the $S_z=1/2$ and $S_z=3/2$
polarizations need to be considered. We start with the $S_z=1/2$ polarization, which is associated
with three spin primitives $|\downarrow \uparrow \uparrow \rangle$,
$|\uparrow \downarrow  \uparrow \rangle$, and $|\uparrow \uparrow \downarrow \rangle$. These
primitives correspond to single Slater determinants which exhibit a breaking of both the total spin
symmetry and of the continuous rotational symmetry. We first proceed with the restoration of the
total spin by noticing that the three spin primitives have a point-group symmetry lower than the
$C_{3}$ symmetry of an equilateral triangle. The $C_{3}$ symmetry, however, can be readily restored
by applying appropriate point-group projection operators according to group theoretical concepts
\cite{cottonbook,wolbarstbook}. This yields the following two different three-determinantal
combinations for the intrinsic part of the many-body wave function,
\begin{equation}
\Phi^{\rm intr}_1 (\gamma_0)= 
|\downarrow \uparrow \uparrow \rangle
+ e^{2\pi i/3} |\uparrow \downarrow \uparrow \rangle
+ e^{-2\pi i/3} |\uparrow \uparrow \downarrow \rangle,
\label{3dete1}
\end{equation}
and
\begin{equation}
\Phi^{\rm intr}_2 (\gamma_0)=
|\downarrow \uparrow \uparrow \rangle
+ e^{-2\pi i/3} |\uparrow \downarrow \uparrow \rangle
+ e^{2\pi i/3} |\uparrow \uparrow \downarrow \rangle.
\label{3dete2}
\end{equation}
Here $\gamma_0=0$ denotes the azimuthal angle of the triangle vertex associated with
the position of the original spin-down fermion in $|\downarrow \uparrow \uparrow \rangle$.
We note that the intrinsic wave functions $\Phi^{\rm intr}_1$
and $\Phi^{\rm intr}_2$ are eigenstates of the square of
the total spin operator ${\hat{\bf S}}^2$  ($\hat{\bf S} = \sum_{i=1}^3
\hat{\bf s}_i$) with quantum number $S=1/2$. This can be verified directly by 
applying to them the ${\hat {\bf S}}^2$ as given in Eq.\ (\ref{s2spin}).

To restore the circular symmetry, one applies the continuous space projection operator
\cite{yann07},
\begin{equation}
2 \pi {\cal P}_L \equiv \int_0^{2 \pi} d\gamma \exp[-i \gamma (\hat{L}-L)]~,
\label{amp}
\end{equation}
where $\hat{L}=\sum_{j=1}^N \hat{l}_j$ is the operator for the total angular momentum. 

The resulting wave function, $\Xi$, has both good total spin and angular momentum quantum numbers;
it is of the form,
\begin{equation}
2 \pi \Xi = \int^{2\pi}_0 d\gamma
\Phi^{\rm intr}_{ 1 {\rm or} 2 } (\gamma) e^{i\gamma L},
\label{rbsi}
\end{equation}
where now the intrinsic wave function [given by Eq.\ (\ref{3dete1}) or
Eq.\ \ref{3dete2})] has an arbitrary azimuthal orientation $\gamma$, which is integrated out.

The operator ${\hat{R}(2\pi/3) \equiv \exp (-i 2\pi{\hat L}/3})$ can be
applied to $\Xi$ in two different ways, namely either on
the intrinsic part $\Phi^{\rm intr}$ or the external part $\exp(i \gamma L)$. 
Using Eq.\ (\ref{3dete1}) and the property $\hat{R}(2\pi/3) 
\Phi^{\rm intr}_1 =\exp (-2\pi i/3)\Phi^{\rm intr}_1$, 
one finds,
\begin{equation}
\hat{R}(2\pi/3) \Xi = \exp (-2\pi i/3) \Xi,
\label{r1rbs}
\end{equation}
from the first alternative, and
\begin{equation}
\hat{R}(2\pi/3) \Xi = \exp (-2\pi L i/3) \Xi,
\label{r2rbs}
\end{equation}
from the second alternative. Now if $\Xi \neq 0$, the only
way that Eqs.\ (\ref{r1rbs}) and (\ref{r2rbs}) can be simultaneously true is
if the condition $\exp [2\pi (L-1) i/3]=1$ is fulfilled. This leads to a first
sequence of magic angular momenta associated with total spin $S=1/2$, i.e.,
\begin{equation}
L = 3 n +1,\; n=0,\pm 1, \pm 2, \pm 3,...
\label{i1}
\end{equation}

Using Eq.\ (\ref{3dete2}) for the intrinsic wave function, and following
similar steps, one can derive a second sequence of magic angular momenta
associated with good total spin $S=1/2$, i.e.,
\begin{equation}
L = 3 n -1,\; n=0,\pm 1, \pm 2, \pm 3,...
\label{i2}
\end{equation}

In the fully polarized case, the spin primitive, $|\uparrow \uparrow \uparrow \;\rangle$, 
is already an eigenstate of $\hat{\bf S}^2$ with quantum number $S=3/2$. Thus only the 
rotational symmetry needs to be restored, that is, the intrinsic wave function is simply 
$\Phi_3^{\rm intr}(\gamma_0) = |\uparrow \uparrow \uparrow \;\rangle$. Since 
$\hat{R}(2\pi/3) \Phi_3^{\rm intr} = \Phi_3^{\rm intr}$, the condition for the allowed 
angular momenta is $\exp [-2\pi L i/3]=1$, which yields the following magic angular momenta,
\begin{equation}
L = 3 n,\; n=0,\pm 1, \pm 2, \pm 3,...
\label{i3}
\end{equation}

We mention again here that only non-negative angular momenta are present in the LLL.

\begin{figure}[t]
\includegraphics[width=7.9cm]{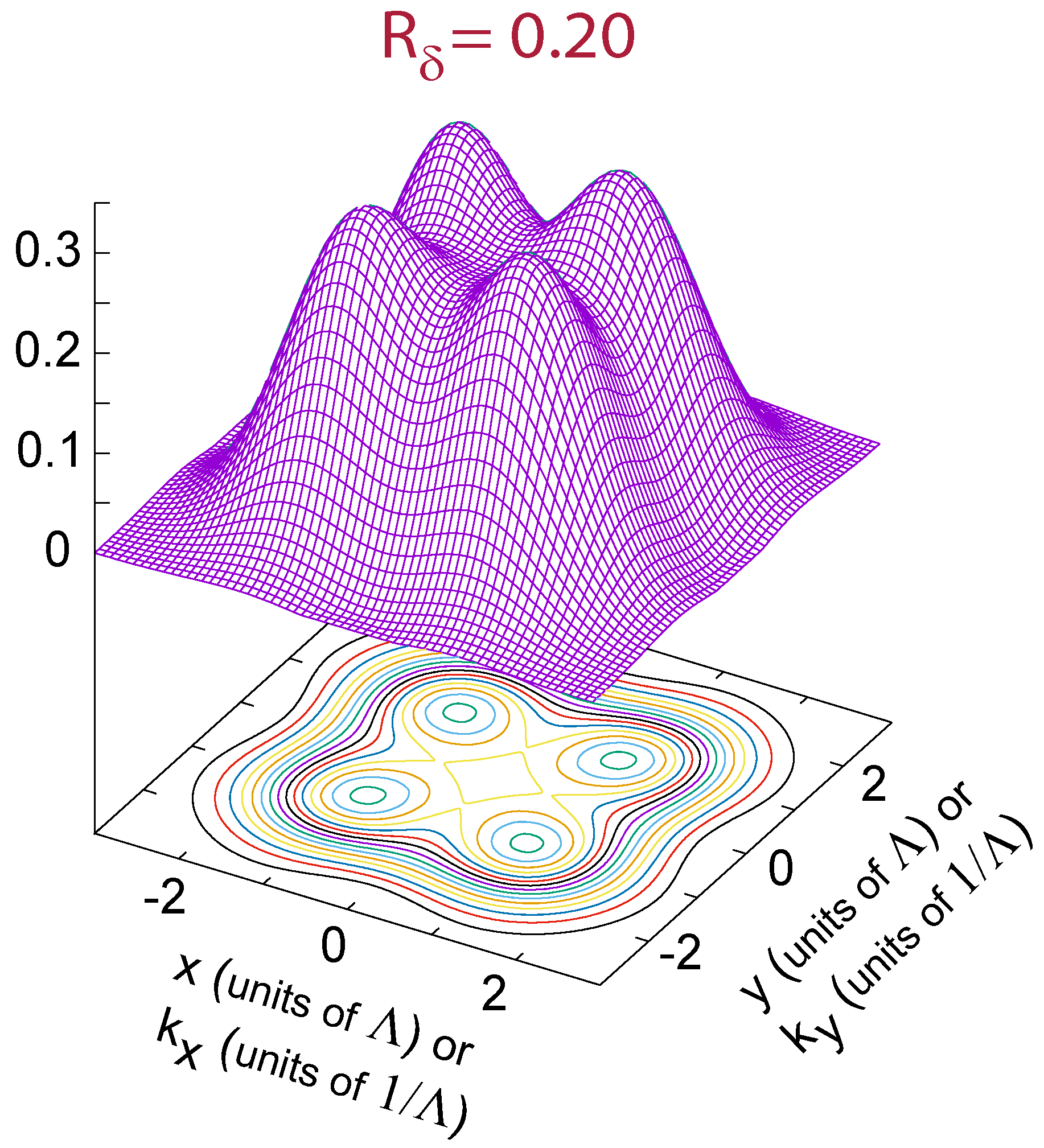}
\caption{
CI single-particle density (both in real space and momentum space)
of the relative ground state of $N=4$ fermions in the spin sector with ($S=0$, $S_z=0$), but
considering $R_\delta=0.20$, in contrast to the value $R_\delta=0.40$ used in the calculations
in the main text. 3D surfaces are plotted.  ${\cal C}=0.004$ at the point $\Omega/\omega=0.9445$
(situated at the avoided crossing between $L=4$ and $L=8$).
The expectation value of the total  angular momentum is $\langle L \rangle=7.8739$, indicating 
that the plotted case is a state with broken rotational symmetry. For the spatial density, the 
vertical axis is in units of $1/\Lambda^2$. For the momentum density, the vertical axis is in 
units of $\Lambda^2$.}
\label{af1}
\end{figure}

\section{The 20 spin primitives for N=6 fermions} 
\label{zetan6}

The 20 spin primitives for $N=6$ fermions are as follows:
\begin{align}
\begin{split}
& Z_1=\alpha(1)\alpha(2)\alpha(3)\beta(4)\beta(5)\beta(6),\\
& Z_2=\alpha(1)\alpha(2)\alpha(4)\beta(3)\beta(5)\beta(6),\\
& Z_3=\alpha(1)\alpha(2)\alpha(5)\beta(3)\beta(4)\beta(6),\\
& Z_4=\alpha(1)\alpha(2)\alpha(6)\beta(3)\beta(4)\beta(5),\\
& Z_5=\alpha(1)\alpha(3)\alpha(4)\beta(2)\beta(5)\beta(6),\\
& Z_6=\alpha(1)\alpha(3)\alpha(5)\beta(2)\beta(4)\beta(6),\\
& Z_7=\alpha(1)\alpha(3)\alpha(6)\beta(2)\beta(4)\beta(5),\\
& Z_8=\alpha(1)\alpha(4)\alpha(5)\beta(2)\beta(3)\beta(6),\\
& Z_9=\alpha(1)\alpha(4)\alpha(6)\beta(2)\beta(3)\beta(5),\\
& Z_{10}=\alpha(1)\alpha(5)\alpha(6)\beta(2)\beta(3)\beta(4),\\
& Z_{11}=\alpha(2)\alpha(3)\alpha(4)\beta(1)\beta(5)\beta(6),\\
& Z_{12}=\alpha(2)\alpha(3)\alpha(5)\beta(1)\beta(4)\beta(6),\\
& Z_{13}=\alpha(2)\alpha(3)\alpha(6)\beta(1)\beta(4)\beta(5),\\
& Z_{14}=\alpha(2)\alpha(4)\alpha(5)\beta(1)\beta(3)\beta(6),\\
& Z_{15}=\alpha(2)\alpha(4)\alpha(6)\beta(1)\beta(3)\beta(5),\\
& Z_{16}=\alpha(2)\alpha(5)\alpha(6)\beta(1)\beta(3)\beta(4),\\
& Z_{17}=\alpha(3)\alpha(4)\alpha(5)\beta(1)\beta(2)\beta(6),\\
& Z_{18}=\alpha(3)\alpha(4)\alpha(6)\beta(1)\beta(2)\beta(5),\\
& Z_{19}=\alpha(3)\alpha(5)\alpha(6)\beta(1)\beta(2)\beta(4),\\
& Z_{20}=\alpha(4)\alpha(5)\alpha(6)\beta(1)\beta(2)\beta(3)
\end{split}
\label{sprn6}
\end{align}

\section{A CI calculation for $N=4$ LLL fermions and $R_\delta=0.20$}
\label{a3}

An example of the formation of a pinned WM for a different value of $R_\delta$,
i.e., $R_\delta=0.20$ is given with Fig.\ \ref{af1}.

\newpage
~~~~~~~~\\
~~~~~~~~\\
~~~~~~~~\\
~~~~~~~~\\
~~~~~~~~\\
~~~~~~~~\\

\makeatletter 
\renewcommand{\thefigure}{SF\@arabic\c@figure}
\renewcommand{\thetable}{ST\@Roman\c@table}
\makeatother

\setcounter{table}{0}

\begin{widetext}

\vspace{3cm}
\begin{center}
{\bf{\Large SUPPLEMENTAL MATERIAL I\\
~~~~\\
Fractional-quantum-Hall physics and higher-order momentum correlations in few 
spinful fermionic contact-interacting ultracold atoms in rotating traps}}\\
~~~~~\\
{\Large Constantine Yannouleas$^\dagger$ and Uzi Landman$^*$\\
{\it School of Physics, Georgia Institute of Technology,
             Atlanta, Georgia 30332-0430}}
\end{center}
~~~~~~\\
~~~~~~\\
~~~~~~\\
~~~~~~\\
~~~~~~\\
~~~~~~\\
~~~~~~\\
~~~~~~\\
\noindent
{\large
{\bf Contents:\\
~~~~~\\
Additional TABLES related to Figs.\ 5(a-f) of the main text, pp. 27-29.}\\
~~~~~~~\\
~~~~~~~\\
~~~~~~~\\
~~~~~~~\\
~~~~~~~\\
~~~~~~~\\
}
$^\dagger$Constantine.Yannouleas@physics.gatech.edu\\
$^*$Uzi.Landman@physics.gatech.edu
\newpage

\begin{table*}[t]
\caption{\label{ts0sz0f4a} {\large
Dominant coefficients, $c(I)$, in the CI expansion of the relative LLL ground state (with 
$\langle L \rangle=7.189$) in the ($S=0$, $S_z=0$) spin sector corresponding to the symmetry-broken 
state whose single-particle density portrayed in Figs.\ 5(a,b) of the main text. 
The CI expansion consists of
$I_{\rm total}=1296$ basis determinants. The index $J$ is introduced to relabel the dominant 
coefficients, with the dominance criterion being $|c(I)|> 0.01$. 48 determinants with total angular 
momenta $L=4$, 8, or 12 participate in this TABLE. Note that $\sum_{i=1}^{48} |c(J)|^2=0.99843$, 
i.e., the corresponding contribution to the normalization constant differs from unity only in the 
third decimal point.}}
\begin{ruledtabular}
\begin{tabular}{r|c|r|c|c||r|c|r|c|c}
$I$&$J$&$c(J)$  &  $(l_1\uparrow,l_2\uparrow,l_3\downarrow,l_4\downarrow)$ & $\sum_{i=1}^4 l_i $ &
$I$&$J$&$c(J)$  &  $(l_1\uparrow,l_2\uparrow,l_3\downarrow,l_4\downarrow)$ & $\sum_{i=1}^4 l_i $ \\ 
\hline
    3&   1&     0.1422510E+00&( 0, 1, 0, 3)&    4&  303&  25&    -0.1891212E-01&( 1, 2, 1, 8)&   12 \\
    9&   2&     0.1926172E+00&( 0, 1, 1, 2)&    4&  304&  26&     0.1504875E+00&( 1, 2, 2, 3)&    8 \\
   13&   3&    -0.1348374E-01&( 0, 1, 1, 6)&    8&  328&  27&    -0.1218959E+00&( 1, 3, 0, 4)&    8 \\
   22&   4&     0.2479879E+00&( 0, 1, 3, 4)&    8&  332&  28&     0.1086694E-01&( 1, 3, 0, 8)&   12 \\
   26&   5&    -0.2194019E-01&( 0, 1, 3, 8)&   12&  334&  29&    -0.3688811E+00&( 1, 3, 1, 3)&    8 \\
   29&   6&     0.1561250E-01&( 0, 1, 4, 7)&   12&  338&  30&     0.2314515E-01&( 1, 3, 1, 7)&   12 \\
   38&   7&    -0.3103776E+00&( 0, 2, 0, 2)&    4&  363&  31&     0.1260920E+00&( 1, 4, 0, 3)&    8 \\
   42&   8&     0.1509984E-01&( 0, 2, 0, 6)&    8&  369&  32&     0.2134452E+00&( 1, 4, 1, 2)&    8 \\
   53&   9&    -0.3058820E+00&( 0, 2, 2, 4)&    8&  433&  33&    -0.1348374E-01&( 1, 6, 0, 1)&    8 \\
   57&  10&     0.2696752E-01&( 0, 2, 2, 8)&   12&  478&  34&     0.2314515E-01&( 1, 7, 1, 3)&   12 \\
   64&  11&    -0.1272792E-01&( 0, 2, 4, 6)&   12&  507&  35&    -0.1107325E-01&( 1, 8, 0, 3)&   12 \\
   73&  12&     0.1422510E+00&( 0, 3, 0, 1)&    4&  513&  36&    -0.1891212E-01&( 1, 8, 1, 2)&   12 \\
   83&  13&     0.1260920E+00&( 0, 3, 1, 4)&    8&  543&  37&     0.2601536E+00&( 2, 3, 0, 3)&    8 \\
   87&  14&    -0.1107325E-01&( 0, 3, 1, 8)&   12&  547&  38&    -0.1633527E-01&( 2, 3, 0, 7)&   12 \\
   88&  15&     0.2601536E+00&( 0, 3, 2, 3)&    8&  549&  39&     0.1504875E+00&( 2, 3, 1, 2)&    8 \\
   92&  16&    -0.1637189E-01&( 0, 3, 2, 7)&   12&  578&  40&    -0.3058820E+00&( 2, 4, 0, 2)&    8 \\
   96&  17&     0.1070732E-01&( 0, 3, 3, 6)&   12&  582&  41&     0.1268128E-01&( 2, 4, 0, 6)&   12 \\
  118&  18&    -0.1218959E+00&( 0, 4, 1, 3)&    8&  687&  42&    -0.1637189E-01&( 2, 7, 0, 3)&   12 \\
  182&  19&     0.1509984E-01&( 0, 6, 0, 2)&    8&  722&  43&     0.2696752E-01&( 2, 8, 0, 2)&   12 \\
  197&  20&     0.1268128E-01&( 0, 6, 2, 4)&   12&  757&  44&     0.2479879E+00&( 3, 4, 0, 1)&    8 \\
  232&  21&    -0.1633527E-01&( 0, 7, 2, 3)&   12&  831&  45&     0.1070732E-01&( 3, 6, 0, 3)&   12 \\
  262&  22&     0.1086694E-01&( 0, 8, 1, 3)&   12&  901&  46&    -0.2194019E-01&( 3, 8, 0, 1)&   12 \\
  289&  23&     0.1926172E+00&( 1, 2, 0, 1)&    4&  974&  47&    -0.1272792E-01&( 4, 6, 0, 2)&   12 \\
  299&  24&     0.2134452E+00&( 1, 2, 1, 4)&    8& 1009&  48&     0.1561250E-01&( 4, 7, 0, 1)&   12 \\
\end{tabular}
\end{ruledtabular}
\end{table*}

\newpage

\begin{table*}[t]
\caption{\label{ts0sz0f4b} {\large
Dominant coefficients, $c(I)$, in the CI expansion of the relative LLL ground state (with 
$\langle L \rangle=8.0364$) in the ($S=0$, $S_z=0$) spin sector corresponding to the symmetry-broken 
state whose single-particle density portrayed in Figs.\ 5(c,d) of the main text. 
The CI expansion consists of 
$I_{\rm total}=1296$ basis determinants. The index $J$ is introduced to relabel the dominant 
coefficients, with the dominance criterion being $|c(I)|> 0.01$. 54 determinants with total angular 
momenta $L=4$, 8, or 12 participate in this TABLE. Note that $\sum_{i=1}^{54} |c(J)|^2=0.99912$, 
i.e., the corresponding contribution to the normalization constant differs from unity only in the 
fourth decimal point.}
}
\begin{ruledtabular}
\begin{tabular}{r|c|r|c|c||r|c|r|c|c}
$I$&$J$&$c(J)$  &  $(l_1\uparrow,l_2\uparrow,l_3\downarrow,l_4\downarrow)$ & $\sum_{i=1}^4 l_i $ &
$I$&$J$&$c(J)$  &  $(l_1\uparrow,l_2\uparrow,l_3\downarrow,l_4\downarrow)$ & $\sum_{i=1}^4 l_i $ \\ 
\hline
    3&   1&    -0.2252529E-01&( 0, 1, 0, 3)&    4&  332&  28&    -0.1409311E-01&( 1, 3, 0, 8)&   12 \\
    9&   2&    -0.2769987E-01&( 0, 1, 1, 2)&    4&  334&  29&     0.4149306E+00&( 1, 3, 1, 3)&    8 \\
   22&   3&    -0.2782560E+00&( 0, 1, 3, 4)&    8&  338&  30&    -0.2986886E-01&( 1, 3, 1, 7)&   12 \\
   26&   4&     0.2823010E-01&( 0, 1, 3, 8)&   12&  347&  31&     0.1136427E-01&( 1, 3, 3, 5)&   12 \\
   29&   5&    -0.2007129E-01&( 0, 1, 4, 7)&   12&  363&  32&    -0.1395374E+00&( 1, 4, 0, 3)&    8 \\
   38&   6&     0.4715954E-01&( 0, 2, 0, 2)&    4&  367&  33&     0.1005852E-01&( 1, 4, 0, 7)&   12 \\
   53&   7&     0.3406431E+00&( 0, 2, 2, 4)&    8&  369&  34&    -0.2401117E+00&( 1, 4, 1, 2)&    8 \\
   57&   8&    -0.3453988E-01&( 0, 2, 2, 8)&   12&  373&  35&     0.1136298E-01&( 1, 4, 1, 6)&   12 \\
   64&   9&     0.1613584E-01&( 0, 2, 4, 6)&   12&  443&  36&     0.1136298E-01&( 1, 6, 1, 4)&   12 \\
   73&  10&    -0.2252529E-01&( 0, 3, 0, 1)&    4&  472&  37&    -0.1001277E-01&( 1, 7, 0, 4)&   12 \\
   83&  11&    -0.1395374E+00&( 0, 3, 1, 4)&    8&  478&  38&    -0.2986886E-01&( 1, 7, 1, 3)&   12 \\
   87&  12&     0.1413700E-01&( 0, 3, 1, 8)&   12&  507&  39&     0.1413700E-01&( 1, 8, 0, 3)&   12 \\
   88&  13&    -0.2935452E+00&( 0, 3, 2, 3)&    8&  513&  40&     0.2436981E-01&( 1, 8, 1, 2)&   12 \\
   92&  14&     0.2115172E-01&( 0, 3, 2, 7)&   12&  543&  41&    -0.2935452E+00&( 2, 3, 0, 3)&    8 \\
   96&  15&    -0.1388384E-01&( 0, 3, 3, 6)&   12&  547&  42&     0.2112242E-01&( 2, 3, 0, 7)&   12 \\
  118&  16&     0.1387186E+00&( 0, 4, 1, 3)&    8&  549&  43&    -0.1691131E+00&( 2, 3, 1, 2)&    8 \\
  122&  17&    -0.1001277E-01&( 0, 4, 1, 7)&   12&  578&  44&     0.3406431E+00&( 2, 4, 0, 2)&    8 \\
  197&  18&    -0.1612048E-01&( 0, 6, 2, 4)&   12&  582&  45&    -0.1612048E-01&( 2, 4, 0, 6)&   12 \\
  227&  19&     0.1005852E-01&( 0, 7, 1, 4)&   12&  687&  46&     0.2115172E-01&( 2, 7, 0, 3)&   12 \\
  232&  20&     0.2112242E-01&( 0, 7, 2, 3)&   12&  693&  47&     0.1214360E-01&( 2, 7, 1, 2)&   12 \\
  262&  21&    -0.1409311E-01&( 0, 8, 1, 3)&   12&  722&  48&    -0.3453988E-01&( 2, 8, 0, 2)&   12 \\
  289&  22&    -0.2769987E-01&( 1, 2, 0, 1)&    4&  757&  49&    -0.2782560E+00&( 3, 4, 0, 1)&    8 \\
  299&  23&    -0.2401117E+00&( 1, 2, 1, 4)&    8&  802&  50&     0.1136427E-01&( 3, 5, 1, 3)&   12 \\
  303&  24&     0.2436981E-01&( 1, 2, 1, 8)&   12&  831&  51&    -0.1388384E-01&( 3, 6, 0, 3)&   12 \\
  304&  25&    -0.1691131E+00&( 1, 2, 2, 3)&    8&  901&  52&     0.2823010E-01&( 3, 8, 0, 1)&   12 \\
  308&  26&     0.1214360E-01&( 1, 2, 2, 7)&   12&  974&  53&     0.1613584E-01&( 4, 6, 0, 2)&   12 \\
  328&  27&     0.1387186E+00&( 1, 3, 0, 4)&    8& 1009&  54&    -0.2007129E-01&( 4, 7, 0, 1)&   12 \\
\end{tabular}
\end{ruledtabular}
\end{table*}

\begin{table*}[t]
\caption{\label{ts0sz0f4c} {\large
Dominant coefficients, $c(I)$, in the CI expansion of the relative LLL ground state (with 
$\langle L \rangle=7.330$) in the ($S=0$, $S_z=0$) spin sector corresponding to the symmetry-broken 
state whose single-particle density portrayed in Figs.\ 5(e,f) of the main text. 
The CI expansion consists of 
$I_{\rm total}=1296$ basis determinants. The index $J$ is introduced to relabel the dominant 
coefficients, with the dominance criterion being $|c(I)|> 0.001$. 20 determinants with total angular 
momenta $L=4$ or 8 participate in this TABLE. Note that $\sum_{i=1}^{20} |c(J)|^2=0.9999939$, i.e., 
the corresponding contribution to the normalization constant differs from unity only in the sixth 
decimal point.}
}
\begin{ruledtabular}
\begin{tabular}{r|c|r|c|c||r|c|r|c|c}
$I$&$J$&$c(J)$  &  $(l_1\uparrow,l_2\uparrow,l_3\downarrow,l_4\downarrow)$ & $\sum_{i=1}^4 l_i $ &
$I$&$J$&$c(J)$  &  $(l_1\uparrow,l_2\uparrow,l_3\downarrow,l_4\downarrow)$ & $\sum_{i=1}^4 l_i $ \\ 
\hline
    3&   1&     0.1255794E+00&( 0, 1, 0, 3)&    4&  299&  11&     0.2212714E+00&( 1, 2, 1, 4)&    8 \\
    9&   2&     0.1729051E+00&( 0, 1, 1, 2)&    4&  304&  12&     0.1564648E+00&( 1, 2, 2, 3)&    8 \\
   22&   3&     0.2555221E+00&( 0, 1, 3, 4)&    8&  328&  13&    -0.1277153E+00&( 1, 3, 0, 4)&    8 \\
   38&   4&    -0.2760651E+00&( 0, 2, 0, 2)&    4&  334&  14&    -0.3832525E+00&( 1, 3, 1, 3)&    8 \\
   53&   5&    -0.3130074E+00&( 0, 2, 2, 4)&    8&  363&  15&     0.1278069E+00&( 1, 4, 0, 3)&    8 \\
   73&   6&     0.1255794E+00&( 0, 3, 0, 1)&    4&  369&  16&     0.2212714E+00&( 1, 4, 1, 2)&    8 \\
   83&   7&     0.1278069E+00&( 0, 3, 1, 4)&    8&  543&  17&     0.2709799E+00&( 2, 3, 0, 3)&    8 \\
   88&   8&     0.2709799E+00&( 0, 3, 2, 3)&    8&  549&  18&     0.1564648E+00&( 2, 3, 1, 2)&    8 \\
  118&   9&    -0.1277153E+00&( 0, 4, 1, 3)&    8&  578&  19&    -0.3130074E+00&( 2, 4, 0, 2)&    8 \\
  289&  10&     0.1729051E+00&( 1, 2, 0, 1)&    4&  757&  20&     0.2555221E+00&( 3, 4, 0, 1)&    8 \\
\end{tabular}
\end{ruledtabular}
\end{table*}

\begin{table*}[t]
\caption{\label{ts0sz0f4d}{\large
Dominant coefficients, $c(I)$, in the CI expansion of the relative LLL ground state (with 
$\langle L \rangle=8.00002$) in the ($S=0$, $S_z=0$) spin sector corresponding to the symmetry-broken
state whose single-particle density portrayed in Figs.\ 5(g,h) of the main text. 
The CI expansion consists of 
$I_{\rm total}=1296$ basis determinants. The index $J$ is introduced to relabel the dominant 
coefficients, with the dominance criterion being $|c(I)|> 0.001$. 16 determinants (15 with total 
angular momentum $L=8$ and 1 with $L=4$) participate in this TABLE. Note that 
$\sum_{i=1}^{16} |c(J)|^2=0.999989$, i.e., the corresponding contribution to the normalization 
constant differs from unity only in the fifth decimal point. The coefficient of the determinant with 
$L=4$ in the TABLE is two orders (in absolute magnitude) smaller than the coefficients of the 15 
determinants with $L=8$.}
}
\begin{ruledtabular}
\begin{tabular}{r|c|r|c|c||r|c|r|c|c}
$I$&$J$&$c(J)$  &  $(l_1\uparrow,l_2\uparrow,l_3\downarrow,l_4\downarrow)$ & $\sum_{i=1}^4 l_i $ &
$I$&$J$&$c(J)$  &  $(l_1\uparrow,l_2\uparrow,l_3\downarrow,l_4\downarrow)$ & $\sum_{i=1}^4 l_i $ \\ 
\hline
   22&   1&    -0.2800321E+00&( 0, 1, 3, 4)&    8&  328&   9&     0.1400277E+00&( 1, 3, 0, 4)&    8 \\
   38&   2&     0.1260197E-02&( 0, 2, 0, 2)&    4&  334&  10&     0.4200807E+00&( 1, 3, 1, 3)&    8 \\
   53&   3&     0.3429776E+00&( 0, 2, 2, 4)&    8&  363&  11&    -0.1400264E+00&( 1, 4, 0, 3)&    8 \\
   83&   4&    -0.1400283E+00&( 0, 3, 1, 4)&    8&  369&  12&    -0.2425468E+00&( 1, 4, 1, 2)&    8 \\
   88&   5&    -0.2970250E+00&( 0, 3, 2, 3)&    8&  543&  13&    -0.2970589E+00&( 2, 3, 0, 3)&    8 \\
  118&   6&     0.1400259E+00&( 0, 4, 1, 3)&    8&  549&  14&    -0.1714872E+00&( 2, 3, 1, 2)&    8 \\
  299&   7&    -0.2425213E+00&( 1, 2, 1, 4)&    8&  578&  15&     0.3430137E+00&( 2, 4, 0, 2)&    8 \\
  304&   8&    -0.1715068E+00&( 1, 2, 2, 3)&    8&  757&  16&    -0.2800782E+00&( 3, 4, 0, 1)&    8 \\
\end{tabular}
\end{ruledtabular}
\end{table*}

\newpage
~~~~~~~~~~~~~~~~\\
~~~~~~~~~~~~~~~~\\
\newpage
~~~~~~~~~~~~~~~~\\
~~~~~~~~~~~~~~~~\\
~~~~~~~~~~~~~~~~\\
\newpage

\begin{verbatim}

*****************************************************************************************
*****************************************************************************************

(* SUPPLEMENTAL MATERIAL II *)
(* TITLE: "Fractional-quantum-Hall physics and higher-order momentum correlations 
   in few spinful fermionic contact-interacting ultracold atoms in rotating traps" *)
(* AUTHORS: Constantine Yannouleas and Uzi Landman *)

(* Coder: Constantine Yannouleas, Atlanta, Georgia, USA, 10aug20 *)

(* ----------------------------------- *)
(* Scripts related to N=4 LLL fermions *) 
(* For N=6, see another note below *)
(* ----------------------------------- *)

(* Analytic expressions for the F_i(z_1,z_2,z_3,z_4) and \tilde{F}_i of Sec. IV D *)
 
ClearAll["Global`*"];

(* Definition of zeta_i, Eq. (22) *)

zeta[1]=a[1]*a[2]*b[3]*b[4]; 
zeta[2]=a[1]*a[3]*b[2]*b[4];
zeta[3]=a[1]*a[4]*b[2]*b[3];
zeta[4]=a[2]*a[3]*b[1]*b[4]; 
zeta[5]=a[2]*a[4]*b[1]*b[3];
zeta[6]=a[3]*a[4]*b[1]*b[2]; 

(* Define matrices entering in LLL basis determinants; the 1/Sqrt[N!] normalization is 
omitted in the Slater determinants here *)
 
mat[l1_,l2_,l3_,l4_]:=
 {{z1^l1*a[1]/Sqrt[Pi*l1!], z2^l1*a[2]/Sqrt[Pi*l1!], z3^l1*a[3]/Sqrt[Pi*l1!], 
   z4^l1*a[4]/Sqrt[Pi*l1!]},
  {z1^l2*a[1]/Sqrt[Pi*l2!], z2^l2*a[2]/Sqrt[Pi*l2!], z3^l2*a[3]/Sqrt[Pi*l2!], 
   z4^l2*a[4]/Sqrt[Pi*l2!]},
  {z1^l3*b[1]/Sqrt[Pi*l3!], z2^l3*b[2]/Sqrt[Pi*l3!], z3^l3*b[3]/Sqrt[Pi*l3!], 
   z4^l3*b[4]/Sqrt[Pi*l3!]},
  {z1^l4*b[1]/Sqrt[Pi*l4!], z2^l4*b[2]/Sqrt[Pi*l4!], z3^l4*b[3]/Sqrt[Pi*l4!], 
   z4^l4*b[4]/Sqrt[Pi*l4!]}};

(* The coefficients in Eq. (32); the index 8 indicates the L=8 total angular momentum *)

c8[1]=c1; c8[16]=c1; c8[4]=c1/2; 
c8[6]=-c1/2; c8[9]=-c1/2; c8[11]=c1/2;
c8[3]=c2; c8[15]=c2; 
c8[8]=-c2/2; c8[14]=-c2/2;
c8[5]=c3; c8[13]=c3;
c8[7]=c4; c8[12]=c4;
c8[10]=c5;

(* The corresponding CI wave function [Eq. (9)] using the 15 determinants in TABLE STIV 
   in SUPPLENENTAL MATERIAL I   *)
 
ciwfL8=c8[1]*Det[mat[0,1,3,4]]+
   c8[3]*Det[mat[0,2,2,4]]+
   c8[4]*Det[mat[0,3,1,4]]+
   c8[5]*Det[mat[0,3,2,3]]+
   c8[6]*Det[mat[0,4,1,3]]+
   c8[7]*Det[mat[1,2,1,4]]+
   c8[8]*Det[mat[1,2,2,3]]+
   c8[9]*Det[mat[1,3,0,4]]+
   c8[10]*Det[mat[1,3,1,3]]+
   c8[11]*Det[mat[1,4,0,3]]+
   c8[12]*Det[mat[1,4,1,2]]+
   c8[13]*Det[mat[2,3,0,3]]+
   c8[14]*Det[mat[2,3,1,2]]+
   c8[15]*Det[mat[2,4,0,2]]+
   c8[16]*Det[mat[3,4,0,1]];

(* The F_i functions associated with ciwf8; see Eq. (39) *)

F[i_]:=Coefficient[ciwfL8,zeta[i]];

(* The tilded F_i functions of Eqs. (40)-(45), denoted here as FF[i] *)

FF[1]=(F[1]+F[3]+F[4]+F[6]-2*F[2]-2*F[5])/Sqrt[12];

FF[2]=(F[1]-F[3]-F[4]+F[6])/2;

FF[3]=(F[6]-F[1])/Sqrt[2];

FF[4]=(F[5]-F[2])/Sqrt[2];

FF[5]=(F[4]-F[3])/Sqrt[2];

FF[6]=(F[1]+F[2]+F[3]+F[4]+F[5]+F[6])/Sqrt[6];

(********************

(* After importing the file into a MATHEMATICA notebook, copy and paste the following 
commands *)

(* Show that the following are vanishing *)

FF[3]//FullSimplify

FF[4]//FullSimplify

FF[5]//FullSimplify

FF[6]//FullSimplify


(* Derive Eq. (47) *)

(* Set the z_i's at the corners of a square *)

z1=z0; z2=z0*I; z3=-z0; z4=-z0*I;

FF[1]

FF[2]//Factor

Clear[z1,z2,z3,z4]

*******************)

(* Derive Eq. (50) in Sec. V B *)

(* The CI wave function [lhs of Eq. (50)] using the 6 determinants in TABLE I *)

ciwfL6S2Sz0=(-Det[mat[0,1,2,3]]+Det[mat[0,2,1,3]]-Det[mat[0,3,1,2]]-
       Det[mat[1,2,0,3]]+Det[mat[1,3,0,2]]-Det[mat[2,3,0,1]])/Sqrt[6];

(******************
(* After importing the file into a MATHEMATICA notebook, copy and paste the following *) 

Clear[z1,z2,z3,z4]; ciwfL6S2Sz0//Factor

*******************)

(* Derive Eq. (51) in Sec. VI A *)

mat2[l1_,l2_,l3_,l4_]:=
 {{z1^l1*a[1]/Sqrt[Pi*l1!], z2^l1*a[2]/Sqrt[Pi*l1!], z3^l1*a[3]/Sqrt[Pi*l1!], 
   z4^l1*a[4]/Sqrt[Pi*l1!]},
  {z1^l2*a[1]/Sqrt[Pi*l2!], z2^l2*a[2]/Sqrt[Pi*l2!], z3^l2*a[3]/Sqrt[Pi*l2!], 
   z4^l2*a[4]/Sqrt[Pi*l2!]},
  {z1^l3*a[1]/Sqrt[Pi*l3!], z2^l3*a[2]/Sqrt[Pi*l3!], z3^l3*a[3]/Sqrt[Pi*l3!], 
   z4^l3*a[4]/Sqrt[Pi*l3!]},
  {z1^l4*a[1]/Sqrt[Pi*l4!], z2^l4*a[2]/Sqrt[Pi*l4!], z3^l4*a[3]/Sqrt[Pi*l4!], 
   z4^l4*a[4]/Sqrt[Pi*l4!]}};

ciwfL6S2Sz2=Det[mat2[0,1,2,3]];

(******************
(* After importing the file into a MATHEMATICA notebook, copy and paste the following *) 

Clear[z1,z2,z3,z4]; ciwfL6S2Sz2//Factor

*******************)
 
(* ----------------------------------- *)
(* Scripts related to N=6 LLL fermions *)
(* ----------------------------------- *)

(* Definition of Zeta_i, Eq. (B1) in Appendix B, denoted as ZZ here *)

list3a=Subsets[Table[i,{i,1,6}],{3}];
list6 = Table[i, {i, 1, 6}];
list3b = Table[list6 /. {list3a[[jj, 1]] -> Nothing, list3a[[jj, 2]] -> Nothing, 
    list3a[[jj, 3]] -> Nothing}, {jj, 1, Length@list3a}];
Do[ZZ[ii] = a[list3a[[ii, 1]]]*a[list3a[[ii, 2]]]*a[list3a[[ii, 3]]]*
   b[list3b[[ii, 1]]]*b[list3b[[ii, 2]]]*b[list3b[[ii, 3]]], {ii, 1, Length@list3a}];

(* Define matrices for N=6 entering in LLL basis determinants; the 1/Sqrt[N!] 
normalization is omitted in the Slater determinants here, along with the factor
1/(Pi^3*Sqrt[l1!l2!l3!l4!l5!l6!]); for the single-particle angular momenta in TABLE II,
this yields a common factor for all 20 dominant Slater determinants *)

mat6[l1_, l2_, l3_, l4_, l5_, l6_] := 
    {{z1^l1*a[1], z2^l1*a[2], z3^l1*a[3], z4^l1*a[4], z5^l1*a[5], z6^l1*a[6]}, 
     {z1^l2*a[1], z2^l2*a[2], z3^l2*a[3], z4^l2*a[4], z5^l2*a[5], z6^l2*a[6]}, 
     {z1^l3*a[1], z2^l3*a[2], z3^l3*a[3], z4^l3*a[4], z5^l3*a[5], z6^l3*a[6]}, 
     {z1^l4*b[1], z2^l4*b[2], z3^l4*b[3], z4^l4*b[4], z5^l4*b[5], z6^l4*b[6]},
     {z1^l5*b[1], z2^l5*b[2], z3^l5*b[3], z4^l5*b[4], z5^l5*b[5], z6^l5*b[6]},
     {z1^l6*b[1], z2^l6*b[2], z3^l6*b[3], z4^l6*b[4], z5^l6*b[5], z6^l6*b[6]}};

(* Derive Eq. (52) in Sec. VI B *)

(* The CI wave function [lhs of Eq. (52)] using the 20 determinants in TABLE II
and taking into consideration the sign of the c(J) coefficients *)

wfnp6 = -Det[mat6[0, 1, 2, 3, 4, 5]] + Det[mat6[0, 1, 3, 2, 4, 5]] - 
  Det[mat6[0, 1, 4, 2, 3, 5]] + Det[mat6[0, 1, 5, 2, 3, 4]] - 
  Det[mat6[0, 2, 3, 1, 4, 5]] + Det[mat6[0, 2, 4, 1, 3, 5]] - 
  Det[mat6[0, 2, 5, 1, 3, 4]] - Det[mat6[0, 3, 4, 1, 2, 5]] + 
  Det[mat6[0, 3, 5, 1, 2, 4]] - Det[mat6[0, 4, 5, 1, 2, 3]] + 
  Det[mat6[1, 2, 3, 0, 4, 5]] - Det[mat6[1, 2, 4, 0, 3, 5]] + 
  Det[mat6[1, 2, 5, 0, 3, 4]] + Det[mat6[1, 3, 4, 0, 2, 5]] - 
  Det[mat6[1, 3, 5, 0, 2, 4]] + Det[mat6[1, 4, 5, 0, 2, 3]] - 
  Det[mat6[2, 3, 4, 0, 1, 5]] + Det[mat6[2, 3, 5, 0, 1, 4]] - 
  Det[mat6[2, 4, 5, 0, 1, 3]] + Det[mat6[3, 4, 5, 0, 1, 2]];

(* Show that wfnp6 is proportional to a Vandermonde determinant for N=6 LLL fermions *)

(* sumZeta/Sqrt[6] below is the spin eigenfunction for N=6 fermions with S=3, S_z=0 *)

sumZeta = Sum[ZZ[i], {i, 1, 20}];

(******************
(* After importing the file into a MATHEMATICA notebook, copy and paste the following *) 

(wfnp6//Expand//FullSimplify)/sumZeta//FullSimplify

*******************)

\end{verbatim}

\end{widetext}


\begin{thebibliography}{99}
\bibitem{tsui82}
D. C. Tsui, H. L. Stormer, and A. C. Gossard,
Two-Dimensional Magnetotransport in the Extreme Quantum Limit,
Phys. Rev. Lett. {\bf 48}, 1559 (1982).
\bibitem{laug83.1}
R. B. Laughlin,
Quantized motion of three two-dimensional electrons in a strong magnetic field,
Phys. Rev. B {\bf 27}, 3383 (1983).
\bibitem{laug83.2}
R. B. Laughlin,
Anomalous Quantum Hall Effect: An Incompressible Quantum Fluid with Fractionally
Charged Excitations,
Phys. Rev. Lett. {\bf 50}, 1395 (1983).
\bibitem{halp83}
B. I. Halperin, 
Theory of the Quantized Hall Conductance, 
Helvetica Physica Acta {\bf 56}, 75 (1983).
\bibitem{jain89}
J. K. Jain,
Composite-fermion approach for the fractional quantum Hall effect,
Phys. Rev. Lett. {\bf 63}, 199 (1989).
\bibitem{yann02}
C. Yannouleas and U. Landman,
Trial wave functions with long-range Coulomb correlations for two-dimensional $N$-electron systems 
in high magnetic fields,
Phys. Rev. B {\bf 66}, 115315 (2002).
\bibitem{wilk00}
N. K. Wilkin and J. M. F. Gunn,
Condensation of “Composite Bosons” in a Rotating BEC,
Phys. Rev. Lett. {\bf 84}, 6 (2000).
\bibitem{coop03}
N. Read and N. R. Cooper,
Free expansion of lowest-Landau-level states of trapped atoms: A wave-function microscope,
Phys. Rev. A {\bf 68}, 035601 (2003).
\bibitem{popp04}
M. Popp, B. Paredes, and J. I. Cirac,
Adiabatic path to fractional quantum Hall states of a few bosonic atoms
Phys. Rev. A {\bf 70}, 053612 (2004).
\bibitem{barb06}
N. Barber\'{a}n, M. Lewenstein, K. Osterloh, and D. Dagnino,
Ordered structures in rotating ultracold Bose gases,
Phys. Rev. A {\bf 73}, 063623 (2006).
\bibitem{baks07}
L. O. Baksmaty, C. Yannouleas, and U. Landman,
-Rapidly rotating boson molecules with long- or short-range repulsion:
An exact diagonalization study,
Phys. Rev. A {\bf 75}, 023620 (2007).
\bibitem{coop08}
N. R. Cooper,
Rapidly rotating atomic gases, 
Advances in Physics, {\bf 57}, 539 (2008).
\bibitem{hazz08}
S. K. Baur, K. R. A. Hazzard, and E. J. Mueller,
Stirring trapped atoms into fractional quantum Hall puddles,
Phys. Rev. A {\bf 78}, 061608(R) (2008).
\bibitem{geme10}
N. Gemelke, E. Sarajlic, and S. Chu,
Rotating Few-body Atomic Systems in the Fractional Quantum Hall Regime,
arXiv:1007.2677.
\bibitem{joch11}
F. Serwane, G. Z\"{u}rn, T. Lompe, T. B. Ottenstein, A. N. Wenz, and S. Jochim,
Deterministic Preparation of a Tunable Few-Fermion System,
Science {\bf 332}, 336 (2011).
\bibitem{joch12}
G. Z\"{u}rn, F. Serwane, T. Lompe, A. N. Wenz, M. G. Ries, J. E. Bohn, and S. Jochim,
Fermionization of Two Distinguishable Fermions,
Phys. Rev. Lett. {\bf 108}, 075303 (2012).
\bibitem{berg19}
A. Bergschneider, V. M. Klinkhamer, J. H. Becher, R. Klemt, L. Palm, G. Z\"{u}rn, S. Jochim, 
and P. M. Preiss,
Experimental characterization of two-particle entanglement through position
and momentum correlations,
Nat. Phys. {\bf 15}, 640 (2019).
\bibitem{prei19}
P. M. Preiss, J. H. Becher, R. Klemt, V. Klinkhamer, A. Bergschneider, N. Defenu, and S. Jochim,
High-Contrast Interference of Ultracold Fermions,
Phys. Rev. Lett. {\bf 122}, 143602 (2019).
\bibitem{bech20}
J. H. Becher, E. Sindici, R. Klemt, S. Jochim, A. J. Daley, and P. M. Preiss,
Measurement of Identical Particle Entanglement and the Influence of Antisymmetrisation,
arXiv:2002.11207.
\bibitem{palm18}
L. Palm,
Exploring fractional quantum hall physics using ultracold fermions in rotating traps,
(Masterarbeit, Heidelberg, 2018)  
\url{http://ultracold.physi.uni-heidelberg.de/files/Masterarbeit_2018_Lukas.pdf}; see also
R.-J. Petzold,
Few ultracold fermions in a two-dimensional trap,
(Masterarbeit, Heidelberg, 2020) 
\url{http://ultracold.physi.uni-heidelberg.de/files/Masterarbeit_2020_RJPetzold.pdf}.
\bibitem{yann03}
C. Yannouleas and U. Landman,
Two-dimensional quantum dots in high magnetic fields: Rotating-electron-molecule versus 
composite-fermion approach,
Phys. Rev. B {\bf 68}, 035326 (2003).
\bibitem{yann04}
C. Yannouleas and U. Landman,
Structural properties of electrons in quantum dots in high magnetic fields: Crystalline character 
of cusp states and excitation spectra,
Phys. Rev. B {\bf 70}, 235319 (2004).
\bibitem{ront06}
M. Rontani, C. Cavazzoni, D. Bellucci, and G. Goldoni,
Full configuration interaction approach to the few-electron problem in artificial atoms,
J. Chem. Phys. {\bf 124}, 124102 (2006).
\bibitem{yues07}
Yuesong Li, C. Yannouleas, and U. Landman,
Three-electron anisotropic quantum dots in variable magnetic fields: Exact results for excitation 
spectra, spin structures, and entanglement,
Phys. Rev. B {\bf 76}, 245310 (2007); Erratum: Phys. Rev. B {\bf 81}, 049902 (2010).
\bibitem{blun10}
S. A. Blundell and S. Chacko,
Isomeric and hybrid isomeric-vibrational states of Wigner molecules,
Phys. Rev. B {\bf 81}, 121104(R) (2010).
\bibitem{szabobook}
A. Szabo and N.S. Ostlund,
{\it Modern Quantum Chemistry: Introduction to Advanced Electronic Structure Theory\/},
Revised first edition (McGraw-Hill, New York, 1989).
\bibitem{yann11}
C. Yannouleas and U. Landman,
Unified microscopic approach to the interplay of pinned-Wigner-solid and liquid behavior of the 
lowest Landau-level states in the neighborhood of $\nu=1/3$.
Phys. Rev. B {\bf 84}, 165327 (2011).
\bibitem{joch18}
A. Bergschneider, V. M. Klinkhamer, J. H. Becher, R. Klemt, G. Z\"{u}rn, P. M. Preiss,
and S. Jochim,
Spin-resolved single-atom imaging of $^6$Li in free space,
Phys. Rev. A {\bf 97}, 063613 (2018).
\bibitem{girv83}
S. M. Girvin and T. Jach,
Interacting electrons in two-dimensional Landau levels: Results for small clusters,
Phys. Rev. B {\bf 28}, 4506 (1983).
\bibitem{maks90}
Quantum dots in a magnetic field: Role of electron-electron interactions,
P. A. Maksym and T. Chakraborty,
Phys. Rev. Lett. {\bf 65}, 108 (1990).
\bibitem{yann07}
C. Yannouleas and U. Landman,
Symmetry breaking and quantum correlations in finite systems: 
studies of quantum dots and ultracold Bose gases and related nuclear and chemical methods,
Rep. Prog. Phys. {\bf 70}, 2067 (2007).
\bibitem{maks96}
P. A. Maksym,
Eckardt frame theory of interacting electrons in quantum dots,
Phys. Rev. B {\bf 53}, 10871 (1996).
\bibitem{roma09}
I. Romanovsky, C. Yannouleas, and U. Landman,
Edge states in graphene quantum dots: Fractional quantum Hall effect analogies and differences
at zero magnetic field,
Phys. Rev. B {\bf 79}, 075311 (2009). 
\bibitem{gira60}
M. Girardeau,
Relationship between Systems of Impenetrable Bosons and Fermions in One Dimension,
J. Math. Phys. {\bf 1}, 516 (1960).
\bibitem{fock}
V. Fock,
Bemerkung zur Quantelung des harmonischen Oszillators im Magnetfeld. 
Z. Phys. {\bf 47}, 446 (1928), \url{https://doi.org/10.1007/BF01390750}.
\bibitem{darw}
C. G. Darwin,
The Diamagnetism of the Free Electron,
C. G. Darwin,
ProarXiv:1412.4529c. Cambridge Philos. Soc. {\bf 27}, 86 (1931),
\url{https://doi.org/10.1017/S0305004100009373}.
\bibitem{note1}
For details regarding the equivalence between applied magnetic field $B$ and the rotational 
frequency $\Omega$, as well as the derivation of the spectrum in Eq.\ (\ref{spfd}), see the 
Appendix in Ref.\ \cite{yann07}; see also Refs.\ \cite{wilk00,coop08}.
\bibitem{note2}
For an instance of a detailed derivation of $H_{\rm LLL}$, see Sec. II.A of Ref.\ \cite{baks07}.
\bibitem{shly00}
D. S. Petrov, M. Holzmann, and G. V. Shlyapnikov,
Bose-Einstein Condensation in Quasi-2D Trapped Gases,
Phys. Rev. Lett. {\bf 84}, 2551 (2000).
\bibitem{note12}
Concerning earlier literature, $R_\delta$ agrees with the parameter $\eta$ in Ref.\ \cite{popp04}.
Furthermore, the parameter $U$ in Ref.\ \cite{hazz08} relates to $R_\delta$ as $U=R_\delta\hbar\omega$. 
\bibitem{astr04}
G. E. Astrakharchik,
Quantum Monte Carlo study of ultracold gases,  
(Ph. D. Dissertation, Trento, 2004),
arXiv:1412.4529
\bibitem{zuer12}
G. Z\"{u}rn,
Few-fermion systems in one dimension,
(Ph. D. Thesis, Heidelberg, 2012)
\url{https://doi.org/10.11588/heidok.00014496}.
\bibitem{juli14}
Contrasting the wide Feshbach resonances in $^6$Li and $^7$Li,
P, S. Julienne and J. M. Hutson,
Phys. Rev. A {\bf 89}, 052715 (2014).
\bibitem{note13}
For a CI calculation with a different value, i.e., $R_\delta=0.2$, which illustrates
explicitly that the formation of the WMs in the LLL is independent of $R_\delta$, see
Fig.\ \ref{af1} in Appendix \ref{a3}.  
\bibitem{arpack}
R.B. Lehoucq, D.C. Sorensen, and C. Yang, 
{\it ARPACK Users' Guide: Solution of Large-Scale Eigenvalue Problems with Implicitly Restarted 
Arnoldi Methods\/} (SIAM, Philadelphia, 1998).
\bibitem{arno51}
W. E. Arnoldi, 
The principle of minimized iterations in the solution of the matrix eigenvalue problem, 
Quarterly of Applied Mathematics {\bf 9}, 17 (1951).
\bibitem{pape99}
G. F. Bertsch and Th. Papenbrock,
Yrast Line for Weakly Interacting Trapped Bosons,
Phys. Rev. Lett. {\bf 83}, 5412 (1999).
\bibitem{note10}
See Tables 2.3 and 2.4 on p. 70 of Ref.\ \cite{szabobook}
\bibitem{altm04}
E. Altman, E. Demler, and M. D. Lukin,
Probing many-body states of ultracold atoms via noise correlations,
Phys. Rev. A {\bf 70}, 013603 (2004).
\bibitem{ruan95}
W. Y. Ruan, Y. Y. Liu, C. G. Bao, and Z. Q. Zhang,
Origin of magic angular momenta in few-electron quantum dots,
Phys. Rev. B {\bf 51}, 7942(R) (1995).
\bibitem{seki96}
T. Seki, Y. Kuramoto, and T. Nishino,
Origin of magic angular momentum in a quantum dot under strong magnetic field,
J. Phys. Soc. Japan {\bf 65}, 3945 (1996).
\bibitem{yann03.2}
C. Yannouleas and U. Landman,
Group theoretical analysis of symmetry breaking in two-dimensional quantum dots,
Phys. Rev. B {\bf 68}, 035325, (2003).
\bibitem{dai07}
Z. Dai, J.-L. Zhu, N. Yang, and Y. Wang,
Spin-dependent rotating Wigner molecules in quantum dots,
Phys. Rev. B {\bf 76}, 085308 (2007).
\bibitem{shi07}
C. Shi, G. S. Jeon, and J. K. Jain,
Composite fermion solid and liquid states in two component quantum dots,
Phys. Rev. B {\bf 75}, 165302 (2007).
\bibitem{yann00}
See the case of two electrons in a 2D parabolic trap, 
C. Yannouleas and U. Landman,
Collective and Independent-Particle Motion in Two-Electron Artificial Atoms,
Phys. Rev. Lett. {\bf 85}, 1726 (2000).
\bibitem{yann99}
C. Yannouleas and U. Landman,
Spontaneous Symmetry Breaking in Single and Molecular Quantum Dots,
Phys. Rev. Lett. {\bf 82}, 5325 (1999); Erratum: Phys. Rev. Lett. {\bf 85}, 2220 (2000).
\bibitem{roma04}
I. Romanovsky, C. Yannouleas, and U. Landman,
Crystalline Boson Phases in Harmonic Traps: Beyond the Gross-Pitaevskii Mean Field,
Phys. Rev. Lett. {\bf 93}, 230405 (2004).
\bibitem{cava07}
U. De Giovannini, F. Cavaliere, R. Cenni, M. Sassetti, and B. Kramer,
Spin and rotational symmetries in unrestricted Hartree-Fock states of quantum dots,
New J. Phys. {\bf 9}, 93 (2007).
\bibitem{note3}
For the use of this term in the context of electrons in 2D semiconductor quantum dots, see Ref.\ 
\cite{yann07}. The variants of ``rotating electron molecule'' \cite{yann07} or
``rotating boson molecule'' \cite{baks07} in the case of bosons have also been employed. 
\bibitem{elle06}
C. Ellenberger, T. Ihn, C. Yannouleas, U. Landman, K. Ensslin, D. Driscoll, and A. C. Gossard,
Excitation Spectrum of Two Correlated Electrons in a Lateral Quantum Dot with Negligible Zeeman
Splitting,
Phys. Rev. Lett. {\bf 96}, 126806 (2006).
\bibitem{maks06}
Y. Nishi, P. A. Maksym, D. G. Austin, T. Hatano, L. P. Kouwenhoven, H. Aoki, and S. Tarucha,
Intermediate low spin states in a few-electron quantum dot in the $\nu\le 1$ regime,
Phys. Rev. B {\bf 74}, 033306 (2006).
\bibitem{kall08}
S. Kalliakos, M. Rontani, V. Pellegrini, C. P. Garc\'{i}a, A. Pinczuk, G. Goldoni, E. Molinari, 
L. N. Pfeiffer, and K. W. West, 
A molecular state of correlated electrons in a quantum dot,
Nature Phys. {\bf 4}, 467 (2008).
\bibitem{mint18}
A. M. Mintairov, J. Kapaldo, J. L. Merz, S. Rouvimov, D. V. Lebedev, N. A. Kalyuzhnyy,
{\it et al.\/},
Control of Wigner localization and electron cavity effects in near-field emission spectra of 
In(Ga)P/GaInP quantum-dot structures,
Phys. Rev. B {\bf 97}, 195443 (2018).
\bibitem{ilan13}
S. Pecker, F. Kuemmeth, A. Secchi, M. Rontani, D. C. Ralph, P. L. McEuen, and S. Ilani,
Observation and spectroscopy of a two-electron Wigner molecule in an ultraclean carbon nanotube,
Nature Phys. {\bf  9}, 576 (2013).
\bibitem{hoen14}
G. H\"{o}nig, G. Callsen, A. Schliwa, S. Kalinowski, Ch. Kindel, S. Kako, 
Y. Arakawa, D. Bimberg, and A. Hoffmann,
Manifestation of unconventional biexciton states in quantum dots,
Nat. Commun. {\bf 5}, 5721 (2014).
\bibitem{note11}
The (1,5) ring configuration for $N=6$ fermions is demonstrated in Sec.\ \ref{fermn6}.
\bibitem{jainbook}
J. K. Jain,
{\it Composite Fermions\/}
(Cambridge University Press, Cambridge, 2007).
\bibitem{supp}
See Supplemental Material for (I) additional Tables related to 
Figs. 5(a-f) (p. 26 below) and (II) MATHEMATICA scripts related 
to Sections IV D, V B, VI A, and VI B (p. 30 below).
\bibitem{land20}
Ch. J. F. van de Ven, C. Gerrit, R. R. Groenenboom, and N. P. Landsman,
Quantum spin systems versus Schr\"{o}dinger operators:
A case study in spontaneous symmetry breaking,
SciPost Phys. {\bf 8}, 022 (2020).
\bibitem{jona81}
G. Jona-Lasinio, F. Martinelli, and E. Scoppola, 
New approach to the semiclassical limit of quantum mechanics, 
Commun. Math. Phys. {\bf 80}, 223 (1981).
\bibitem{simo85}
B. Simon, 
Semiclassical analysis of low lying eigenvalues. IV. The flea on the elephant, 
J. Funct. Anal. {\bf 63}, 123 (1985).
\bibitem{ande52}
P. W. Anderson, 
An approximate quantum theory of the antiferromagnetic ground state,
Phys. Rev. {\bf 86}, 694 (1952).
\bibitem{math}
Wolfram Research, Inc., Mathematica, Version 12.1, Champaign, IL (2020).
\bibitem{yann16}
C. Yannouleas, B. B. Brandt, and U. Landman,
Ultracold few fermionic atoms in needle-shaped double wells: spin chains and resonating spin 
clusters from microscopic Hamiltonians emulated via antiferromagnetic Heisenberg and $t-J$ models,
New J. Phys. {\bf 18}, 073018 (2016).
\bibitem{fert96}
A. H. MacDonald, H. A. Fertig, and L. Brey,
Skyrmions without Sigma Models in Quantum Hall Ferromagnets,
Phys. Rev. Lett. {\bf 76}, 2153 (1996).
\bibitem{girv96}
S. M. Girvin and A. H. MacDonald,
Multicomponent Quantum Hall Systems: The Sum of Their Parts and More,
in {\it Perspectives in Quantum Hall Effects: Novel Quantum Liquids in Low-Dimensional
  Semiconductor Structures\/}, Book Editor(s): S. Das Sarma and A. Pinczuk
(WILEY-VCH Verlag, Wiley Online Library, 1996).
\bibitem{tong16}
D. Tong,
Lectures on the Quantum Hall Effect,
arXiv:1606.06687v2.
\bibitem{note4}
The corresponding filling factor is $\nu=2/(p+q)$, i.e., $\nu=1$ for $p=q=1$.
\bibitem{note5}
It turns out that this state becomes also the global ground state in all spin sectors for 
$\Omega/\omega \gtrsim 0.77 $.
\bibitem{wilk00.2}
R. A. Smith and N. K. Wilkin,
Exact eigenstates for repulsive bosons in two dimensions,
Phys. Rev. A {\bf 62}, 061602(R) (2000).
\bibitem{pape01}
Th. Papenbrock and G. F. Bertsch,
Rotational spectra of weakly interacting Bose-Einstein condensates,
Phys. Rev. A {\bf 63}, 023616 (2001).
\bibitem{yann10}
C. Yannouleas and U. Landman,
Quantal molecular description and universal aspects of the spectra of bosons and fermions 
in the lowest Landau level,
Phys. Rev. A {\bf 81}, 023609 (2010).
\bibitem{mr91}
G. Moore and N. Read,
Nonabelions in the fractional quantum hall effect,
Nucl. Phys. B {\bf 360}, 362 (1991).
\bibitem{note6}
See Sec.\ III D and Fig. 4 in Ref.\ \cite{yann10}
\bibitem{note7}
For a similar effect of the Pauli exclusion principle regarding the formation of a Wigner molecule
in the case of $N=2$ ultracold fermionic atoms confined in a single quasi-1D well, see the
comparison of 2nd-order correlations between the singlet and triplet states in Sec.\ III A of
B. B. Brandt, C. Yannouleas, and U. Landman, 
Two-point momentum correlations of few ultracold quasi-one-dimensional trapped fermions: 
Diffraction patterns,
Phys. Rev. A {\bf 96}, 053632 (2017); 
compare also Fig.\ 2(e) (singlet) and Fig.\ 2(g) (triplet) in 
B. B. Brandt, C. Yannouleas, and U. Landman,
Double-Well Ultracold-Fermions Computational Microscopy:
Wave-Function Anatomy of Attractive-Pairing 
and Wigner-Molecule Entanglement and Natural Orbitals,
Nano Lett. {\bf 15}, 7105 (2015).
Naturally, unlike the LLL case, a strong repulsion (for separating the particles) is needed for the 
formation of the singlet-state Wigner molecule in a nonrotating trap.  
Without addressing the fermionization mapping, the limiting case (referred to as a Pauli crystal)
of a  Wigner molecule associated with fully polarized ultracold fermions in a static 2D harmonic
trap has been discussed in 
M. Holten, L. Bayha, K. Subramanian, C. Heintze, Ph. M. Preiss, and S. Jochim,
Observation of Pauli Crystals,
arXiv:2005.03929,
and in 
M. Gajda, J. Mostowski, T. Sowi\'{n}ski, and M. Za{\l}uska-Kotur,
Single-shot imaging of trapped Fermi gas,
{\it  EPL} {\bf 115}, 20012 (2016).
\bibitem{note9}
A direct visualization of this trend is seen at the mean-field level, where symmetry-broken wave
functions are obtained as the value of these parameters increases \cite{yann99}.
Restoration \cite{yann07,roma04,cava07} of the broken rotational and spin symmetries via projection
techniques produces symmetry-preserving wave functions that are comparable to the CI many-body ones
\cite{yann07,roma04,cava07}; for the simpler case of two electrons, see C. Yannouleas and U.
Landman, Strongly correlated wavefunctions for artificial atoms and molecules, J. Phys.: Condens.
Matter {\bf 14}, L591 (2002). See also Yuesong Li, C. Yannouleas, and U. Landman,
From a few to many electrons in quantum dots under strong magnetic fields: Properties of 
rotating electron molecules with multiple rings, Phys. Rev. B {\bf 73}, 075301 (2006).
\bibitem{lewe07}
K. Osterloh, N. Barber\'{a}n, and M. Lewenstein,
Strongly Correlated States of Ultracold Rotating Dipolar Fermi Gases,
Phys. Rev. Lett. {\bf 99}, 160403 (2007).
\bibitem{schw04}
V. Schweikhard, I. Coddington, P. Engels, V. P. Mogendorff, and E. A. Cornell,
Rapidly Rotating Bose-Einstein Condensates in and near the Lowest Landau Level, 
Phys. Rev. Lett. {\bf 92}, 040404 (2004).
\bibitem{zwie05}
M. W. Zwierlein, J. R. Abo-Shaeer, A. Schirotzek, C. H. Schunck, and W. Ketterle,
Vortices and superfluidity in a strongly interacting Fermi gas, 
Nature {\bf 435}, 1047 (2005).
\bibitem{clem19}
C. Carcy, H. Cayla, A. Tenart, A. Aspect, M. Mancini, and D. Cl\'{e}ment, 
Momentum-Space Atom Correlations in a Mott Insulator, 
Phys. Rev. X {\bf 9}, 041028 (2019).
\bibitem{deur15}
S. Murmann, F. Deuretzbacher, G. Z\"{u}rn, J. Bjerlin, S. M. Reimann, 
L. Santos, T. Lompe, and S. Jochim,   
Antiferromagnetic Heisenberg Spin Chain of a Few Cold Atoms in  a One-Dimensional Trap,  
Phys. Rev. Lett. {\bf 115}, 215301 (2015).
\bibitem{bran17}
B. B. Brandt, C. Yannouleas, and U. Landman, 
Bottom-up configuration-interaction emulations of ultracold fermions in entangled and
tunnel-coupled two-dimensional optical plaquettes: Building blocks of unconventional
superconductivity, 
Phys. Rev. A {\bf 95}, 043617 (2017).
\bibitem{yann20}
C. Yannouleas and U. Landman,
All-order momentum correlations of three ultracold bosonic atoms confined in triple-well traps: 
Signatures of emergent many-body quantum phase transitions and analogies with three-photon 
quantum-optics interference,
Phys. Rev. A {\bf 101}, 063614 (2020).
\bibitem{note8}
For details, especially concerning the group-theoretical aspects, see Ref.\ \cite{yann03.2}.
\bibitem{cottonbook}
F. A. Cotton,
{\it Chemical Applications of Group Theory\/}
(Wiley, New York, 1990).
\bibitem{wolbarstbook}
A. B. Wolbarst,
{\it Symmetry and Quantum Systems\/}
(Van Nostrand Reinold, New York, 1977).
\end{thebibliography}
\end{document}